%% file: LowReheat_sterile.tex
\documentclass[a4paper,11pt,nofootinbib,superscriptaddress]{revtex4-1}

\usepackage[dvipdfmx]{graphicx}
\usepackage{color}
\usepackage{amsmath,amssymb}
\usepackage{verbatim}
\usepackage{bm}
\usepackage{version}
\usepackage{here}

\numberwithin{equation}{section}

\begin{document}

\title{MeV-scale reheating temperature and cosmological production of\\ light sterile neutrinos}


\author{Takuya Hasegawa}
\affiliation{\small The Graduate University of Advanced Studies (Sokendai), Tsukuba 305-0801, Japan}
\affiliation{\small Theory Center, IPNS, KEK, Tsukuba 305-0801, Japan}
\author{Nagisa Hiroshima}
\affiliation{\small Theory Center, IPNS, KEK, Tsukuba 305-0801, Japan}
\affiliation{\small RIKEN Interdisciplinary Theoretical and Mathematical Sciences (iTHEMS), Wako, Saitama 351-0198, Japan}
\affiliation{\small Department of Physics, University of Toyama, 3190 Gofuku, Toyama 930-8555, Japan}
\author{Kazunori Kohri}
\affiliation{\small The Graduate University of Advanced Studies (Sokendai), Tsukuba 305-0801, Japan}
\affiliation{\small Theory Center, IPNS, KEK, Tsukuba 305-0801, Japan}
\affiliation{\small Kavli IPMU (WPI), UTIAS, The University of Tokyo, Kashiwa, Chiba 277-8583, Japan}
\author{Rasmus S. L. Hansen}
\affiliation{\small Max-Planck-Institut f\,$\ddot{u}$r Kernphysik, Saupfercheckweg 1, 69117 Heidelberg, Germany}
\affiliation{\small Department of Physics and Astronomy, University of Aarhus, Ny Munkegade 120, DK–8000 Aarhus C, Denmark}
\affiliation{\small Niels Bohr International Academy and DARK, Niels Bohr Institute, Blegdamsvej 17, 2100 Copenhagen, Denmark}
\author{Thomas Tram}
\affiliation{\small Department of Physics and Astronomy, University of Aarhus, Ny Munkegade 120, DK–8000 Aarhus C, Denmark}
\affiliation{\small Aarhus Institute of Advanced Studies (AIAS), Aarhus University, DK–8000 Aarhus C, Denmark}
\author{Steen Hannestad}
\affiliation{\small Department of Physics and Astronomy, University of Aarhus, Ny Munkegade 120, DK–8000 Aarhus C, Denmark}


\date{\today}

%
\begin{abstract}
 We investigate how sterile neutrinos with a range of masses influence cosmology 
 in MeV-scale reheating temperature scenarios. By computing the production of sterile
 neutrinos through the combination of mixing and scattering in the early Universe,
 we find that light sterile neutrinos, with masses and mixings as inferred from
 short-baseline neutrino oscillation experiments, are consistent with big-bang
 nucleosynthesis (BBN) and cosmic microwave background (CMB) radiation for the reheating
 temperature of ${\cal O}(1)$~MeV if the parent particle responsible for reheating decays 
 into electromagnetic components (radiative decay). In contrast, if the parent particle 
 mainly decays into hadrons (hadronic decay), the bound from BBN becomes more stringent. 
 In this case, the existence of the light sterile neutrinos can be cosmologically excluded, 
 depending on the mass and the hadronic branching ratio of the parent particle.
\end{abstract}

\maketitle

\include{intro}
\include{formalism}
\include{result_nus}

\include{result_bbn}
\include{result_bbncmb}
\include{conclusion}

\include{acknowledgement}
\bibliography{ref}
\bibliographystyle{jhep}

\end{document}

%% file: intro.tex
\section{Introduction} \label{sec:intro}
The anomaly in short-baseline (SBL) neutrino experiments is a long-standing problem 
in the neutrino sector. Since the LSND collaboration reported a $3.8$-$\sigma$ anomaly 
in their results in the 1990s~\cite{Athanassopoulos_1997}, various experimental projects 
have been performed to investigate the origin. The MiniBooNE collaboration found the 
similar anomaly in both the neutrino and anti-neutrino modes~\cite{AguilarArevalo_2010}, 
and it remains after the update in the experiment~\cite{MiniBooNE_2018}. 
In addition to the accelerator neutrino oscillation experiments, similar anomalies have
been found in other types of experiments, {\it e.g.}  reactor neutrino experiments such 
as Daya Bay~\cite{DayaBay_2016} and Double Chooz~\cite{DoubleChooz_2011}, or Gallium 
experiments such as SAGE~\cite{SAGE_1996,SAGE_1998,SAGE_2005} and GALLEX~\cite{Gallex_1995,Gallex_1997,Gallex_2010}.

The existence of the eV-scale sterile neutrino produced through the mixing
with active neutrinos is a well-motivated scenario to explain the anomaly.
This scenario has been tested in different kinds of experiments. In contrast
to the appearance experiments, disappearance experiments such as MINOS/MINOS+~\cite{MINOS_2017}
and NO$\nu$A~\cite{NOvA_2017} reported the results disfavoring the existence
of such a sterile neutrino. The IceCube collaboration also investigated a 
signature of the conversion from active to sterile neutrinos in the atmospheric 
neutrino spectrum and gave a strong constraint on the parameter space of the 
light sterile neutrino for the SBL neutrino anomaly~\cite{IceCube_2016}. 
The origin of the anomaly is still under debate, and future experimental programs 
such as the SBN experiment~\cite{SBN_2019} and the JSNS$^2$ experiment~\cite{JSNS2_2017} are
expected to unveil the origin.

Cosmological observations are another important probe of sterile neutrinos.
If such light sterile neutrinos exist and have an appreciable mixing with active
neutrinos, they are abundantly produced in the early Universe and affect the 
big-bang nucleosynthesis (BBN) and the cosmic microwave background (CMB) radiation.
In Refs.~\cite{Hamann_2011,Hannestad_2012,Gariazzo_2019,Hagstotz_2020}, it was shown 
that the light sterile neutrino inferred from the SBL anomaly is completely thermalized well-before the 
onset of BBN or the last scattering of CMB. This means the existence of the light 
sterile neutrino is strongly excluded from BBN and CMB. The tension could be 
alleviated by suppressing the thermalization of sterile neutrinos. 
Several scenarios have been proposed as the suppression mechanism: large chemical 
potentials of active neutrinos~\cite{Abazajian_2004,Hannestad_2012,Mirizzi_2012,Saviano_2013},
self-interaction or non-standard interaction of sterile
neutrinos~\cite{Hannestad_2013_2,Saviano_2014,Archidiacono_2014,Archidiacono_2016,Chu_2015,Chu_2018}, 
or low reheating temperature of the 
Universe~\cite{Gelmini_2004,Gelmini_2008,Gelmini_2019,Gelmini_2019b,Yaguna_2007,Abazajian_2017}.

In this paper, we focus on the thermalization of sterile neutrinos 
in the Universe with an MeV-scale reheating temperature to solve the 
tension between the light sterile neutrino and cosmology.
Since the sterile neutrino production through the weak interaction of 
active neutrinos effectively finishes when the cosmic temperature 
becomes $\sim~{\cal O}$(1)~MeV, the MeV-scale reheating temperature 
leads to the incomplete thermalization of sterile neutrinos, 
which offers the solution to the problem.

The reheating temperature of the Universe is much lower than that of 
the standard cosmology if there exists a long-lived massive particle 
and it causes reheating of the Universe. The existence of such 
particles is naturally expected in varieties of extensions of the 
standard model of particle physics. For example, curvaton, gravitino, 
flaton, modulus, or dilaton are well-motivated candidates for this particle. 
If such a long-lived particle dominates the energy density in the early epoch, 
the Universe experiences the early matter-dominated era before the ordinary 
radiation-dominated epoch, which modifies the initial condition of the 
standard cosmology. 

This solution to the tension has been proposed in Ref.~\cite{Gelmini_2004}, 
and Refs.~\cite{Gelmini_2019,Gelmini_2019b,Yaguna_2007} later revisited the 
same scenario. In order to probe this scenario, it is necessary to simultaneously solve  
reheating of the Universe and the thermalization of neutrinos to accurately 
compute the abundance of sterile neutrinos. This is because 
most of the active and sterile neutrinos are produced during reheating, 
and matter effects on the thermalization of sterile neutrinos, cannot be 
neglected. However, such a computation is technically difficult, and 
Refs.~\cite{Gelmini_2004,Gelmini_2019,Gelmini_2019b,Yaguna_2007} assumed a 
simplified picture where sterile neutrinos are produced via vacuum oscillations 
after the completion of reheating. Also, the reheating temperature is 
fixed to be 5~MeV by hand in the studies. If sterile neutrinos are completely 
absent from the thermal bath, the lower bound on the reheating temperature 
is known to be almost 5~MeV~\cite{Salas_2015,Hasegawa_2019a}, but this is 
not true if sterile neutrinos exist, and it contributes to the energy 
density of the Universe. Therefore, we should not fix the reheating temperature 
to the typical value in advance. 

Ref.~\cite{Yaguna_2007} later updated the sterile neutrino production 
in the MeV-scale reheating scenario by calculating the semi-classical Boltzmann equation 
with effective collision terms, which include the matter effects, and provided a more 
detailed analysis of the thermalization of sterile neutrinos. It is however necessary 
to calculate the original quantum kinetic equation (QKE) instead of the semi-classical 
Boltzmann equation, to correctly follow the sterile neutrino thermalization unless the 
off-diagonal components of the collision term for neutrinos ({\it i.e.} the collisional 
damping term) dominate those of the neutrino Hamiltonian~\cite{Bell_1999,Venumadhav_2016}. 

The purpose of this study is to revisit the sterile neutrino thermalization in the cosmological model with 
an MeV-scale reheating temperature and refine the cosmological constraint on sterile 
neutrinos obtained in the previous studies~\cite{Gelmini_2004,Gelmini_2019,Gelmini_2019b} 
by performing a detailed computation of QKE and BBN. The main focus of this study is the 
eV-scale sterile neutrinos, motivated by the SBL neutrino anomaly.

The structure of this paper is as follows. In Sec.~\ref{sec:dynamics} we introduce our 
formulation for calculating the production of active and sterile neutrinos during reheating. 
In Secs.~\ref{sec:results_reheating} and \ref{sec:results_bbn} we show our numerical results 
of the neutrino thermalization and BBN, respectively. In Sec.~\ref{sec:constraint} we summarize 
the constraint on sterile neutrinos obtained from cosmological observations and ground-based 
experiments. Sec.~V\hspace{-.1em}I is devoted to the conclusion. 

%% file: formalism.tex
\section{Sterile neutrino production during reheating} \label{sec:dynamics}

In this section, we explain the dynamics of cosmological models with late-time entropy 
production, which results in the MeV-scale reheating temperature. Also, we introduce 
key equations for calculating the production of sterile neutrinos during reheating.

We assume that a long-lived massive particle $\phi$ is responsible for reheating. 
In this case, the decay of $\phi$ induces the late-time entropy production and the 
subsequent dramatic particle production of the standard-model particles. 
Photons and charged leptons are rapidly thermalized through the electromagnetic interaction 
during reheating, while active neutrinos are slowly produced through 
the weak interaction. If the reheating temperature of the Universe is lower than 
the QCD scale $\sim 100$~MeV and the radiation-dominated epoch therefore realizes 
after the hadronization, active neutrinos are solely produced in the annihilation 
process of charged leptons $l + \bar{l} \rightarrow \nu_\alpha + \bar{\nu}_\alpha$ ($\alpha = e, \mu, \tau$), 
where $l$ and $\bar{l}$ denote the charged leptons and corresponding anti-particles, 
respectively. Sterile neutrinos are generated from active neutrinos through the 
flavor mixing as reheating proceeds. Therefore, we need to consider both 
neutrino collisions and neutrino oscillations in the thermalization calculations 
of active and sterile neutrinos.~\footnote{
  Even if a primordial component of sterile neutrinos exists before 
  reheating, such a component is completely diluted by the entropy 
  production associated with reheating. Also, we do not consider 
  any other exotic interactions among the standard-model particles 
  and sterile neutrinos. Therefore, sterile neutrinos are produced 
  only through the active-sterile neutrino oscillation.
} 
Ref.~\cite{Dodelson_1993} provided an analytical expression to estimate the sterile 
neutrino abundance produced in the non-resonant active-sterile mixing. The production 
rate of sterile neutrinos has a sharp peak at temperature $T_{\rm max}$:
\begin{equation}
 T_{\rm max} \sim 13\,{\rm MeV} 
\left(\frac{m_{\rm s}}{1\,{\rm eV}} \right)^{1/3},\label{eq:tmax}
\end{equation}
where $m_{\rm s}$ is the mass of the sterile neutrino. Hence, the abundance of 
sterile neutrinos is strongly suppressed compared to those obtained in the 
standard cosmology case if the reheating temperature is lower than $T_{\rm max}$.

The states of active and sterile neutrinos are expressed in terms of a one-body-irreducible
density matrix, which is expressed in an $N_f \times N_f$ Hermitian matrix, where
$N_f$ is the number of neutrino flavors to mix. In this study, we adopt the so-called 
1+1 mixing scheme in which one sterile neutrino species mixes with one active species.
This approximation is reasonable when the mixing of the sterile neutrino with one 
active species dominates the mixing with the other active species. 
For the current calculation, we assume sterile neutrinos to mix with electron neutrinos, 
and we assume that $\mu$ neutrinos ($\nu_\mu$) and $\tau$ neutrinos ($\nu_\tau$) decouple 
from the neutrino oscillations.
Under these assumptions, the states of spectator neutrinos, namely $\nu_\mu$ and $\nu_\tau$, 
are degenerate, and it is unnecessary to separately calculate dynamical equations for each.
This is because the cosmic temperature is always below ${\cal O}$(1)~MeV after 
reheating for $T_{\rm RH} \sim {\cal O}(1)$~MeV, and muons and $\tau$ leptons, which 
are heavier than the cosmic temperature, do not exist in the thermal bath of the Universe. 
In the following, quantities of the spectator neutrinos are multiplied by a factor of two 
for summing up contributions from $\nu_\mu$ and $\nu_\tau$.

Since we assume the 1+1 mixing, the density matrix of neutrinos with energy $E$ 
can be expressed as a 2$\times$2 matrix: 
\begin{equation}
 \varrho_{\bm p}(t) \equiv \varrho (E, t) = \left(
\begin{array}{cc}
 \varrho_{\rm aa} & \varrho_{\rm as} \\
 \varrho^*_{\rm as} & \varrho_{\rm ss}
\end{array} \right)\,.\label{eq:densmatrix}
\end{equation}
The energy of active and sterile neutrinos $E$ is replaced with their 
absolute momentum $p$, {\it i.e.} $E \rightarrow p \equiv |\bm{p}|$, where 
$\bm{p}$ is the three-momenta of neutrinos. This is because masses of the 
active neutrinos are known to be sub-eV scale~\cite{PDG2018} and safely neglected 
in a thermal bath of $T \sim {\cal O}(1)$ MeV. In addition, we restrict 
ourselves to the mass range of sterile neutrinos below $10$~keV so that 
they are always relativistic before their production effectively finishes 
at around a temperature of the neutrino decoupling $T \sim T_{\rm dec}$.~\footnote{
  This limitation is mandatory because non-relativistic neutrinos do not 
  oscillate into another flavor~\cite{Akhmedov_2017}, and we cannot rely 
  on the QKE for neutrinos (Eq.~\eqref{eq:QKE}) in such cases. 
}
In Eq.~\eqref{eq:densmatrix}, the diagonal elements of the density matrix 
correspond to the distribution functions of active and sterile neutrinos, 
{\it i.e.} $\varrho_{\rm aa} = f_{\rm a}$ and $\varrho_{\rm ss} = f_{\rm s}$, 
while the off-diagonal elements correspond to a quantum coherence between them. 

The time evolution of the density matrix is governed by the momentum-dependent 
quantum kinetic equation (QKE) in the following~\cite{McKellar_1994,Sigl_1993}:
\begin{equation}
 \frac{d\varrho_{\bm p}(t)}{dt} = 
  \left(\frac{\partial}{\partial t} - H\, p 
  \frac{\partial}{\partial p}\right) \varrho_{\bm p} (t)
 = -i\,[{\mathcal H_\nu},\varrho_{\bm p}(t)] + C[\varrho_{\bm p}(t),t]\,,
\label{eq:QKE}
\end{equation}
where $C[\varrho_{\bm p}(t),t]$ is the collision term for the active-mixed neutrinos, 
$H$ is the Hubble parameter, and ${\mathcal H_\nu}$ in the commutator is the neutrino 
Hamiltonian.

In this study, we neglect the neutrino chemical potentials.~\footnote{
This is a reasonable assumption since its effect on the neutrino oscillation 
can be safely neglected for the chemical potentials of ${\cal O}(10^{-10})$, 
as is naturally attained in the standard mechanism of baryogenesis associated 
with the sphaleron process. For those interested in the effect of the neutrino 
chemical potentials on the sterile neutrino thermalization, see {\it e.g.} 
Refs.~\cite{Shi_1998,Hannestad_2012,Hannestad_2013,Mirizzi_2012,Saviano_2013}.
}
Then, it is unnecessary to follow the time evolution of anti-neutrinos separately 
from corresponding neutrinos, and the neutrino Hamiltonian on the right-hand side 
of Eq.~\eqref{eq:QKE} is reduced to
\begin{equation}
 {\mathcal H_\nu} = \frac{{\sf M}^2}{2p} -\frac{8\sqrt{2}\,G_{\rm F}p}{3} 
  \left[ \frac{\bm E_{\rm CC}}{m^2_W} + \frac{{\bm E}_{\rm NC}}{m^2_Z} \right] \,,
\label{eq:nuhamil}
\end{equation}
where $G_F$ is the Fermi coupling constant, and $m_W$ ($m_Z$) is the mass of 
$W$ ($Z$) boson. On the right-hand side of Eq.~\eqref{eq:nuhamil}, the first term corresponds to
the vacuum oscillation of neutrinos. 
The mass matrix ${\sf M}$ in the flavor basis is related to that in the mass basis
$\mathcal M$ as ${\sf M^2} = U {\mathcal M^2} U^\dagger$ where $U$ is the
flavor-mixing matrix. The mass matrix ${\mathcal M}$ for the 1+1 mixing 
has the explicit form of 
\begin{equation}
{\mathcal M^2} = \left(
\begin{array}{cc}
m_1^2 & 0 \\
0 & m_2^2
\end{array} \right),\ \ 
  U = \left(
\begin{array}{cc}
\cos\theta & -\sin\theta \\
\sin\theta & \cos\theta
\end{array} \right),
\end{equation}
where $m_1$ and $m_2\,(> m_1)$ are the mass eigenvalues for active 
and sterile neutrinos, whereas $\theta$ is the active-sterile mixing 
angle in a vacuum. Throughout this paper, we consider the normal mass ordering for sterile 
neutrinos $m_2 > m_1$, which is favored in cosmological observations~\cite{Planck_2018}. 
The flavor-mixing matrix $U$ uniquely determines the relation between 
the mass and the flavor eigenstates as 
\begin{eqnarray}
|\nu_{\rm a}\rangle &=& \cos\theta\,|\nu_1\rangle-\sin\theta\,|\nu_2\rangle \,,\\
|\nu_{\rm s}\rangle &=& \sin\theta\,|\nu_1\rangle+\cos\theta\,|\nu_2\rangle \,,
\end{eqnarray}
where $|\nu_{\rm a}\rangle$ and $|\nu_{\rm s}\rangle$ are flavor eigenstates 
of active and sterile neutrinos, while $|\nu_1\rangle$ and $|\nu_2\rangle$ 
are the mass eigenstates of lighter and heavier states, respectively. The 
second and third terms in Eq.~\eqref{eq:nuhamil} correspond to the matter 
effects induced by the coherent scatterings of the active-mixed neutrinos with electrons 
$\nu_{\rm a} + e^\pm \rightarrow \nu_{\rm a} + e^\pm$. The matter effect modifies 
the relation between the mass and the flavor eigenstates. Particularly, the 
second (third) term arises from the charged- (neutral-) current interaction 
of $\nu_{\rm a}\ (= \nu_e)$ with electrons, where ${\bm E_{\rm CC}} \equiv {\rm diag}(\rho_e, 0)$ 
and ${\bm E_{\rm NC}} \equiv {\rm diag}(\rho_{\nu_{\rm a}}, 0)$ with $\rho_e$ 
and $\rho_{\nu_{\rm a}}$ the energy densities of electrons and the 
active-mixed neutrinos, respectively.~\footnote{
  In a thermal bath of $T \sim {\cal O}(1)$~MeV, abundances of muons and 
  $\tau$ leptons are much smaller than that of electrons due to the Boltzmann 
  suppression. Therefore, we do not consider contributions from muons or 
  $\tau$ leptons to the scatterings.
}   

The collision term of the QKE, Eq.~\eqref{eq:QKE}, is written as
\begin{eqnarray}
C[\varrho_{\bm p}(t),t] = \left(
\begin{array}{cc}
R_{\nu_{\rm a}} & -D \varrho_{\rm as}\\\
-D \varrho^*_{\rm as} & 0
\end{array}
\right)\,, 
\end{eqnarray}
where $R_{\nu_{\rm a}}$ is the production rate of the active-mixed neutrinos,
and $D$ is the collisional-damping factor, which gives the decoherence 
between states of $\nu_{\rm a}$ and $\nu_{\rm s}$. We take into account 
the production of active neutrinos from the electron-pair annihilation, 
the neutrino-electron scattering, and the neutrino self-interaction. 
These processes are summarized in Table~I of Ref.~\cite{Hannestad_2015}. 
For each reaction process, we analytically reduce the dimension of momentum 
integrals from nine to two, without imposing any simplifying assumptions in 
the same way as in Ref.~\cite{Hannestad_2015}.

For numerical implementations, we expand the density matrix with Pauli matrices 
${\bm \sigma} = (\sigma_x, \sigma_y, \sigma_z)$ and convert the QKE 
into a set of scalar equations:
\begin{equation}
 \varrho_{\bm p}(t) = \left(
\begin{array}{cc}
 \varrho_{\rm aa} & \varrho_{\rm as} \\
 \varrho^*_{\rm as} & \varrho_{\rm ss}
\end{array} \right)
 = \frac{1}{2}\,[\, P_0\,\sigma_0 + \boldsymbol{P}\,\cdot {\bm \sigma}\,] \,.
\end{equation}
where $P_0$ and $\boldsymbol{P} = (P_x,P_y,P_z)$ are expansion coefficients 
referred to as the polarization vectors and $\sigma_0=\bm{1}$ is the identity matrix. 
Since the diagonal components of the density matrix correspond to the distribution 
functions for the active-mixed and sterile neutrinos, we have
\begin{equation}
 f_{\nu_{\rm a}} = \frac{1}{2}(P_0 + P_z),\ f_{\nu_{\rm s}} = \frac{1}{2}(P_0 - P_z) \,.
\end{equation}
The QKE~(Eq.~\eqref{eq:QKE}) is rewritten with polarization vectors as  
\begin{eqnarray}
 \dot{\boldsymbol{P}} &=& \overrightarrow{\mathcal{H}} \times \boldsymbol{P} 
  - D \,(P_x\, \mathbf{x}+P_y\, \mathbf{y})+\dot{P}_0\,\mathbf{z} \,,
  \label{eq:QKE_pola}\\
 \dot{P}_0 &=& R_{\nu_{\rm a}} \,,
  \label{eq:QKE_pola2}
\end{eqnarray}
where $\overrightarrow{\mathcal{H}} = (\mathcal{H}_x, \mathcal{H}_y, \mathcal{H}_z)$
is the neutrino Hamiltonian. We define $P_{\nu_a} \equiv P_0 + P_z$ and $P_{\nu_{\rm s}} \equiv P_0 - P_z$ 
and rewrite the above equation into 
\begin{eqnarray}
 \dot{P}_{\nu_{\rm a}} &=& \mathcal{H}_x\, P_y + R_{\nu_{\rm a}} \,, \label{eq:QKE_pola_P_nu_e}\\
 \dot{P}_{\nu_{\rm s}} &=& - \mathcal{H}_x\, P_y \,, \label{eq:QKE_pola_P_nu_x}\\
 \dot{P_x} &=&  - \mathcal{H}_z\, P_y - D \,P_x\,, \label{eq:QKE_pola_Px}\\
 \dot{P_y} &=&    \mathcal{H}_z\, P_x - \frac{1}{2}\,\mathcal{H}_x\, (P_{\nu_{\rm a}} 
  - P_{\nu_{\rm s}}) - D \,P_y\,.\label{eq:QKE_pola_Py}
\end{eqnarray}
Given the squared-mass difference between the mass eigenstates 
$\delta m^2 \equiv m_2^2 - m_1^2$ and the mixing angle in a vacuum $\theta$, 
each component of the neutrino Hamiltonian is explicitly expressed as 
\begin{eqnarray}
 \mathcal{H}_x &=& \frac{\delta m^2}{2 p} \sin 2 \theta \label{eq:Hx}\,,\\
 \mathcal{H}_y &=& 0 \label{eq:Hy}\,,\\
 \mathcal{H}_z &=& - \frac{\delta m^2}{2 p} \cos 2 \theta + \mathcal{H}_{\rm mat} \label{eq:Hz}\,.
\end{eqnarray}
The matter effect appears as the potential term $\mathcal{H}_{\rm mat}$, which can be written as 
\begin{eqnarray}
 \mathcal{H}_{\rm mat} &=& - \frac{8\sqrt{2}}{3} G_{\rm F}\, p\left[ \frac{\rho_e}{m_W^2} 
						+ \frac{\rho_{\nu_{\rm a}}}{m_Z^2} \right], \nonumber\\ 
 &=& - \frac{4\sqrt{2}}{3\,\pi^2} G_{\rm F}\, p \left[ \frac{g_e}{m_W^2} \int_0^\infty dp'\, p'^2 
     \frac{E_e}{\exp(E_e/T_\gamma)+1} + \frac{g_\nu}{m_Z^2}\int^{\infty}_{0} dp'\,p'^3 f_{\nu_{\rm a}} \right].
\end{eqnarray}
In the above expression, $T_\gamma$ is the photon temperature, 
and $E_e = \sqrt{p^2 + m_e^2}$ is the energy of electrons. 
Also, $g_e = 4$ is the statistical degree of freedom of electrons 
and $g_\nu = 2$ is that for each flavor of neutrinos. The first 
and second terms in the bracket correspond to the charged- and 
neutral-current interactions of $\nu_e$ with electrons, respectively. 

The active-spectator neutrinos $\nu_{\rm sp}$ are irrelevant to the 
neutrino oscillation. Therefore, their time evolution can be described 
by the momentum-dependent classical Boltzmann equation:
\begin{equation}
 \frac{d f_{\nu_{\rm sp}}(t)}{d t} = \left(\frac{\partial}{\partial t} 
  - H\, p \frac{\partial}{\partial p}\right)f_{\nu_{\rm sp}}(t) = C[f_{\nu_{\rm sp}}(t),t] \,, 
\label{eq:spBoltzmann}
\end{equation}
where $f_{\nu_{\rm sp}}$ is the distribution function of the active-spectator neutrino, 
and $C[f_{\nu_{\rm sp}}(t),t]$ is the collision term for $\nu_{\rm sp}$, whose expression 
is given by the same equation as the production rate for $\nu_{\rm a}$, but with $f_{\nu_{\rm sp}}$.

In order to calculate the thermalization of active and sterile neutrinos in 
the expanding Universe, it is necessary to solve the Friedman equation,  
\begin{equation}
 H \equiv \frac{\dot{a}}{a} = \sqrt{\frac{8\pi G\rho}{3}}\,,\label{eq:Friedmann}
\end{equation}
to give the time evolution of the scale factor $a(t)$. The total energy density 
$\rho$ is written as
\begin{eqnarray} 
 \rho &=& \rho_\gamma + \rho_e + \rho_\nu + \rho_\phi \ \nonumber \\
 &=& \frac{\pi^2}{15}T_\gamma^4 + \frac{g_e}{2\pi^2} \int_0^\infty dp \,p^2 
  \frac{E_e}{\exp(E_e/T_\gamma)+1} \nonumber \\ && \ \ \ \ \ \ \ \ \ \ + 
  \frac{g_\nu}{2\pi^2}\int^{\infty}_{0} dp\,p^3 (f_{\nu_{\rm a}} + 2 f_{\nu_{\rm sp}} 
  + f_{\nu_{\rm s}}) + \rho_\phi.
  \label{eq:tot_ene}
\end{eqnarray}
In the above expression, $\rho_\gamma$, $\rho_e$, $\rho_\nu$, and $\rho_\phi$ 
are the energy densities of photons, electrons, neutrinos, and the parent particle, respectively. 
All flavors of neutrinos contribute to the total energy density of neutrinos, 
{\it i.e.} $\rho_\nu = \rho_{\nu_a} + \rho_{\nu_{\rm sp}} + \rho_{\nu_{\rm s}}$. 

The evolution of $\rho_\phi$ can be obtained by solving the integrated
Boltzmann equation for $\phi$:
\begin{equation}
 \frac{d\rho_\phi}{dt} = -\Gamma_\phi \rho_\phi - 3H \rho_\phi \,,
\end{equation}
where $\Gamma_\phi$ is the decay rate of $\phi$, and the lifetime of $\phi$ 
is given by its inverse, {\it i.e.} $\tau_\phi = \Gamma_\phi^{-1}$. 
This equation can be integrated analytically for the non-relativistic 
particle~$\phi$, and we obtain
\begin{equation}
 \frac{\rho_\phi}{s} = \frac{\rho_{\phi,0}}{s_0}\ e^{-\Gamma_\phi t}\,,
\label{eq:PhiBoltzmann}
\end{equation}
where $\rho_{\phi,0}$ and $s_0$ are the energy density of $\phi$ and 
the total entropy density at the initial time $t_0$, respectively. 
In Eq.~\eqref{eq:PhiBoltzmann}, we have assumed $\rho_{\phi,0}$ 
dominates the energy densities of other background particles, 
{\it i.e.} $\rho_{\phi,0} >> (\rho_\gamma + \rho_e + \rho_\nu)_{t=t_0}$.

The energy and entropy injected from the decay of the parent particle $\phi$ during 
reheating are taken into account by solving the energy conservation equation:  
\begin{equation}
\frac { d \rho } { d t } = - 3 H ( \rho + P ) \,.
\label{eq:eneconserv}
\end{equation}
The total pressure $P$ can be expressed as 
\begin{eqnarray}
 P &=& P_\gamma + P_e + P_\nu  \nonumber \\
 &=& \frac{\pi^2}{45}T_\gamma^4 + \frac{g_e}{6\pi^2} \int_0^\infty dp \, 
  \frac{p^4}{E_e}\frac{1}{\exp(E_e/T_\gamma)+1} \nonumber \\ && \ \ \ \ \ \ \ \ \ \ \ \ \ \ 
  + \frac{g_\nu}{6\pi^2}\int^{\infty}_{0} dp\,p^3 
  (f_{\nu_{\rm a}} + 2 f_{\nu_{\rm sp}} + f_{\nu_{\rm s}}) \,.
\end{eqnarray}
Since all electromagnetic particles are instantaneously thermalized
during reheating, they have a common temperature
$T_\gamma$. Hence, we can rewrite Eq.~(2.26) into the 
differential equation for the time evolution of $T_\gamma$:
\begin{equation}
 \frac{dT_\gamma}{dt} =
  - \frac{-\Gamma_\phi\rho_\phi+4H(\rho_\gamma+\rho_\nu) + 3H(\rho_e+P_e) + 
  \frac{d\rho_\nu}{dt}}{\frac{\partial \rho_\gamma}{\partial T_\gamma} |_{a(t)} 
  + \frac{\partial \rho_e}{\partial T_\gamma} |_{a(t)}} \,,
\label{eq:tgamevol}
\end{equation}
where $\Gamma_\phi$ and $T_{\rm RH}$ are uniquely related through the relation: 
\begin{equation}
 \Gamma_\phi = 3\, H(T_{\rm RH})\,. 
  \label{eq:taurh_def}
\end{equation}
The energy density is dominated by radiation after reheating, and hence 
the Hubble expansion rate can be written as 
\begin{equation}
 H = \sqrt{\frac{g^* \pi^2}{90}} \frac{T_{\rm RH}^2}{m_{\rm pl}}\,, 
  \label{eq:hubbletrh}
\end{equation}
where $m_{\rm pl} \sim 2.4 \times 10^{18}$~GeV is the reduced Planck mass, 
and $g^* = 10.75$ is the canonical value of the relativistic degrees of 
freedom at the cosmic temperature of ${\cal O}$(1)~MeV. Substituting Eq.~(2.30) into Eq.~(2.29) 
yields the one-to-one correspondence between 
the reheating temperature and the lifetime of the parent particle $\phi$:
\begin{equation}
 T_{\rm RH} \sim 0.7~{\rm MeV} \,\left(\frac{\Gamma_\phi}{\rm sec^{-1}}\right)^{1/2}\,.
\label{eq:trh_rela} 
\end{equation}
This gives a reasonable estimate of the cosmic temperature at which 
the radiation-dominated epoch attains. 

The results of the thermalization of active and sterile neutrinos are obtained 
by simultaneously solving Eqs.~\eqref{eq:QKE_pola_P_nu_e}--\eqref{eq:QKE_pola_Py}, 
\eqref{eq:spBoltzmann}, \eqref{eq:Friedmann}, \eqref{eq:PhiBoltzmann}, (2.28). 
For this purpose, we use the LASAGNA code~\cite{Hannestad_2012,Hannestad_2013}, 
which is an efficient ordinary differential equation solver optimized for the sterile
neutrino production in the early Universe, with suitable modifications. Also, we utilize 
the SuperLU MT package~\cite{Li_2005,Demmel_1999} to make use of the multicore CPU servers 
for numerical performance.

%% file: result_nus.tex
\section{Numerical result: sterile neutrino thermalization} \label{sec:results_reheating}

The abundance of neutrinos is often described in terms of the effective number 
of neutrino species $N_{\rm eff}$. In the case of $\nu_e$--$\nu_{\rm s}$ mixing, it is defined as 
\begin{equation}
 N_{\rm eff} = N_{{\rm eff},\, \nu_{\rm a}} + N_{{\rm eff},\, \nu_{\rm s}} + 2\,N_{{\rm eff},\, \nu_{\rm sp}} 
 = \rho_{\nu_{\rm a}}/\rho_{\nu_{\rm std}} + \rho_{\nu_{\rm s}}/\rho_{\nu_{\rm std}} 
 + 2\,\rho_{\nu_{\rm sp}}/\rho_{\nu_{\rm std}} \,,
 \label{eq:neff_def}
\end{equation}
where $N_{{\rm eff},\,\nu_\alpha}$ ($\alpha = e,\,{\rm s},\,{\rm sp}$)
is the contribution of each neutrino species to the total
$N_{\rm eff}$, and $\rho_{\nu_{\rm std}}$ is the energy density of one
species of neutrinos in the standard big-bang model.~\footnote{
  Here we normalize the contribution of $\nu_{\rm \alpha}$ ($\alpha = e, \mu, \tau$) 
  to the effective number of neutrino species $N_{{\rm eff},\, \nu_\alpha}$ in units 
  of the energy density of $\nu_e$ in the standard big-bang model, 
  $\rho_{\nu_{e,\,{\rm std}}}$. Another choice for the normalization is to take the 
  standard energy density of $\nu_\mu$ or $\nu_\tau$, but the difference between
  $\rho_{\nu_\alpha}/\rho_{\nu_{e,\,{\rm std}}}$ and 
  $\rho_{\nu_\alpha}/\rho_{\nu_{\mu,\,{\rm std}}}$ 
  (or $\rho_{\nu_\alpha}/\rho_{\nu_{\tau,\,{\rm std}}}$) should be quite
  small ($< 1\%$), and hence negligible. The difference in $N_{\rm eff}$ due to 
  its definition is therefore irrelevant to our final results.
} 
The factor of two in front of $N_{{\rm eff},\, \nu_{\rm sp}}$ accounts 
for the contribution from $\nu_\mu$ and $\nu_\tau$. By definition, 
$N_{{\rm eff},\,\nu_\alpha} = 1$ corresponds to the full thermalization 
of $\nu_\alpha$, {\it i.e.} the energy spectrum of $\nu_\alpha$ 
can be expressed as the thermal (Fermi-Dirac) distribution.

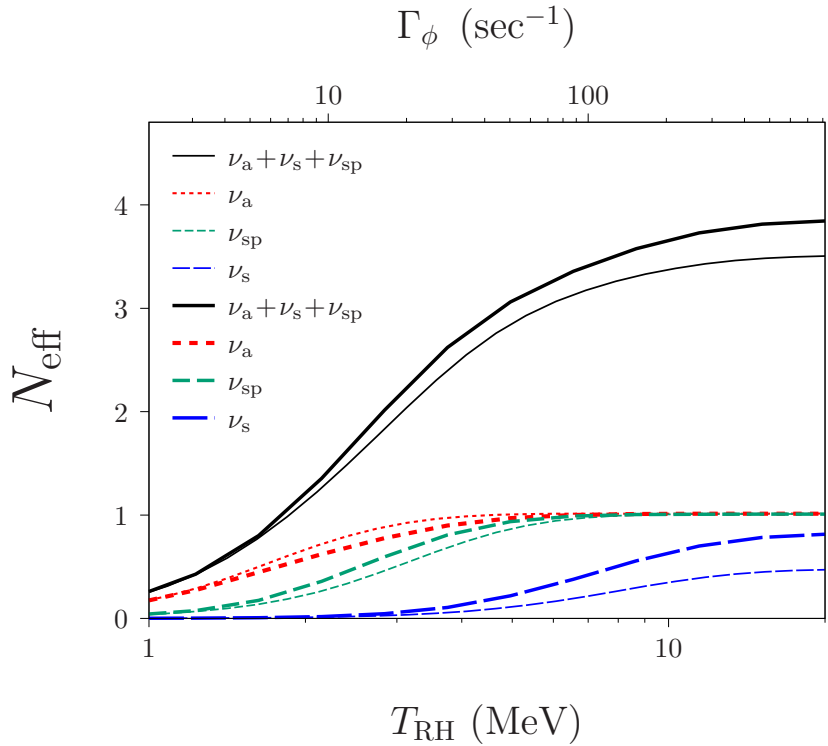
\begin{figure}[!t]
\begin{center}
\vspace{-1.2cm}
\resizebox{0.90\linewidth}{!}{\hspace{1.5cm}\input{Trh_vs_Neff}}
\end{center}
\vspace{1.0cm}
 \caption{
 Effective number of neutrino species
 $N_{\rm eff} = N_{{\rm eff},\,\nu_{\rm a}} + N_{{\rm eff},\,\nu_{\rm s}} 
 + 2\,N_{{\rm eff},\,\nu_{\rm sp}}$ 
 as a function of the reheating temperature $T_{\rm RH}$, for the case of $\nu_e$--$\nu_{\rm s}$ mixing.
 The narrow (bold) line corresponds to the case without (with) the neutrino self-interaction. 
 The black line is for $N_{\rm eff}$, whereas the red, green, and blue lines are for 
 $N_{{\rm eff},\,\nu_{\rm a}}$, $N_{{\rm eff},\,\nu_{\rm sp}}$, 
 and $N_{{\rm eff},\,\nu_{\rm s}}$, respectively. 
 } \label{fig:trh_vs_neff}
\end{figure}

Figure~\ref{fig:trh_vs_neff} shows the relation between the reheating temperature 
$T_{\rm RH}$ and the effective number of the neutrino species $N_{\rm eff}$. 
The mass difference $\delta m^2$ and the mixing angle $\theta$ between the 
active and sterile neutrinos are fixed to 
$(\delta m^2,\sin^2 2\theta) = (1.29~{\rm eV}^2\,,\ 0.035)$, which is the 
best-fit value obtained from data analysis of $\nu_e$ disappearance experiments 
in Ref.~\cite{Dentler_2018}. 
In the figure, both $N_{\rm eff}$ and $N_{{\rm eff},\,\nu_\alpha}$ increase with
$T_{\rm RH}$ since the active neutrino production via
$e^- + e^+ \rightarrow \nu_\alpha + \bar{\nu}_\alpha$ is more efficient
at a higher temperature, and neutrinos have more time to be produced during 
reheating. The abundance of sterile neutrinos decreases for 
$T_{\rm RH} < T_{\rm max}$ ($\sim 13$~MeV for $m_s
\sim 1$~eV) and vanishes at $T_{\rm RH} < 1$~MeV due to the decoupling 
of active neutrinos from the thermal plasma.
An inequality $N_{{\rm eff},\, \nu_{\rm a}} > N_{{\rm eff},\, \nu_{\rm sp}}$
always holds because the neutrino interaction with the background electrons 
is stronger for $\nu_e$ than $\nu_\mu$ or $\nu_\tau$. The figure also reveals 
that the neutrino self-interaction enhances the production efficiency of the 
sterile neutrino. This is because the neutrino self-interaction increases the 
collisional-damping rate $D$ and increase the effective production rate of 
sterile neutrinos, which approximately scales as $D\,\sin^2 2\theta_{\rm M}$, 
where $\theta_{\rm M}$ is the mixing angle in a medium~\cite{Kainulainen_1990}.
The effect of the neutrino self-interaction was neglected in Ref.~\cite{Yaguna_2007}, 
or approximately considered in Refs.~\cite{Gelmini_2004,Gelmini_2019,Gelmini_2019b}. 

\begin{figure}[!t]
\vspace{0.8cm}
\begin{center}
\vspace{-4.0cm}
\resizebox{0.95\linewidth}{!}{\hspace{1.5cm}\input{Neff_evolution_vs_tgam}}
\end{center}
\vspace{0.7cm}
 \caption{
 Temperature evolution of $N_{{\rm eff},\,{\nu_\alpha}}$ ($\alpha={\rm a,s,sp}$), 
 for the case of $\nu_e$--$\nu_{\rm s}$ mixing. 
 The left panel is for $N_{{\rm eff},\,\nu_{\rm a}}$, the middle panel is 
 for $N_{{\rm eff},\,\nu_{\rm sp}}$, and the right panel is for $N_{{\rm eff},\,\nu_{\rm s}}$. 
 In each panel, the red long-dashed line is for $T_{\rm RH} = 1$~MeV, the blue 
 middle-dashed line is for $T_{\rm RH} = 2$~MeV, the green short-dashed line is 
 for $T_{\rm RH} = 5$~MeV, and the black solid line is for $T_{\rm RH} = 10$~MeV. 
 } \label{fig:neff_evolution}
\end{figure}
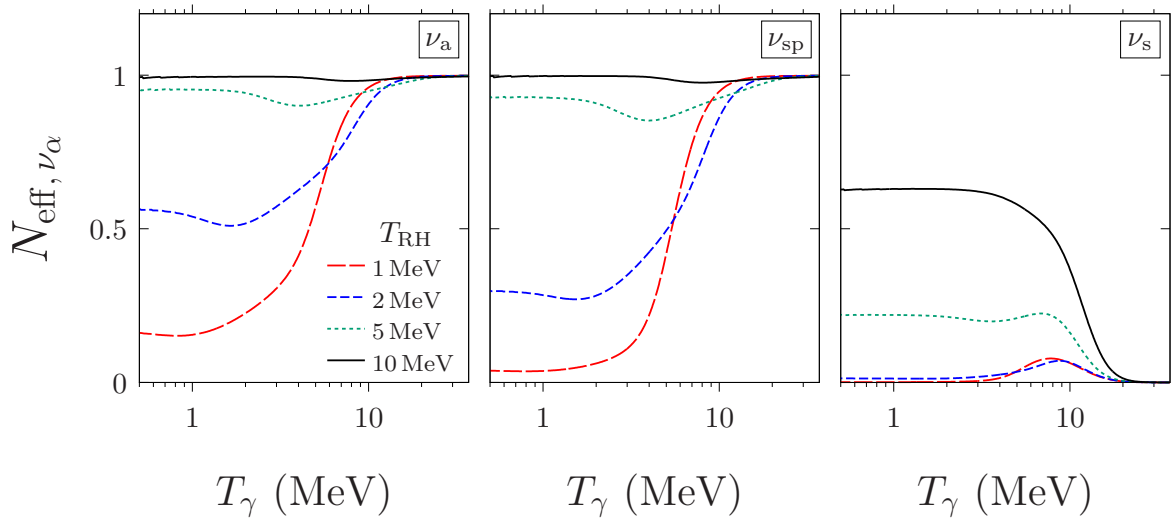

Figure~\ref{fig:neff_evolution} shows the time evolution of $N_{{\rm eff},\,\nu_\alpha}$ 
for typical values of $T_{\rm RH}$. The competition between the dilution of neutrinos due 
to the entropy production induced by the decay of $\phi$ and the production of 
neutrinos determines the behavior of $N_{{\rm eff},\,\nu_\alpha}$. Since neutrinos 
are only weakly produced in the thermal bath of photons and electrons, the former 
is dominant for $T_\gamma > T_{\rm RH}$, and the effect takes its maximum at 
$T_\gamma \sim T_{\rm RH}$, when the cosmic time is comparable to the lifetime of $\phi$. 
This leads the local minimum of $N_{{\rm eff},\,\nu_\alpha}$ in Fig.~\ref{fig:neff_evolution}. 
For $T_\gamma < T_{\rm RH}$, neutrino production becomes dominant compared to the 
dilution effect, which is negligible after the decay of $\phi$. 
The value of $N_{{\rm eff},\,\nu_\alpha}$ increases for $T_\gamma < T_{\rm RH}$ 
until neutrinos decouple from other particles at around $T_\gamma \sim 1$~MeV. 
This explains the behavior of the time evolution of $N_{\rm eff}$. Also, 
Fig.~\ref{fig:neff_evolution} shows that eV-scale sterile neutrinos start 
to be produced through neutrino oscillation at around $T_\gamma \sim 13$~MeV 
($= T_{\rm max}$), as discussed in the previous section (see Eq.~\eqref{eq:tmax}). 

In Fig.~3, we plot the energy spectrum of each 
flavor of neutrinos for the typical values of $T_{\rm RH}$. We evaluate the 
spectra at $T_\gamma = 10^{-2}$~MeV, which is much later than the electron-pair 
annihilation and the neutrino decoupling. It can be seen from Fig.~3 that a peak 
position of the neutrino energy spectra is shifted to lower than $p/T_\gamma \sim 3.15$, 
which corresponds to the thermal value for fermions. This is because the
photon temperature $T_\gamma$ increases by a factor of $(11/4)^{1/3} \sim 1.4$
compared to those of neutrinos after the annihilation of electrons. 
 
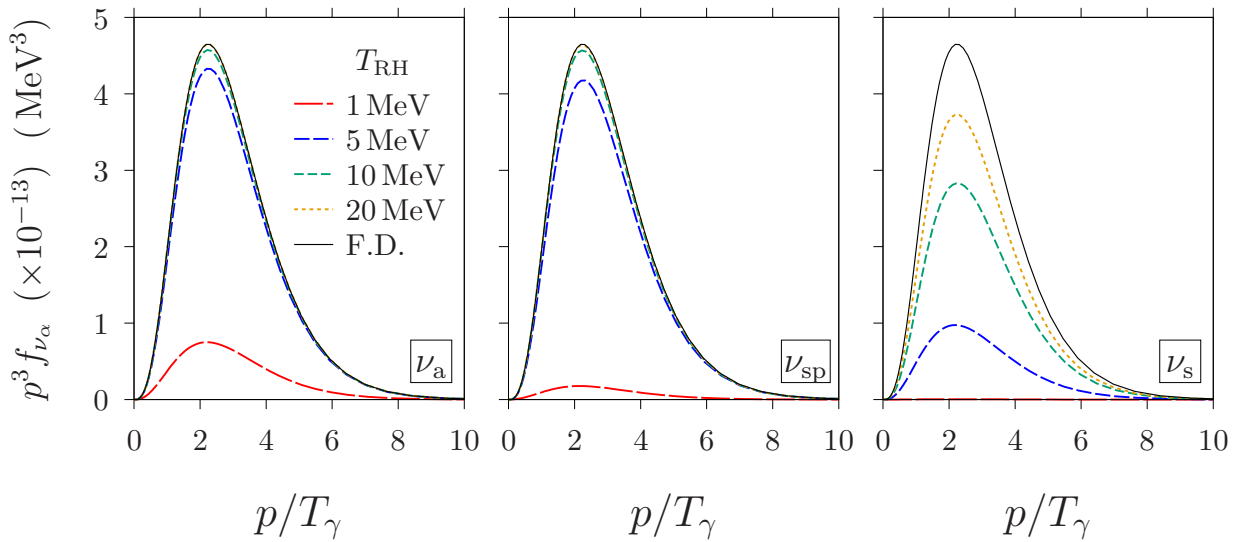
\begin{figure}[!t]
\vspace{0.8cm}
\begin{center}
\vspace{-4cm}
\resizebox{0.95\linewidth}{!}{\hspace{1cm}\input{fnu_final_vs_T_RH}}
\end{center}
\vspace{0.7cm}
 \caption{
 Dependence of the final energy spectra of neutrinos on the reheating temperature, 
 evaluated at $T_\gamma = 10^{-2}$~MeV, for the case of $\nu_e$--$\nu_{\rm s}$ mixing.
 The effect of the neutrino self-interaction is included in the calculation. 
 The $x$-axis is the neutrino energy $p$ divided by the photon temperature $T_\gamma$, 
 and the $y$-axis is the differential energy spectrum of neutrinos.
 The left panel is for $\nu_{\rm a}$, the middle panel is for $\nu_{\rm sp}$,
 and the right panel is for $\nu_{\rm s}$. In each panel, the red long-dashed 
 line is for $T_{\rm RH} = 1$~MeV, the blue middle-dashed
 line is for $T_{\rm RH} = 5$~MeV, the green short-dashed line is for
 $T_{\rm RH} = 10$~MeV, and the black solid line is for $T_{\rm RH} = 20$~MeV. 
 In the left and middle panels, the spectrum for $T_{\rm RH} = 20$~MeV is almost 
 the same as that of the Fermi-Dirac distribution $f_{\rm FD} = 1/\exp[p/T_\nu + 1]$, 
 where $T_\nu = (4/11)^{1/3}\, T_\gamma \sim T_\gamma$/1.4.
 }\label{fig:nuspectrum}
\end{figure}

The averaged energy of neutrinos is expressed by the distortion parameter 
$R_{{\rm dist},\,\nu_\alpha}$, defined by 
\begin{equation}
 R_{{\rm dist},\,\nu_\alpha}
  = \frac{1}{3.15\,T_{\nu_\alpha,\,{\rm eff}}}\frac{\rho_{\nu_\alpha}}{n_{\nu_\alpha}}\,,
\label{eq:R_dist}
\end{equation}
where $T_{\nu_\alpha,\,{\rm eff}}\,(= [4\pi^2 n_{\nu_\alpha}/3\zeta(3)]^{1/3})$
is the effective temperature for each flavor of neutrinos. The thermal spectrum 
corresponds to $R_{{\rm dist},\,\nu_\alpha} = 1$ by definition. 
In Fig.~4, we show the dependence of the distortion parameter 
$R_{\rm dist}$ on the reheating temperature. The figure reveals that $R_{\rm dist}$ 
increases as the reheating temperature decreases, and particularly it goes to unity 
for active neutrinos at $T_{\rm RH} > 10$~MeV. The production mechanism 
of active neutrinos is responsible for this feature. Active neutrinos are produced 
only from the electron-pair annihilation, and each neutrino in the final state has 
energy larger than the electron mass $m_e \sim 0.5$~MeV. Therefore, the value of 
$R_{{\rm dist},\,\nu_\alpha}$ becomes larger than unity due to the large contribution 
from neutrinos produced when the electron is still relativistic. The scattering 
rate of the process $e^\pm + \nu_\alpha \rightarrow e^\pm + \nu_\alpha$ ($\alpha={\rm a,s,sp}$), 
is not sufficient to fully equilibrate the neutrino spectrum since it is of the order 
of ${\cal O}(G_F^2)$, which is the same as that of the neutrino-pair production. 
As seen in Fig.~4, the distortion parameter for the spectator 
neutrino is always larger than that of the active-mixed neutrino since we consider the 
mixing between $\nu_e$ and $\nu_s$.  The scattering between the background electron 
and the electron neutrino is more frequent than those of the other active neutrino species.
Also, the energy spectrum of sterile neutrinos is heavily distorted compared to 
active neutrinos since electrons do not interact with sterile neutrinos.
We note that the value of $R_{{\rm dist},\,\nu_{\rm s}} (\sim 1.6)$ for $T_{\rm RH} = 5$~MeV
is ${\cal O}$(10)\% larger than that obtained in Ref.~\cite{Gelmini_2019}, $R_{{\rm dist},\,\nu_{\rm s}} \sim 1.3$.
This is possibly due to assumptions which they adopted to simplify the thermalization
calculation of sterile neutrinos, namely the sterile neutrino abundance is negligible 
compared to the thermal abundance $f_{\nu_{\rm s}} << f_{\rm FD} = 1/\exp[p/T_\gamma + 1]$, 
and the sterile neutrino production only takes place at $T_\gamma < T_{\rm RH}$.
 
\begin{figure}[!t]
\vspace{0.3cm}
\begin{center}
\resizebox{0.9\linewidth}{!}{\input{Trh_vs_R_dist}}
\end{center}
 \caption{
 Dependence of the distortion parameter for each flavor of neutrinos 
 $R_{{\rm dist},\,\nu_\alpha}$ ($\alpha =$\,a,\,s,\,sp) on the reheating 
 temperature $T_{\rm RH}$, for the case of $\nu_e$--$\nu_{\rm s}$ mixing. 
 The red short-dashed line is for $\nu_{\rm a}$, 
 the greed middle-dashed line is for $\nu_{\rm sp}$, 
 and the blue long-dashed line is for $\nu_{\rm sp}$. $R_{\rm dist} = 1$ 
 corresponds to the thermal Fermi-Dirac spectrum.
 } \label{fig:dist_factor}
\end{figure}

Figure~5 shows the mass dependence of $N_{\rm eff}$. 
As can be seen from Fig.~5, 
the mass dependence is negligible for $T_{\rm RH} \sim {\mathcal O}(1)$~MeV, which justifies the 
assumption in Ref.~\cite{Yaguna_2007}. The reason is that the matter effect is almost negligible 
at a low temperature of ${\mathcal O}(1)$~MeV, and the effective production rate of the 
sterile neutrino is therefore given by $\sim D\,\sin^2 2\theta$ 
for both cases of $m_{\rm s} = 1$~eV and 1~keV. 
The damping rate $D$ is associated only with active neutrinos, and it has 
no sensitivity to the sterile neutrino property. 
 
\begin{figure}[!t]
\vspace{1cm}
\begin{center}
\vspace{-3cm}
\resizebox{0.9\linewidth}{!}{\hspace{1cm}\input{Trh_vs_Neff_mass_dep}}
\end{center}
\vspace{0.5cm}
 \caption{
 Mass dependence of the effective number of neutrino species for all flavors of neutrinos (left) 
 and sterile neutrinos (right) for each value of the active-sterile mixing $\sin^22\theta$, 
 for the case of $\nu_e$--$\nu_{\rm s}$ mixing
 The solid lines are for $m_{\rm s} = 1$~eV, while dashed lines are for $m_{\rm s}=$1~keV. 
 The red, blue, and black lines correspond to $\sin^22\theta$ = $10^{-3}$, $10^{-2}$, 
 and $10^{-1}$, respectively. There are no apparent mass dependences for the cases of $\sin^22\theta = 10^{-1}$. 
 } \label{fig:neffmassdep}
\end{figure}
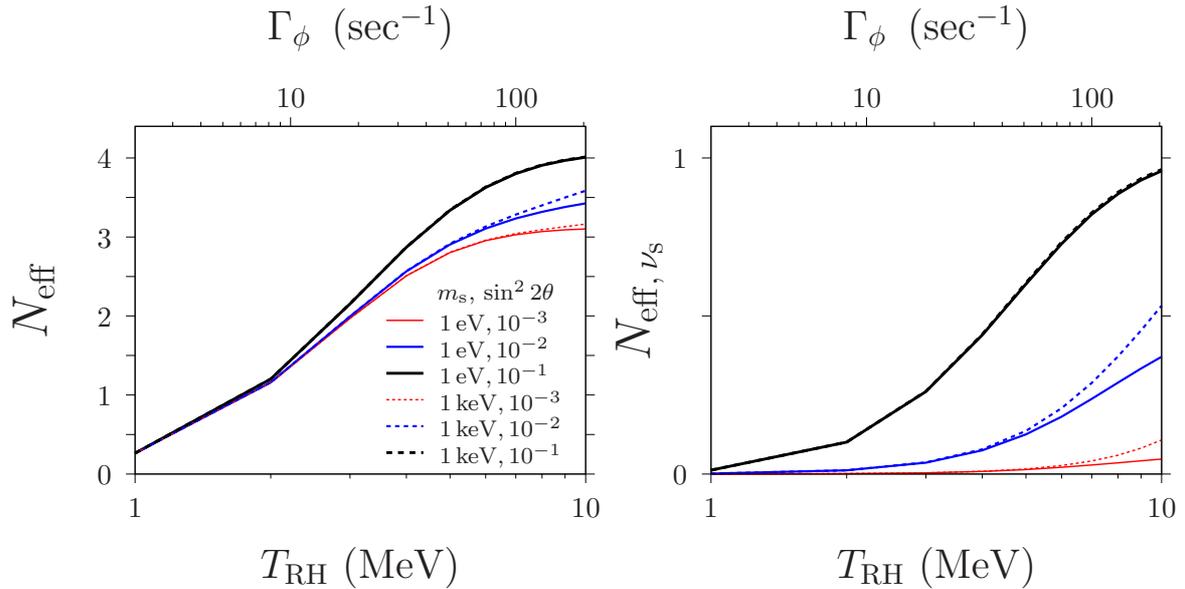

%% file: Trh_vs_Neff.tex
\begingroup
  \makeatletter
  \providecommand\color[2][]{%
    \GenericError{(gnuplot) \space\space\space\@spaces}{%
      Package color not loaded in conjunction with
      terminal option `colourtext'%
    }{See the gnuplot documentation for explanation.%
    }{Either use 'blacktext' in gnuplot or load the package
      color.sty in LaTeX.}%
    \renewcommand\color[2][]{}%
  }%
  \providecommand\includegraphics[2][]{%
    \GenericError{(gnuplot) \space\space\space\@spaces}{%
      Package graphicx or graphics not loaded%
    }{See the gnuplot documentation for explanation.%
    }{The gnuplot epslatex terminal needs graphicx.sty or graphics.sty.}%
    \renewcommand\includegraphics[2][]{}%
  }%
  \providecommand\rotatebox[2]{#2}%
  \@ifundefined{ifGPcolor}{%
    \newif\ifGPcolor
    \GPcolortrue
  }{}%
  \@ifundefined{ifGPblacktext}{%
    \newif\ifGPblacktext
    \GPblacktexttrue
  }{}%
  \let\gplgaddtomacro\g@addto@macro
  \gdef\gplbacktext{}%
  \gdef\gplfronttext{}%
  \makeatother
  \ifGPblacktext
    \def\colorrgb#1{}%
    \def\colorgray#1{}%
  \else
    \ifGPcolor
      \def\colorrgb#1{\color[rgb]{#1}}%
      \def\colorgray#1{\color[gray]{#1}}%
      \expandafter\def\csname LTw\endcsname{\color{white}}%
      \expandafter\def\csname LTb\endcsname{\color{black}}%
      \expandafter\def\csname LTa\endcsname{\color{black}}%
      \expandafter\def\csname LT0\endcsname{\color[rgb]{1,0,0}}%
      \expandafter\def\csname LT1\endcsname{\color[rgb]{0,1,0}}%
      \expandafter\def\csname LT2\endcsname{\color[rgb]{0,0,1}}%
      \expandafter\def\csname LT3\endcsname{\color[rgb]{1,0,1}}%
      \expandafter\def\csname LT4\endcsname{\color[rgb]{0,1,1}}%
      \expandafter\def\csname LT5\endcsname{\color[rgb]{1,1,0}}%
      \expandafter\def\csname LT6\endcsname{\color[rgb]{0,0,0}}%
      \expandafter\def\csname LT7\endcsname{\color[rgb]{1,0.3,0}}%
      \expandafter\def\csname LT8\endcsname{\color[rgb]{0.5,0.5,0.5}}%
    \else
      \def\colorrgb#1{\color{black}}%
      \def\colorgray#1{\color[gray]{#1}}%
      \expandafter\def\csname LTw\endcsname{\color{white}}%
      \expandafter\def\csname LTb\endcsname{\color{black}}%
      \expandafter\def\csname LTa\endcsname{\color{black}}%
      \expandafter\def\csname LT0\endcsname{\color{black}}%
      \expandafter\def\csname LT1\endcsname{\color{black}}%
      \expandafter\def\csname LT2\endcsname{\color{black}}%
      \expandafter\def\csname LT3\endcsname{\color{black}}%
      \expandafter\def\csname LT4\endcsname{\color{black}}%
      \expandafter\def\csname LT5\endcsname{\color{black}}%
      \expandafter\def\csname LT6\endcsname{\color{black}}%
      \expandafter\def\csname LT7\endcsname{\color{black}}%
      \expandafter\def\csname LT8\endcsname{\color{black}}%
    \fi
  \fi
    \setlength{\unitlength}{0.0500bp}%
    \ifx\gptboxheight\undefined%
      \newlength{\gptboxheight}%
      \newlength{\gptboxwidth}%
      \newsavebox{\gptboxtext}%
    \fi%
    \setlength{\fboxrule}{0.5pt}%
    \setlength{\fboxsep}{1pt}%
\begin{picture}(7200.00,5040.00)%
      \csname LTb\endcsname
      \put(3600,4760){\makebox(0,0){\strut{}}}%
    \gplgaddtomacro\gplbacktext{%
      \csname LTb\endcsname
      \put(1035,100){\makebox(0,0)[r]{\strut{}$0$}}%
      \put(1035,845){\makebox(0,0)[r]{\strut{}$1$}}%
      \put(1035,1591){\makebox(0,0)[r]{\strut{}$2$}}%
      \put(1035,2336){\makebox(0,0)[r]{\strut{}$3$}}%
      \put(1035,3082){\makebox(0,0)[r]{\strut{}$4$}}%
      \put(1182,-103){\makebox(0,0){\strut{}$1$}}%
      \put(4898,-103){\makebox(0,0){\strut{}$10$}}%
      \put(2464,3881){\makebox(0,0){\strut{}$10$}}%
      \put(4323,3881){\makebox(0,0){\strut{}$100$}}%
    }%
    \gplgaddtomacro\gplfronttext{%
      \csname LTb\endcsname
      \put(403,1889){\rotatebox{-270}{\makebox(0,0){\strut{}\huge $N_{\rm eff}$}}}%
      \put(3599,-663){\makebox(0,0){\strut{}\Large $T_{\rm RH}$ (MeV)}}%
      \put(3599,4357){\makebox(0,0){\strut{}\Large $\Gamma_\phi$\ \,(${\rm sec^{-1}}$)}}%
      \csname LTb\endcsname
      \put(1713,3436){\makebox(0,0)[l]{\strut{}\,$\nu_{\rm a}\!+\!\nu_{\rm s}\!+\!\nu_{\rm sp}$}}%
      \csname LTb\endcsname
      \put(1713,3166){\makebox(0,0)[l]{\strut{}\,$\nu_{\rm a}$}}%
      \csname LTb\endcsname
      \put(1713,2896){\makebox(0,0)[l]{\strut{}\,$\nu_{\rm sp}$}}%
      \csname LTb\endcsname
      \put(1713,2626){\makebox(0,0)[l]{\strut{}\,$\nu_{\rm s}$}}%
      \csname LTb\endcsname
      \put(1713,2356){\makebox(0,0)[l]{\strut{}\,$\nu_{\rm a}\!+\!\nu_{\rm s}\!+\!\nu_{\rm sp}$}}%
      \csname LTb\endcsname
      \put(1713,2086){\makebox(0,0)[l]{\strut{}\,$\nu_{\rm a}$}}%
      \csname LTb\endcsname
      \put(1713,1816){\makebox(0,0)[l]{\strut{}\,$\nu_{\rm sp}$}}%
      \csname LTb\endcsname
      \put(1713,1546){\makebox(0,0)[l]{\strut{}\,$\nu_{\rm s}$}}%
    }%
    \gplbacktext
    \put(0,0){\includegraphics{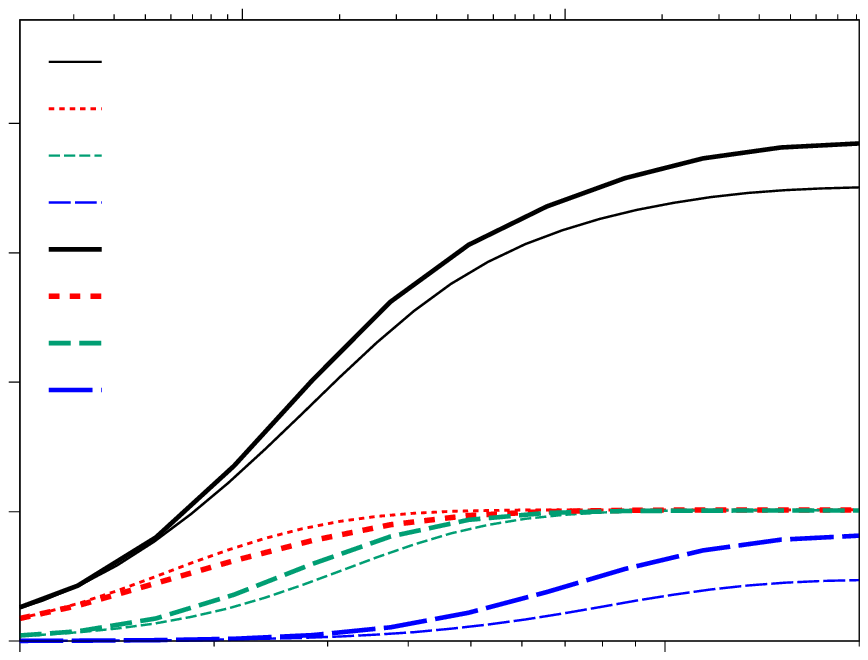}}%
    \gplfronttext
  \end{picture}%
\endgroup

%% file: Neff_evolution_vs_tgam.tex
\begingroup
  \makeatletter
  \providecommand\color[2][]{%
    \GenericError{(gnuplot) \space\space\space\@spaces}{%
      Package color not loaded in conjunction with
      terminal option `colourtext'%
    }{See the gnuplot documentation for explanation.%
    }{Either use 'blacktext' in gnuplot or load the package
      color.sty in LaTeX.}%
    \renewcommand\color[2][]{}%
  }%
  \providecommand\includegraphics[2][]{%
    \GenericError{(gnuplot) \space\space\space\@spaces}{%
      Package graphicx or graphics not loaded%
    }{See the gnuplot documentation for explanation.%
    }{The gnuplot epslatex terminal needs graphicx.sty or graphics.sty.}%
    \renewcommand\includegraphics[2][]{}%
  }%
  \providecommand\rotatebox[2]{#2}%
  \@ifundefined{ifGPcolor}{%
    \newif\ifGPcolor
    \GPcolortrue
  }{}%
  \@ifundefined{ifGPblacktext}{%
    \newif\ifGPblacktext
    \GPblacktexttrue
  }{}%
  \let\gplgaddtomacro\g@addto@macro
  \gdef\gplbacktext{}%
  \gdef\gplfronttext{}%
  \makeatother
  \ifGPblacktext
    \def\colorrgb#1{}%
    \def\colorgray#1{}%
  \else
    \ifGPcolor
      \def\colorrgb#1{\color[rgb]{#1}}%
      \def\colorgray#1{\color[gray]{#1}}%
      \expandafter\def\csname LTw\endcsname{\color{white}}%
      \expandafter\def\csname LTb\endcsname{\color{black}}%
      \expandafter\def\csname LTa\endcsname{\color{black}}%
      \expandafter\def\csname LT0\endcsname{\color[rgb]{1,0,0}}%
      \expandafter\def\csname LT1\endcsname{\color[rgb]{0,1,0}}%
      \expandafter\def\csname LT2\endcsname{\color[rgb]{0,0,1}}%
      \expandafter\def\csname LT3\endcsname{\color[rgb]{1,0,1}}%
      \expandafter\def\csname LT4\endcsname{\color[rgb]{0,1,1}}%
      \expandafter\def\csname LT5\endcsname{\color[rgb]{1,1,0}}%
      \expandafter\def\csname LT6\endcsname{\color[rgb]{0,0,0}}%
      \expandafter\def\csname LT7\endcsname{\color[rgb]{1,0.3,0}}%
      \expandafter\def\csname LT8\endcsname{\color[rgb]{0.5,0.5,0.5}}%
    \else
      \def\colorrgb#1{\color{black}}%
      \def\colorgray#1{\color[gray]{#1}}%
      \expandafter\def\csname LTw\endcsname{\color{white}}%
      \expandafter\def\csname LTb\endcsname{\color{black}}%
      \expandafter\def\csname LTa\endcsname{\color{black}}%
      \expandafter\def\csname LT0\endcsname{\color{black}}%
      \expandafter\def\csname LT1\endcsname{\color{black}}%
      \expandafter\def\csname LT2\endcsname{\color{black}}%
      \expandafter\def\csname LT3\endcsname{\color{black}}%
      \expandafter\def\csname LT4\endcsname{\color{black}}%
      \expandafter\def\csname LT5\endcsname{\color{black}}%
      \expandafter\def\csname LT6\endcsname{\color{black}}%
      \expandafter\def\csname LT7\endcsname{\color{black}}%
      \expandafter\def\csname LT8\endcsname{\color{black}}%
    \fi
  \fi
    \setlength{\unitlength}{0.0500bp}%
    \ifx\gptboxheight\undefined%
      \newlength{\gptboxheight}%
      \newlength{\gptboxwidth}%
      \newsavebox{\gptboxtext}%
    \fi%
    \setlength{\fboxrule}{0.5pt}%
    \setlength{\fboxsep}{1pt}%
\begin{picture}(7200.00,5040.00)%
      \csname LTb\endcsname
      \put(3600,4760){\makebox(0,0){\strut{}}}%
    \gplgaddtomacro\gplbacktext{%
      \csname LTb\endcsname
      \put(-12,504){\makebox(0,0)[r]{\strut{}$0$}}%
      \put(-12,1554){\makebox(0,0)[r]{\strut{}$0.5$}}%
      \put(-12,2603){\makebox(0,0)[r]{\strut{}$1$}}%
      \put(431,273){\makebox(0,0){\strut{}$1$}}%
      \put(1625,273){\makebox(0,0){\strut{}$10$}}%
      \put(1701,1537){\makebox(0,0)[l]{\strut{}\small $T_{\rm RH}$}}%
    }%
    \gplgaddtomacro\gplfronttext{%
      \csname LTb\endcsname
      \put(-628,1763){\rotatebox{-270}{\makebox(0,0){\strut{}\LARGE $N_{{\rm eff},\,\nu_\alpha}$}}}%
      \put(1187,-259){\makebox(0,0){\strut{}\Large $T_\gamma$ (MeV)}}%
      \csname LTb\endcsname
      \put(1648,1303){\makebox(0,0)[l]{\strut{}\scriptsize 1\,MeV}}%
      \csname LTb\endcsname
      \put(1648,1087){\makebox(0,0)[l]{\strut{}\scriptsize 2\,MeV}}%
      \csname LTb\endcsname
      \put(1648,871){\makebox(0,0)[l]{\strut{}\scriptsize 5\,MeV}}%
      \csname LTb\endcsname
      \put(1648,655){\makebox(0,0)[l]{\strut{}\scriptsize 10\,MeV}}%
      \csname LTb\endcsname
      \put(2102,2872){\makebox(0,0){\strut{}$\nu_{\rm a}$}}%
    }%
    \gplgaddtomacro\gplbacktext{%
      \csname LTb\endcsname
      \put(2364,504){\makebox(0,0)[r]{\strut{} }}%
      \put(2364,1554){\makebox(0,0)[r]{\strut{} }}%
      \put(2364,2603){\makebox(0,0)[r]{\strut{} }}%
      \put(2807,273){\makebox(0,0){\strut{}$1$}}%
      \put(4001,273){\makebox(0,0){\strut{}$10$}}%
    }%
    \gplgaddtomacro\gplfronttext{%
      \csname LTb\endcsname
      \put(3563,-259){\makebox(0,0){\strut{}\Large $T_\gamma$ (MeV)}}%
      \csname LTb\endcsname
      \put(4456,2872){\makebox(0,0){\strut{}$\nu_{\rm sp}$}}%
    }%
    \gplgaddtomacro\gplbacktext{%
      \csname LTb\endcsname
      \put(4740,504){\makebox(0,0)[r]{\strut{} }}%
      \put(4740,1554){\makebox(0,0)[r]{\strut{} }}%
      \put(4740,2603){\makebox(0,0)[r]{\strut{} }}%
      \put(5183,273){\makebox(0,0){\strut{}$1$}}%
      \put(6377,273){\makebox(0,0){\strut{}$10$}}%
    }%
    \gplgaddtomacro\gplfronttext{%
      \csname LTb\endcsname
      \put(5939,-259){\makebox(0,0){\strut{}\Large $T_\gamma$ (MeV)}}%
      \csname LTb\endcsname
      \put(6854,2872){\makebox(0,0){\strut{}$\nu_{\rm s}$}}%
    }%
    \gplbacktext
    \put(0,0){\includegraphics{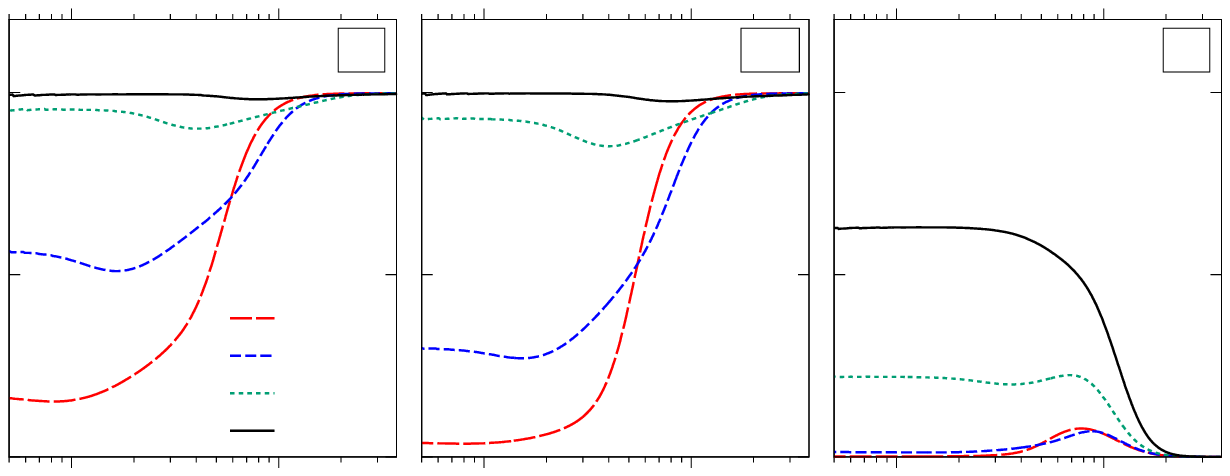}}%
    \gplfronttext
  \end{picture}%
\endgroup

%% file: fnu_final_vs_T_RH.tex
\begingroup
  \makeatletter
  \providecommand\color[2][]{%
    \GenericError{(gnuplot) \space\space\space\@spaces}{%
      Package color not loaded in conjunction with
      terminal option `colourtext'%
    }{See the gnuplot documentation for explanation.%
    }{Either use 'blacktext' in gnuplot or load the package
      color.sty in LaTeX.}%
    \renewcommand\color[2][]{}%
  }%
  \providecommand\includegraphics[2][]{%
    \GenericError{(gnuplot) \space\space\space\@spaces}{%
      Package graphicx or graphics not loaded%
    }{See the gnuplot documentation for explanation.%
    }{The gnuplot epslatex terminal needs graphicx.sty or graphics.sty.}%
    \renewcommand\includegraphics[2][]{}%
  }%
  \providecommand\rotatebox[2]{#2}%
  \@ifundefined{ifGPcolor}{%
    \newif\ifGPcolor
    \GPcolortrue
  }{}%
  \@ifundefined{ifGPblacktext}{%
    \newif\ifGPblacktext
    \GPblacktexttrue
  }{}%
  \let\gplgaddtomacro\g@addto@macro
  \gdef\gplbacktext{}%
  \gdef\gplfronttext{}%
  \makeatother
  \ifGPblacktext
    \def\colorrgb#1{}%
    \def\colorgray#1{}%
  \else
    \ifGPcolor
      \def\colorrgb#1{\color[rgb]{#1}}%
      \def\colorgray#1{\color[gray]{#1}}%
      \expandafter\def\csname LTw\endcsname{\color{white}}%
      \expandafter\def\csname LTb\endcsname{\color{black}}%
      \expandafter\def\csname LTa\endcsname{\color{black}}%
      \expandafter\def\csname LT0\endcsname{\color[rgb]{1,0,0}}%
      \expandafter\def\csname LT1\endcsname{\color[rgb]{0,1,0}}%
      \expandafter\def\csname LT2\endcsname{\color[rgb]{0,0,1}}%
      \expandafter\def\csname LT3\endcsname{\color[rgb]{1,0,1}}%
      \expandafter\def\csname LT4\endcsname{\color[rgb]{0,1,1}}%
      \expandafter\def\csname LT5\endcsname{\color[rgb]{1,1,0}}%
      \expandafter\def\csname LT6\endcsname{\color[rgb]{0,0,0}}%
      \expandafter\def\csname LT7\endcsname{\color[rgb]{1,0.3,0}}%
      \expandafter\def\csname LT8\endcsname{\color[rgb]{0.5,0.5,0.5}}%
    \else
      \def\colorrgb#1{\color{black}}%
      \def\colorgray#1{\color[gray]{#1}}%
      \expandafter\def\csname LTw\endcsname{\color{white}}%
      \expandafter\def\csname LTb\endcsname{\color{black}}%
      \expandafter\def\csname LTa\endcsname{\color{black}}%
      \expandafter\def\csname LT0\endcsname{\color{black}}%
      \expandafter\def\csname LT1\endcsname{\color{black}}%
      \expandafter\def\csname LT2\endcsname{\color{black}}%
      \expandafter\def\csname LT3\endcsname{\color{black}}%
      \expandafter\def\csname LT4\endcsname{\color{black}}%
      \expandafter\def\csname LT5\endcsname{\color{black}}%
      \expandafter\def\csname LT6\endcsname{\color{black}}%
      \expandafter\def\csname LT7\endcsname{\color{black}}%
      \expandafter\def\csname LT8\endcsname{\color{black}}%
    \fi
  \fi
    \setlength{\unitlength}{0.0500bp}%
    \ifx\gptboxheight\undefined%
      \newlength{\gptboxheight}%
      \newlength{\gptboxwidth}%
      \newsavebox{\gptboxtext}%
    \fi%
    \setlength{\fboxrule}{0.5pt}%
    \setlength{\fboxsep}{1pt}%
\begin{picture}(7200.00,5040.00)%
      \csname LTb\endcsname
      \put(3600,4760){\makebox(0,0){\strut{}}}%
    \gplgaddtomacro\gplbacktext{%
      \csname LTb\endcsname
      \put(-92,504){\makebox(0,0)[r]{\strut{}$0$}}%
      \put(-92,1008){\makebox(0,0)[r]{\strut{}$1$}}%
      \put(-92,1512){\makebox(0,0)[r]{\strut{}$2$}}%
      \put(-92,2015){\makebox(0,0)[r]{\strut{}$3$}}%
      \put(-92,2519){\makebox(0,0)[r]{\strut{}$4$}}%
      \put(-92,3023){\makebox(0,0)[r]{\strut{}$5$}}%
      \put(72,245){\makebox(0,0){\strut{}$0$}}%
      \put(504,245){\makebox(0,0){\strut{}$2$}}%
      \put(936,245){\makebox(0,0){\strut{}$4$}}%
      \put(1367,245){\makebox(0,0){\strut{}$6$}}%
      \put(1799,245){\makebox(0,0){\strut{}$8$}}%
      \put(2231,245){\makebox(0,0){\strut{}$10$}}%
      \put(1519,2746){\makebox(0,0)[l]{\strut{}$T_{\rm RH}$}}%
    }%
    \gplgaddtomacro\gplfronttext{%
      \csname LTb\endcsname
      \put(-607,1763){\rotatebox{-270}{\makebox(0,0){\strut{}\large $p^3 f_{\nu_\alpha}\ \,(\times 10^{-13})\ \ ({\rm\,MeV}^3)$}}}%
      \put(1151,-259){\makebox(0,0){\strut{}\Large $p/T_\gamma$}}%
      \csname LTb\endcsname
      \put(1422,2454){\makebox(0,0)[l]{\strut{}\small \,1\,MeV}}%
      \csname LTb\endcsname
      \put(1422,2222){\makebox(0,0)[l]{\strut{}\small \,5\,MeV}}%
      \csname LTb\endcsname
      \put(1422,1990){\makebox(0,0)[l]{\strut{}\small \,10\,MeV}}%
      \csname LTb\endcsname
      \put(1422,1758){\makebox(0,0)[l]{\strut{}\small \,20\,MeV}}%
      \csname LTb\endcsname
      \put(1422,1526){\makebox(0,0)[l]{\strut{}\small \,F.D.}}%
      \csname LTb\endcsname
      \put(2015,756){\makebox(0,0){\strut{}\large $\nu_{\rm a}$}}%
    }%
    \gplgaddtomacro\gplbacktext{%
      \csname LTb\endcsname
      \put(2356,504){\makebox(0,0)[r]{\strut{} }}%
      \put(2356,1008){\makebox(0,0)[r]{\strut{} }}%
      \put(2356,1512){\makebox(0,0)[r]{\strut{} }}%
      \put(2356,2015){\makebox(0,0)[r]{\strut{} }}%
      \put(2356,2519){\makebox(0,0)[r]{\strut{} }}%
      \put(2356,3023){\makebox(0,0)[r]{\strut{} }}%
      \put(2520,245){\makebox(0,0){\strut{}$0$}}%
      \put(2952,245){\makebox(0,0){\strut{}$2$}}%
      \put(3384,245){\makebox(0,0){\strut{}$4$}}%
      \put(3815,245){\makebox(0,0){\strut{}$6$}}%
      \put(4247,245){\makebox(0,0){\strut{}$8$}}%
      \put(4679,245){\makebox(0,0){\strut{}$10$}}%
    }%
    \gplgaddtomacro\gplfronttext{%
      \csname LTb\endcsname
      \put(3599,-259){\makebox(0,0){\strut{}\Large $p/T_\gamma$}}%
      \csname LTb\endcsname
      \put(4463,756){\makebox(0,0){\strut{}\large $\nu_{\rm sp}$}}%
    }%
    \gplgaddtomacro\gplbacktext{%
      \csname LTb\endcsname
      \put(4804,504){\makebox(0,0)[r]{\strut{} }}%
      \put(4804,1008){\makebox(0,0)[r]{\strut{} }}%
      \put(4804,1512){\makebox(0,0)[r]{\strut{} }}%
      \put(4804,2015){\makebox(0,0)[r]{\strut{} }}%
      \put(4804,2519){\makebox(0,0)[r]{\strut{} }}%
      \put(4804,3023){\makebox(0,0)[r]{\strut{} }}%
      \put(4968,245){\makebox(0,0){\strut{}$0$}}%
      \put(5400,245){\makebox(0,0){\strut{}$2$}}%
      \put(5832,245){\makebox(0,0){\strut{}$4$}}%
      \put(6263,245){\makebox(0,0){\strut{}$6$}}%
      \put(6695,245){\makebox(0,0){\strut{}$8$}}%
      \put(7127,245){\makebox(0,0){\strut{}$10$}}%
    }%
    \gplgaddtomacro\gplfronttext{%
      \csname LTb\endcsname
      \put(6047,-259){\makebox(0,0){\strut{}\Large $p/T_\gamma$}}%
      \csname LTb\endcsname
      \put(6911,756){\makebox(0,0){\strut{}\large $\nu_{\rm s}$}}%
    }%
    \gplbacktext
    \put(0,0){\includegraphics{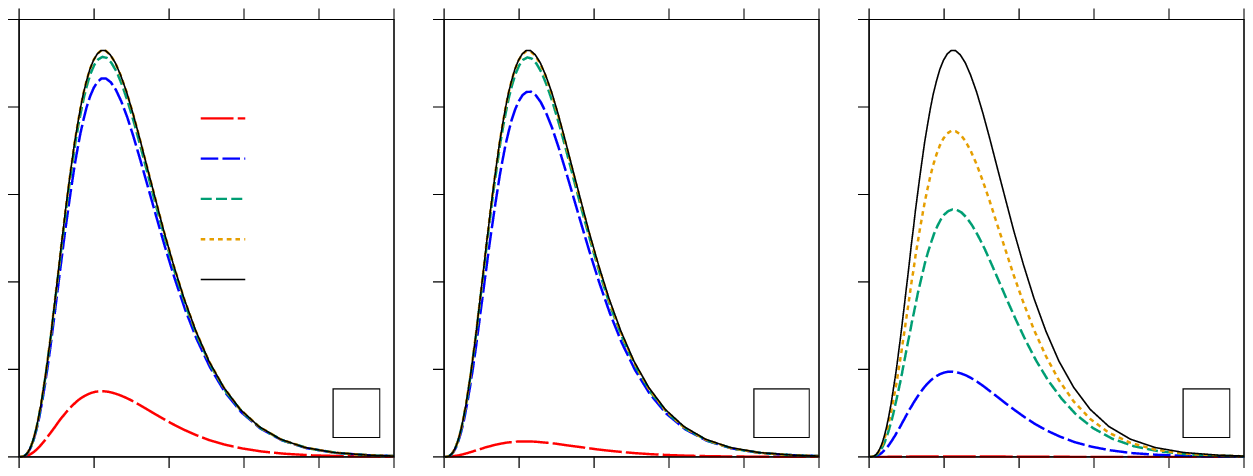}}%
    \gplfronttext
  \end{picture}%
\endgroup

%% file: Trh_vs_R_dist.tex
\begingroup
  \makeatletter
  \providecommand\color[2][]{%
    \GenericError{(gnuplot) \space\space\space\@spaces}{%
      Package color not loaded in conjunction with
      terminal option `colourtext'%
    }{See the gnuplot documentation for explanation.%
    }{Either use 'blacktext' in gnuplot or load the package
      color.sty in LaTeX.}%
    \renewcommand\color[2][]{}%
  }%
  \providecommand\includegraphics[2][]{%
    \GenericError{(gnuplot) \space\space\space\@spaces}{%
      Package graphicx or graphics not loaded%
    }{See the gnuplot documentation for explanation.%
    }{The gnuplot epslatex terminal needs graphicx.sty or graphics.sty.}%
    \renewcommand\includegraphics[2][]{}%
  }%
  \providecommand\rotatebox[2]{#2}%
  \@ifundefined{ifGPcolor}{%
    \newif\ifGPcolor
    \GPcolortrue
  }{}%
  \@ifundefined{ifGPblacktext}{%
    \newif\ifGPblacktext
    \GPblacktexttrue
  }{}%
  \let\gplgaddtomacro\g@addto@macro
  \gdef\gplbacktext{}%
  \gdef\gplfronttext{}%
  \makeatother
  \ifGPblacktext
    \def\colorrgb#1{}%
    \def\colorgray#1{}%
  \else
    \ifGPcolor
      \def\colorrgb#1{\color[rgb]{#1}}%
      \def\colorgray#1{\color[gray]{#1}}%
      \expandafter\def\csname LTw\endcsname{\color{white}}%
      \expandafter\def\csname LTb\endcsname{\color{black}}%
      \expandafter\def\csname LTa\endcsname{\color{black}}%
      \expandafter\def\csname LT0\endcsname{\color[rgb]{1,0,0}}%
      \expandafter\def\csname LT1\endcsname{\color[rgb]{0,1,0}}%
      \expandafter\def\csname LT2\endcsname{\color[rgb]{0,0,1}}%
      \expandafter\def\csname LT3\endcsname{\color[rgb]{1,0,1}}%
      \expandafter\def\csname LT4\endcsname{\color[rgb]{0,1,1}}%
      \expandafter\def\csname LT5\endcsname{\color[rgb]{1,1,0}}%
      \expandafter\def\csname LT6\endcsname{\color[rgb]{0,0,0}}%
      \expandafter\def\csname LT7\endcsname{\color[rgb]{1,0.3,0}}%
      \expandafter\def\csname LT8\endcsname{\color[rgb]{0.5,0.5,0.5}}%
    \else
      \def\colorrgb#1{\color{black}}%
      \def\colorgray#1{\color[gray]{#1}}%
      \expandafter\def\csname LTw\endcsname{\color{white}}%
      \expandafter\def\csname LTb\endcsname{\color{black}}%
      \expandafter\def\csname LTa\endcsname{\color{black}}%
      \expandafter\def\csname LT0\endcsname{\color{black}}%
      \expandafter\def\csname LT1\endcsname{\color{black}}%
      \expandafter\def\csname LT2\endcsname{\color{black}}%
      \expandafter\def\csname LT3\endcsname{\color{black}}%
      \expandafter\def\csname LT4\endcsname{\color{black}}%
      \expandafter\def\csname LT5\endcsname{\color{black}}%
      \expandafter\def\csname LT6\endcsname{\color{black}}%
      \expandafter\def\csname LT7\endcsname{\color{black}}%
      \expandafter\def\csname LT8\endcsname{\color{black}}%
    \fi
  \fi
    \setlength{\unitlength}{0.0500bp}%
    \ifx\gptboxheight\undefined%
      \newlength{\gptboxheight}%
      \newlength{\gptboxwidth}%
      \newsavebox{\gptboxtext}%
    \fi%
    \setlength{\fboxrule}{0.5pt}%
    \setlength{\fboxsep}{1pt}%
\begin{picture}(7200.00,5040.00)%
    \gplgaddtomacro\gplbacktext{%
      \csname LTb\endcsname
      \put(1516,981){\makebox(0,0)[r]{\strut{}$1$}}%
      \put(1516,1419){\makebox(0,0)[r]{\strut{}$2$}}%
      \put(1516,1857){\makebox(0,0)[r]{\strut{}$3$}}%
      \put(1516,2296){\makebox(0,0)[r]{\strut{}$4$}}%
      \put(1516,2734){\makebox(0,0)[r]{\strut{}$5$}}%
      \put(1516,3172){\makebox(0,0)[r]{\strut{}$6$}}%
      \put(1516,3610){\makebox(0,0)[r]{\strut{}$7$}}%
      \put(1516,4048){\makebox(0,0)[r]{\strut{}$8$}}%
      \put(1663,728){\makebox(0,0){\strut{}$1$}}%
      \put(4920,728){\makebox(0,0){\strut{}$10$}}%
      \put(2787,4339){\makebox(0,0){\strut{}$10$}}%
      \put(4415,4339){\makebox(0,0){\strut{}$100$}}%
    }%
    \gplgaddtomacro\gplfronttext{%
      \csname LTb\endcsname
      \put(816,2547){\rotatebox{-270}{\makebox(0,0){\strut{}\huge $R_{{\rm dist},\,\nu}$}}}%
      \put(3781,196){\makebox(0,0){\strut{}\LARGE $T_{\rm RH}$ (MeV)}}%
      \put(3781,4815){\makebox(0,0){\strut{}\LARGE $\Gamma_\phi$\ \,(${\rm sec^{-1}}$)}}%
      \csname LTb\endcsname
      \put(5005,2345){\makebox(0,0)[l]{\strut{}\,\LARGE $\nu_{\rm a}$}}%
      \csname LTb\endcsname
      \put(5005,1940){\makebox(0,0)[l]{\strut{}\,\LARGE $\nu_{\rm sp}$}}%
      \csname LTb\endcsname
      \put(5005,1535){\makebox(0,0)[l]{\strut{}\,\LARGE $\nu_{\rm s}$}}%
    }%
    \gplbacktext
    \put(0,0){\includegraphics{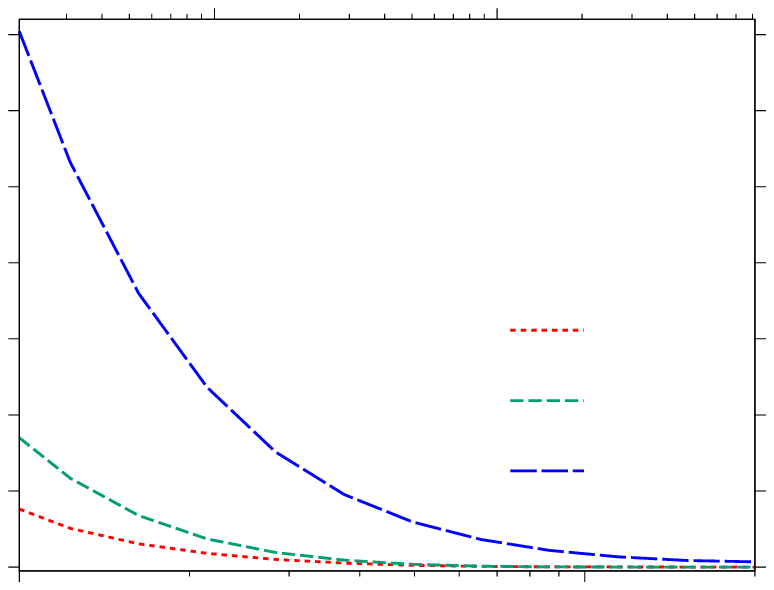}}%
    \gplfronttext
  \end{picture}%
\endgroup

%% file: Trh_vs_Neff_mass_dep.tex
\begingroup
  \makeatletter
  \providecommand\color[2][]{%
    \GenericError{(gnuplot) \space\space\space\@spaces}{%
      Package color not loaded in conjunction with
      terminal option `colourtext'%
    }{See the gnuplot documentation for explanation.%
    }{Either use 'blacktext' in gnuplot or load the package
      color.sty in LaTeX.}%
    \renewcommand\color[2][]{}%
  }%
  \providecommand\includegraphics[2][]{%
    \GenericError{(gnuplot) \space\space\space\@spaces}{%
      Package graphicx or graphics not loaded%
    }{See the gnuplot documentation for explanation.%
    }{The gnuplot epslatex terminal needs graphicx.sty or graphics.sty.}%
    \renewcommand\includegraphics[2][]{}%
  }%
  \providecommand\rotatebox[2]{#2}%
  \@ifundefined{ifGPcolor}{%
    \newif\ifGPcolor
    \GPcolortrue
  }{}%
  \@ifundefined{ifGPblacktext}{%
    \newif\ifGPblacktext
    \GPblacktexttrue
  }{}%
  \let\gplgaddtomacro\g@addto@macro
  \gdef\gplbacktext{}%
  \gdef\gplfronttext{}%
  \makeatother
  \ifGPblacktext
    \def\colorrgb#1{}%
    \def\colorgray#1{}%
  \else
    \ifGPcolor
      \def\colorrgb#1{\color[rgb]{#1}}%
      \def\colorgray#1{\color[gray]{#1}}%
      \expandafter\def\csname LTw\endcsname{\color{white}}%
      \expandafter\def\csname LTb\endcsname{\color{black}}%
      \expandafter\def\csname LTa\endcsname{\color{black}}%
      \expandafter\def\csname LT0\endcsname{\color[rgb]{1,0,0}}%
      \expandafter\def\csname LT1\endcsname{\color[rgb]{0,1,0}}%
      \expandafter\def\csname LT2\endcsname{\color[rgb]{0,0,1}}%
      \expandafter\def\csname LT3\endcsname{\color[rgb]{1,0,1}}%
      \expandafter\def\csname LT4\endcsname{\color[rgb]{0,1,1}}%
      \expandafter\def\csname LT5\endcsname{\color[rgb]{1,1,0}}%
      \expandafter\def\csname LT6\endcsname{\color[rgb]{0,0,0}}%
      \expandafter\def\csname LT7\endcsname{\color[rgb]{1,0.3,0}}%
      \expandafter\def\csname LT8\endcsname{\color[rgb]{0.5,0.5,0.5}}%
    \else
      \def\colorrgb#1{\color{black}}%
      \def\colorgray#1{\color[gray]{#1}}%
      \expandafter\def\csname LTw\endcsname{\color{white}}%
      \expandafter\def\csname LTb\endcsname{\color{black}}%
      \expandafter\def\csname LTa\endcsname{\color{black}}%
      \expandafter\def\csname LT0\endcsname{\color{black}}%
      \expandafter\def\csname LT1\endcsname{\color{black}}%
      \expandafter\def\csname LT2\endcsname{\color{black}}%
      \expandafter\def\csname LT3\endcsname{\color{black}}%
      \expandafter\def\csname LT4\endcsname{\color{black}}%
      \expandafter\def\csname LT5\endcsname{\color{black}}%
      \expandafter\def\csname LT6\endcsname{\color{black}}%
      \expandafter\def\csname LT7\endcsname{\color{black}}%
      \expandafter\def\csname LT8\endcsname{\color{black}}%
    \fi
  \fi
    \setlength{\unitlength}{0.0500bp}%
    \ifx\gptboxheight\undefined%
      \newlength{\gptboxheight}%
      \newlength{\gptboxwidth}%
      \newsavebox{\gptboxtext}%
    \fi%
    \setlength{\fboxrule}{0.5pt}%
    \setlength{\fboxsep}{1pt}%
\begin{picture}(7200.00,5040.00)%
      \csname LTb\endcsname
      \put(3600,4760){\makebox(0,0){\strut{}}}%
    \gplgaddtomacro\gplbacktext{%
      \csname LTb\endcsname
      \put(-92,504){\makebox(0,0)[r]{\strut{}0}}%
      \put(-92,779){\makebox(0,0)[r]{\strut{}}}%
      \put(-92,1054){\makebox(0,0)[r]{\strut{}1}}%
      \put(-92,1329){\makebox(0,0)[r]{\strut{}}}%
      \put(-92,1604){\makebox(0,0)[r]{\strut{}2}}%
      \put(-92,1878){\makebox(0,0)[r]{\strut{}}}%
      \put(-92,2153){\makebox(0,0)[r]{\strut{}3}}%
      \put(-92,2428){\makebox(0,0)[r]{\strut{}}}%
      \put(-92,2703){\makebox(0,0)[r]{\strut{}4}}%
      \put(72,273){\makebox(0,0){\strut{}$1$}}%
      \put(3181,273){\makebox(0,0){\strut{}$10$}}%
      \put(1145,3126){\makebox(0,0){\strut{}$10$}}%
      \put(2699,3126){\makebox(0,0){\strut{}$100$}}%
      \put(2155,1786){\makebox(0,0)[l]{\strut{}\scriptsize $m_{\rm s},\, \sin^2 2\theta$}}%
    }%
    \gplgaddtomacro\gplfronttext{%
      \csname LTb\endcsname
      \put(-607,1713){\rotatebox{-270}{\makebox(0,0){\strut{}\LARGE $N_{\rm eff}$}}}%
      \put(1626,-147){\makebox(0,0){\strut{}\Large $T_{\rm RH}$ (MeV)}}%
      \put(1626,3602){\makebox(0,0){\strut{}\Large $\Gamma_\phi$\ \,(${\rm sec^{-1}}$)}}%
      \csname LTb\endcsname
      \put(2109,1573){\makebox(0,0)[l]{\strut{}\,\scriptsize $ \,1 \,{\rm eV}, 10^{-3}$}}%
      \csname LTb\endcsname
      \put(2109,1389){\makebox(0,0)[l]{\strut{}\,\scriptsize $ \,1 \,{\rm eV}, 10^{-2}$}}%
      \csname LTb\endcsname
      \put(2109,1205){\makebox(0,0)[l]{\strut{}\,\scriptsize $ \,1 \,{\rm eV}, 10^{-1}$}}%
      \csname LTb\endcsname
      \put(2109,1021){\makebox(0,0)[l]{\strut{}\,\scriptsize $ \,1 \,{\rm keV}, 10^{-3}$}}%
      \csname LTb\endcsname
      \put(2109,837){\makebox(0,0)[l]{\strut{}\,\scriptsize $ \,1 \,{\rm keV}, 10^{-2}$}}%
      \csname LTb\endcsname
      \put(2109,653){\makebox(0,0)[l]{\strut{}\,\scriptsize $ \,1 \,{\rm keV}, 10^{-1}$}}%
    }%
    \gplgaddtomacro\gplbacktext{%
      \csname LTb\endcsname
      \put(3882,504){\makebox(0,0)[r]{\strut{}0}}%
      \put(3882,1604){\makebox(0,0)[r]{\strut{}}}%
      \put(3882,2703){\makebox(0,0)[r]{\strut{}1}}%
      \put(4046,273){\makebox(0,0){\strut{}$1$}}%
      \put(7155,273){\makebox(0,0){\strut{}$10$}}%
      \put(5119,3126){\makebox(0,0){\strut{}$10$}}%
      \put(6673,3126){\makebox(0,0){\strut{}$100$}}%
    }%
    \gplgaddtomacro\gplfronttext{%
      \csname LTb\endcsname
      \put(3551,1713){\rotatebox{-270}{\makebox(0,0){\strut{}\LARGE $N_{{\rm eff},\,\nu_{\rm s}}$}}}%
      \put(5600,-147){\makebox(0,0){\strut{}\Large $T_{\rm RH}$ (MeV)}}%
      \put(5600,3602){\makebox(0,0){\strut{}\Large $\Gamma_\phi$\ \,(${\rm sec^{-1}}$)}}%
    }%
    \gplbacktext
    \put(0,0){\includegraphics{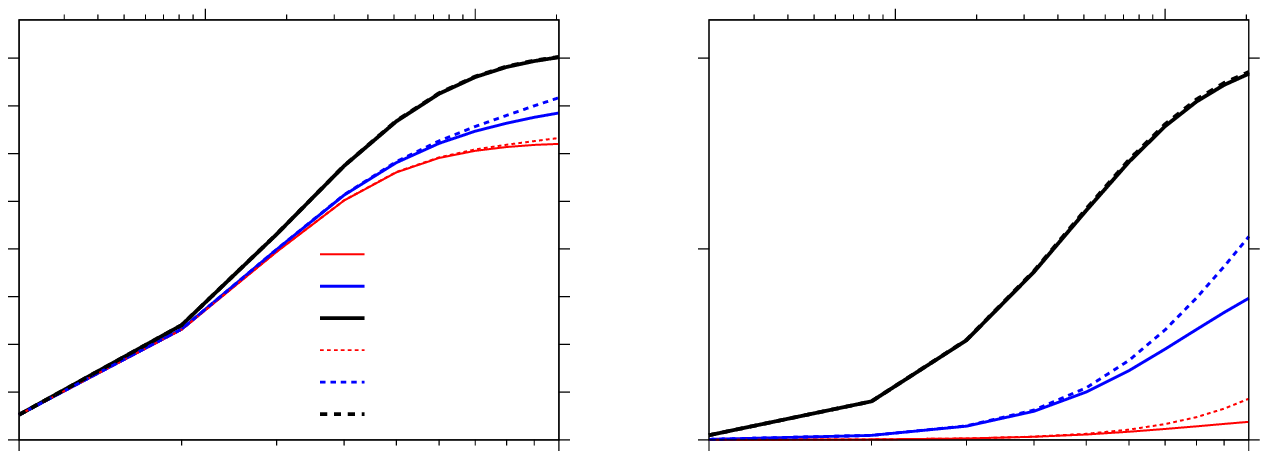}}%
    \gplfronttext
  \end{picture}%
\endgroup

%% file: result_bbn.tex
\section{Numerical result: Big Bang Nucleosynthesis} \label{sec:results_bbn}
The thermalization of neutrinos is closely associated with BBN.  In
this section, we introduce the formalism for the calculation of BBN
after a brief introduction to the role of neutrinos in BBN.  For the
detail of the theoretical framework, we refer the reader to 
Ref.~\cite{Hasegawa_2019a}. The numerical results of the BBN 
calculation in cosmological models with MeV-scale reheating temperature 
is presented in the latter part of this section.

\subsection{Neutrino thermalization and neutron-to-proton ratio}
Neutrinos affect the light-element abundances synthesized in BBN 
since they are involved in the exchange reactions between protons and neutrons:
\begin{subequations}
\begin{align}
 n &\leftrightarrow p + e^- + \bar{\nu}_e\,,\label{eq:np_rate1b}\\
 e^+ + n &\leftrightarrow p + \bar{\nu}_e\,,\label{eq:np_rate2b}\\
 \nu_e + n &\leftrightarrow p + e^-\,,\label{eq:np_rate3b}
\end{align}
\end{subequations}
which set the neutron-to-proton ratio $(n/p)$ before the nucleosynthesis.
The neutron-to-proton ratio is one of the most important parameters in BBN, 
which determines the final abundances of light elements. In particular, 
the mass fraction of $^4$He, which is denoted as $Y_p$, is written in a 
simple analytical form of $Y_p \sim 1/\{1+(n/p)_{\rm bbn}^{-1}\}$. 
The neutron-to-proton ratio $(n/p)_{\rm bbn}$ is the value just before 
the deuterium bottleneck opens and the synthesis of light elements becomes 
effective, which corresponds to the cosmic time of $t_{\rm bbn} \sim 200$~sec 
(or the temperature of $T_{\rm bbn} \sim 80$~keV). The freeze-out value of 
the ratio $(n/p)_f$ is related to $(n/p)_{\rm bbn}$ as 
\begin{equation}
 (n/p)_{\rm bbn} = (n/p)_f\,e^{-t_{\rm bbn}/\tau_n},
\end{equation}
where $\tau_n = 880.2 \pm 1.0\ \,{\rm sec}\ (68\%\,{\rm C.L.})$ is the 
neutron lifetime~\cite{PDG2018}. The freeze-out value is $(n/p)_f \sim 1/6$ 
in the standard big-bang model, where all active neutrinos are fully 
thermalized well-before BBN, and hence $(n/p)_{\rm bbn} \sim 1/7$ and $Y_p \sim 0.25$~\cite{PDG2018}. 

As discussed in Secs.~\ref{sec:dynamics} and \ref{sec:results_reheating}, neutrinos 
are not completely thermalized in cosmological models with MeV-scale reheating temperature. 
The incomplete thermalization of neutrinos changes both the freeze-out 
of the processes~\eqref{eq:np_rate1b}--\eqref{eq:np_rate3b} and the Hubble expansion 
rate. Therefore, we expect a different $(n/p)_f$ value from that attained in the 
standard big-bang model in this case. Refs.~\cite{Kawasaki_1999, Kawasaki_2000} 
provided a comprehensive discussion of the mechanism. The abundances of the other 
light elements such as D, $^3$He, $^6$Li, and $^7$Li are very sensitive to the 
production abundance of $^4$He, which is the second most abundant element in the 
Universe. Consequently, light-element abundances in the low-reheating scenario 
are different from those of the standard big-bang model. Also, among the neutrino 
species only $\nu_e$ is relevant to the processes~\eqref{eq:np_rate1b}--\eqref{eq:np_rate3b}. 
For that reason, light-element abundances are highly sensitive to the $\nu_e$ spectrum, 
and any physics changing the neutrino flavors in the early Universe such as the neutrino 
self-interaction and neutrino oscillation plays an important role in the synthesis of 
light elements. 

\subsection{Observational abundances}
The abundances of deuterium and helium in the current Universe are 
measured with ${\cal O}(1)$\% accuracy. The baryon-to-proton ratio 
$\eta_B$, which is the only free parameter of the standard theory of BBN, 
is also determined with high precision using these measurements~\cite{PDG2018}.
In this study, we adopt the primordial mass fraction of $^4$He reported in Ref.~\cite{Aver_2015}:
\begin{equation}
 Y_p = 0.2449 \pm 0.0040\ \ (68\%\,{\rm C.L.}) \,,
\label{eq:YpAber2015}
\end{equation}
which was obtained from the observation of the recombination line of
metal-poor stars in the extra-galactic HI\hspace{-.1em}I regions or
blue compact galaxies.~\footnote{ 
  A slightly large value of the $^4$He
  abundance was reported in Ref.~\cite{Izotov:2014fga},
  $Y_p = 0.2551 \pm 0.0022\ (68\%\,{\rm C.L.})$.  Since the authors of
  Ref.~\cite{Aver_2015} reanalyzed the same dataset used in
  Ref.~\cite{Izotov:2014fga}, it should be reasonable to adopt the
  value reported in Ref.~\cite{Aver_2015}.  } 
For deuterium, we
adopt~\cite{Zavarygin_2018}
\begin{equation}
 {\rm D/H} = (2.545 \pm 0.025) \times 10^{-5}\ \ (68\%\,{\rm C.L.}) \,,
\label{eq:Dobs}
\end{equation}
which was determined using the absorption spectra in high-redshift
metal-poor quasar absorption systems.~\footnote{ The authors of
  Ref.~\cite{Cooke:2017cwo} reported
  ${\rm D/H} = (2.527 \pm 0.030) \times 10^{-5}\ (68\%\,{\rm C.L.})$,
  which is similar to (\ref{eq:Dobs}).  Since the difference of the
  mean values between this value and (\ref{eq:Dobs}) falls within a
  $1$-$\sigma$ error, our result does not change even if we adopted this
  value to be the primordial D/H.}

\subsection{Numerical calculation}
We numerically solve the code of the reaction network for light
elements based on the Kawano code~\cite{Kawano_1992} with the updated
nuclear reaction rates reported in
Refs.~\cite{Kernan:1994je,Angulo:1999zz,Serpico:2004gx,Pisanti:2007hk,Cyburt:2008up}
(see Ref.~\cite{Kawasaki_2018} for more details).  The contribution of
$\phi$ is accounted for in the Friedman equation,
Eq.~\eqref{eq:Friedmann}, and the energy conservation equation,
Eq.~(2.26). 
Also, we pre-evaluate the energy densities of active and sterile
neutrinos together with weak reaction rates of the
processes~\eqref{eq:np_rate1b}--\eqref{eq:np_rate3b} with the LASAGNA
code and interpolate the data in the BBN code.  In the standard BBN,
the baryon-to-photon ratio $\eta_B$ is the only free parameter of the
theory.  Currently, the value of $\eta_B$ is precisely determined from
the observation of CMB in the Planck
collaboration~\cite{Planck_2018}. In this study, we adopt the Planck
bound on $\eta_B$ for the base $\Lambda$CDM model extended by two
additional parameters, namely the effective number of neutrino
species, $N_{\rm eff}$, and the effective mass of sterile neutrinos,
$m^{\rm eff}_{\rm s}$:~\footnote{
 The thermalization of neutrinos and light-element abundances affect the 
 recombination, and this is therefore an approximate treatment. 
 Ref.~\cite{Planck_2018} assumes that three-flavors of active neutrinos 
 are completely thermalized and adopt the value of $Y_p$ in the standard 
 big-bang model.
}
\begin{equation}
 \eta_B = (6.14 \pm 0.04) \times 10^{-10}\ \ (68\%\,{\rm C.L.})\,,
\label{eq:baryon_asym}
\end{equation}
as a prior for the calculation of BBN.

The light-element abundance is sensitive to the decay mode of the parent 
particle $\phi$~\cite{Kawasaki_2000,Hannestad_2004,Hasegawa_2019a}. 
In this study, we assume that the parent particle $\phi$ decays into both radiation 
({\it i.e.} photons and charged leptons) and hadrons ({\it i.e.} quarks and gluons) 
as in Refs.~\cite{Kawasaki_2000,Hasegawa_2019a}. For the case of the 
direct decay of the parent particle into active neutrino pairs 
$\phi \rightarrow \nu_\alpha + \bar{\nu}_\alpha$ ($\alpha = e, \mu, \tau$), 
see Ref.~\cite{Hannestad_2004}.

The effect of the particle injection from the parent particle $\phi$ is 
more significant in the case where the hadronic branching ratio $B_h \neq 0$.
In the case where the parent particle decays exclusively into radiation,
the thermal bath of photons and electrons is instantaneously produced by 
cascade reactions through electromagnetic interaction, and active 
neutrinos are gradually produced from the thermal bath through the weak 
interaction. For $B_h \neq 0$, mesons, baryons, and their anti-particles 
are copiously produced from quarks and gluons after the hadronization, and they induce additional 
exchange reactions between neutrons and protons. Consequently, it affects 
the freeze-out value of the neutron-to-proton ratio $(n/p)_f$ and leads to 
different outcomes of BBN~\cite{Reno_1988,Kawasaki_2000,Hasegawa_2019a}.
In our calculation, we take into account the hadronic effect induced by 
charged pions ($\pi^\pm$) and nucleons ($n, \bar{n}, p, \bar{p}$) injected 
from the decay of $\phi$ and neglect the effect of other hadronic particles 
for conservative treatment. We use the thermal reaction rates for the 
hadronic processes interchanging neutrons and protons. This treatment 
is justified since most of the injected hadrons are instantaneously stopped 
by inverse Compton-like scattering or Coulomb interaction with background 
particles ({\it i.e.} mainly photons and electrons) and reach equilibrium~\cite{Kohri:2001jx,Kawasaki:2004qu}. 
Also, we evaluate the number of hadrons produced in the hadronic decay using 
the Pythia~8.2 code~\cite{Sjostrand_2006,Sjostrand_2007}, 
assuming the decays of $\phi$ into the $u\bar{u}$ pair as a specific process 
for the quark production.
 
\begin{figure}[!t]
\vspace{0.5cm}
\begin{center}
\vspace{-3cm}
\resizebox{15cm}{!}{\hspace{1.5cm}\input{Trh_vs_Yp_D_radiative}}
\end{center}
\vspace{0.7cm}
 \caption{
 Mass fraction of $^4$He, $Y_p$, and the deuterium-to-hydrogen ratio, 
 D/H, as functions of the reheating temperature $T_{\rm RH}$ for the cases corresponding to the 
 100\% radiative decay of the parent particle $\phi$, with the assumption of $\nu_e$--$\nu_{\rm s}$ mixing. 
 The value of the baryon-to-photon ratio is fixed to $\eta_B = 6.14 \times 10^{-10}$
 in the figure. The red solid line is for sterile neutrinos with
 the best-fit mixing parameters $(\delta m^2,\sin^2 2\theta) = (1.29~{\rm eV}^2\,,\ 0.035)$ 
 reported in Ref.~\cite{Dentler_2018}, and the 
 case without sterile neutrinos is plotted by the black dotted line
 for reference. The 2-$\sigma$ observational bounds on $Y_p$
 (Ref.~\cite{Aver_2015}) and D/H (Ref.~\cite{Zavarygin_2018}) are also
 shown by the gray-shaded regions.  
 } \label{fig:trh_vs_D_and_Yp_radiative}
\end{figure}
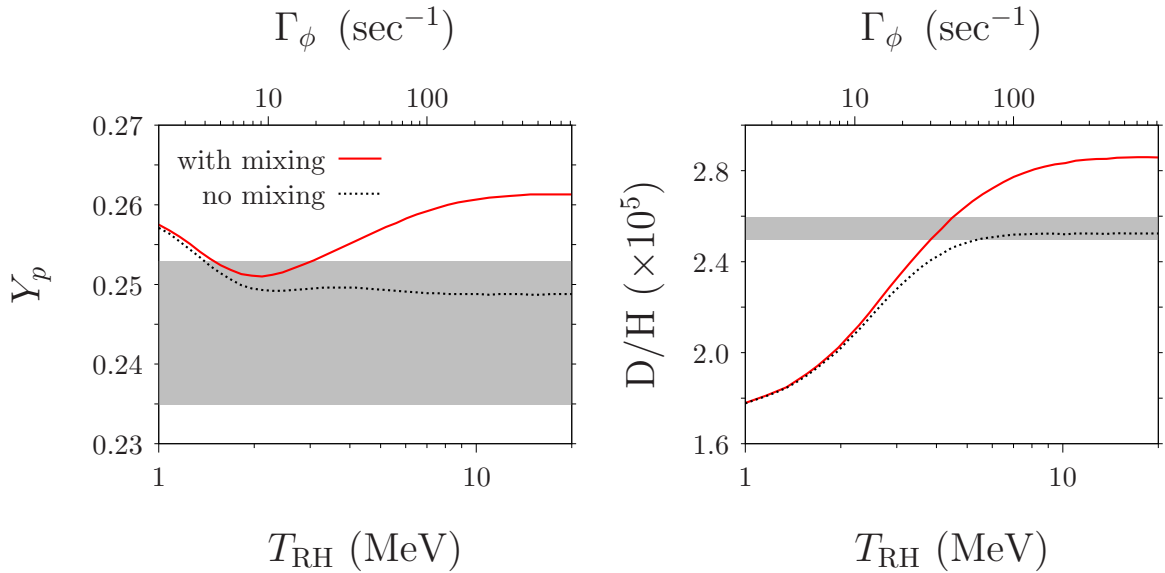
 
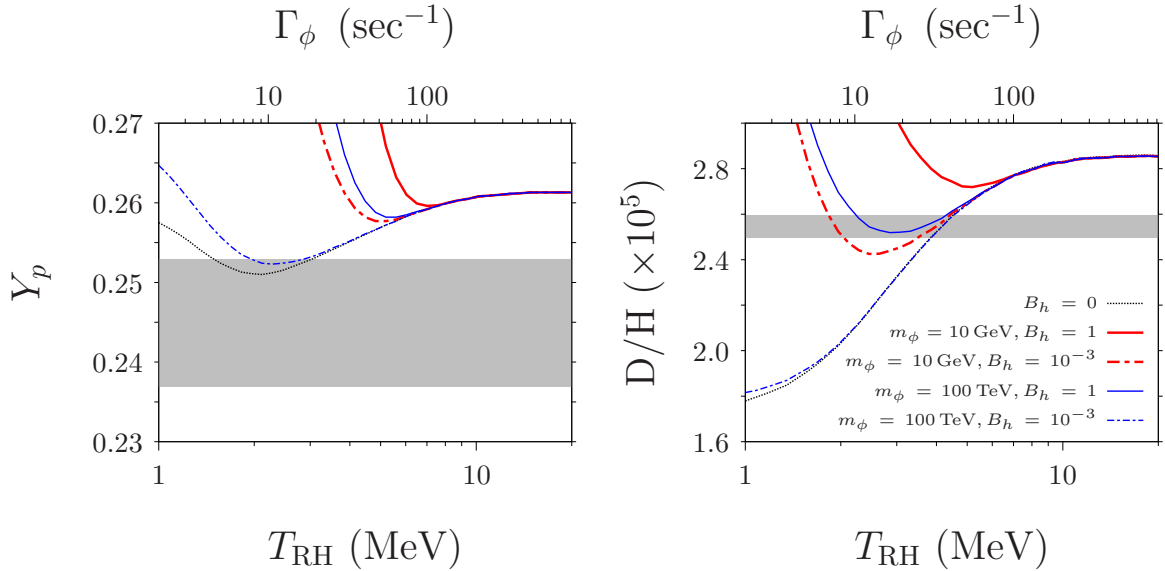
\begin{figure}[!t]
\vspace{0.5cm}
\begin{center}
\vspace{-3cm}
\resizebox{15cm}{!}{\hspace{1.5cm}\input{Trh_vs_Yp_D_hadronic}}
\end{center}
\vspace{0.7cm}
 \caption{
 Same as Fig.~\ref{fig:trh_vs_D_and_Yp_radiative}, but for the cases with the 
 hadronic decay of the parent particle $\phi$. The case of $B_h = 0$ (black dotted) 
 corresponds to the 100\% radiative decay of $\phi$.
 } \label{fig:trh_vs_D_and_Yp_hadronic}
\end{figure}

Figure~\ref{fig:trh_vs_D_and_Yp_radiative} shows the abundances of helium ($Y_p$) 
and deuterium (D/H) as a function of the reheating temperature $T_{\rm RH}$ 
for the 100\% radiative decay cases. The mixing parameters of sterile 
neutrinos are fixed to the best-fit values in Ref.~\cite{Dentler_2018}, 
as in the previous section.
Also, the baryon-to-photon ratio $\eta_B$ is set to be the median value of Eq.~\eqref{eq:baryon_asym}. 
As seen in Fig.~\ref{fig:trh_vs_D_and_Yp_radiative}, 
the light-element abundance increases with the reheating temperature.
This is because sterile neutrinos are more abundantly produced for a
large $T_{\rm RH}$ (see Fig.~\ref{fig:trh_vs_neff}), and therefore the
expansion rate $H$ increases. This leads to the early decoupling of
the exchange reactions between protons and
neutrons~\eqref{eq:np_rate1b}--\eqref{eq:np_rate3b}.  Consequently,
the freeze-out temperature $T_c$ increases and more neutrons remain
unburnt, which increases $(n/p)_f$ and light-element abundances, {\it i.e.} $Y_p$ and D/H. 
In Fig.~\ref{fig:trh_vs_D_and_Yp_radiative}, 
a similar dependence of $Y_p$ and D/H on the reheating temperature 
can be seen even for the case without sterile neutrinos. For small
$T_{\rm RH}$ an incomplete thermalization of active neutrinos
decreases the expansion rate, and $Y_p$ and D/H decrease due to the
effect. The increase of $Y_p$ for small $T_{\rm RH}$ is caused by a
decrease in the weak rates responsible for the interconversion between
protons and neutrons, $\Gamma_{np}$. This accelerates the decoupling
of the processes~\eqref{eq:np_rate1b}--\eqref{eq:np_rate3b} and plays 
a role in increasing the freeze-out value of the neutron-to-proton 
ratio $(n/p)_f$ and $Y_p$. The increase and decrease of $Y_p$ due to
small $T_{\rm RH}$ are competing, but the former dominates the latter
for large $T_{\rm RH}$ and the opposite is true for small $T_{\rm RH}$ 
(see Refs.~\cite{Kawasaki_2000,Hasegawa_2019a} for more detail).

Figure~7 shows the results of the hadronic decay cases. 
It can be seen that light-element abundances increase for $T_{\rm RH} \lesssim 10$~MeV due 
to the hadronic decay effect compared to those for the 100\% radiative decay cases. 
This is because the injection of the high-energy hadrons induces additional exchange 
reactions between protons and neutrons, $p + N \leftrightarrow n + N'$, where $N$ and $N'$ 
are mesons or baryons, to equilibrate the number densities of protons and 
neutrons~\cite{Reno_1988,Kawasaki_2000,Hasegawa_2019a,Kohri:2001jx}. 
This is not true for $T_{\rm RH} \gtrsim 10$~MeV because the hadronic decay 
occurs much before the decoupling of the processes~\eqref{eq:np_rate1b}--\eqref{eq:np_rate3b}, 
and the neutron-to-proton ratio is subsequently equilibrated by the weak 
processes again, which erases the hadronic decay effect. 

%% file: Trh_vs_Yp_D_radiative.tex
\begingroup
  \makeatletter
  \providecommand\color[2][]{%
    \GenericError{(gnuplot) \space\space\space\@spaces}{%
      Package color not loaded in conjunction with
      terminal option `colourtext'%
    }{See the gnuplot documentation for explanation.%
    }{Either use 'blacktext' in gnuplot or load the package
      color.sty in LaTeX.}%
    \renewcommand\color[2][]{}%
  }%
  \providecommand\includegraphics[2][]{%
    \GenericError{(gnuplot) \space\space\space\@spaces}{%
      Package graphicx or graphics not loaded%
    }{See the gnuplot documentation for explanation.%
    }{The gnuplot epslatex terminal needs graphicx.sty or graphics.sty.}%
    \renewcommand\includegraphics[2][]{}%
  }%
  \providecommand\rotatebox[2]{#2}%
  \@ifundefined{ifGPcolor}{%
    \newif\ifGPcolor
    \GPcolortrue
  }{}%
  \@ifundefined{ifGPblacktext}{%
    \newif\ifGPblacktext
    \GPblacktexttrue
  }{}%
  \let\gplgaddtomacro\g@addto@macro
  \gdef\gplbacktext{}%
  \gdef\gplfronttext{}%
  \makeatother
  \ifGPblacktext
    \def\colorrgb#1{}%
    \def\colorgray#1{}%
  \else
    \ifGPcolor
      \def\colorrgb#1{\color[rgb]{#1}}%
      \def\colorgray#1{\color[gray]{#1}}%
      \expandafter\def\csname LTw\endcsname{\color{white}}%
      \expandafter\def\csname LTb\endcsname{\color{black}}%
      \expandafter\def\csname LTa\endcsname{\color{black}}%
      \expandafter\def\csname LT0\endcsname{\color[rgb]{1,0,0}}%
      \expandafter\def\csname LT1\endcsname{\color[rgb]{0,1,0}}%
      \expandafter\def\csname LT2\endcsname{\color[rgb]{0,0,1}}%
      \expandafter\def\csname LT3\endcsname{\color[rgb]{1,0,1}}%
      \expandafter\def\csname LT4\endcsname{\color[rgb]{0,1,1}}%
      \expandafter\def\csname LT5\endcsname{\color[rgb]{1,1,0}}%
      \expandafter\def\csname LT6\endcsname{\color[rgb]{0,0,0}}%
      \expandafter\def\csname LT7\endcsname{\color[rgb]{1,0.3,0}}%
      \expandafter\def\csname LT8\endcsname{\color[rgb]{0.5,0.5,0.5}}%
    \else
      \def\colorrgb#1{\color{black}}%
      \def\colorgray#1{\color[gray]{#1}}%
      \expandafter\def\csname LTw\endcsname{\color{white}}%
      \expandafter\def\csname LTb\endcsname{\color{black}}%
      \expandafter\def\csname LTa\endcsname{\color{black}}%
      \expandafter\def\csname LT0\endcsname{\color{black}}%
      \expandafter\def\csname LT1\endcsname{\color{black}}%
      \expandafter\def\csname LT2\endcsname{\color{black}}%
      \expandafter\def\csname LT3\endcsname{\color{black}}%
      \expandafter\def\csname LT4\endcsname{\color{black}}%
      \expandafter\def\csname LT5\endcsname{\color{black}}%
      \expandafter\def\csname LT6\endcsname{\color{black}}%
      \expandafter\def\csname LT7\endcsname{\color{black}}%
      \expandafter\def\csname LT8\endcsname{\color{black}}%
    \fi
  \fi
    \setlength{\unitlength}{0.0500bp}%
    \ifx\gptboxheight\undefined%
      \newlength{\gptboxheight}%
      \newlength{\gptboxwidth}%
      \newsavebox{\gptboxtext}%
    \fi%
    \setlength{\fboxrule}{0.5pt}%
    \setlength{\fboxsep}{1pt}%
\begin{picture}(7200.00,5040.00)%
      \csname LTb\endcsname
      \put(3600,4760){\makebox(0,0){\strut{}}}%
    \gplgaddtomacro\gplbacktext{%
      \csname LTb\endcsname
      \put(-43,504){\makebox(0,0)[r]{\strut{}0.23}}%
      \put(-43,787){\makebox(0,0)[r]{\strut{}}}%
      \put(-43,1071){\makebox(0,0)[r]{\strut{}0.24}}%
      \put(-43,1354){\makebox(0,0)[r]{\strut{}}}%
      \put(-43,1637){\makebox(0,0)[r]{\strut{}0.25}}%
      \put(-43,1921){\makebox(0,0)[r]{\strut{}}}%
      \put(-43,2204){\makebox(0,0)[r]{\strut{}0.26}}%
      \put(-43,2488){\makebox(0,0)[r]{\strut{}}}%
      \put(-43,2771){\makebox(0,0)[r]{\strut{}0.27}}%
      \put(72,273){\makebox(0,0){\strut{}$1$}}%
      \put(2313,273){\makebox(0,0){\strut{}$10$}}%
      \put(845,2974){\makebox(0,0){\strut{}$10$}}%
      \put(1965,2974){\makebox(0,0){\strut{}$100$}}%
    }%
    \gplgaddtomacro\gplfronttext{%
      \csname LTb\endcsname
      \put(-844,1637){\rotatebox{-270}{\makebox(0,0){\strut{}\Large $Y_p$}}}%
      \put(1529,-259){\makebox(0,0){\strut{}\Large $T_{\rm RH}$ (MeV)}}%
      \put(1529,3450){\makebox(0,0){\strut{}\Large $\Gamma_\phi$\ \,(${\rm sec^{-1}}$)}}%
      \csname LTb\endcsname
      \put(1281,2533){\makebox(0,0)[r]{\strut{}\small with mixing\,}}%
      \csname LTb\endcsname
      \put(1281,2283){\makebox(0,0)[r]{\strut{}\small no mixing\,}}%
    }%
    \gplgaddtomacro\gplbacktext{%
      \csname LTb\endcsname
      \put(4097,504){\makebox(0,0)[r]{\strut{}1.6}}%
      \put(4097,828){\makebox(0,0)[r]{\strut{}}}%
      \put(4097,1152){\makebox(0,0)[r]{\strut{}2.0}}%
      \put(4097,1476){\makebox(0,0)[r]{\strut{}}}%
      \put(4097,1799){\makebox(0,0)[r]{\strut{}2.4}}%
      \put(4097,2123){\makebox(0,0)[r]{\strut{}}}%
      \put(4097,2447){\makebox(0,0)[r]{\strut{}2.8}}%
      \put(4097,2771){\makebox(0,0)[r]{\strut{}}}%
      \put(4212,273){\makebox(0,0){\strut{}$1$}}%
      \put(6453,273){\makebox(0,0){\strut{}$10$}}%
      \put(4985,2974){\makebox(0,0){\strut{}$10$}}%
      \put(6105,2974){\makebox(0,0){\strut{}$100$}}%
    }%
    \gplgaddtomacro\gplfronttext{%
      \csname LTb\endcsname
      \put(3497,1637){\rotatebox{-270}{\makebox(0,0){\strut{}\Large D/H $(\times 10^5)$}}}%
      \put(5669,-259){\makebox(0,0){\strut{}\Large $T_{\rm RH}$ (MeV)}}%
      \put(5669,3450){\makebox(0,0){\strut{}\Large $\Gamma_\phi$\ \,(${\rm sec^{-1}}$)}}%
    }%
    \gplbacktext
    \put(0,0){\includegraphics{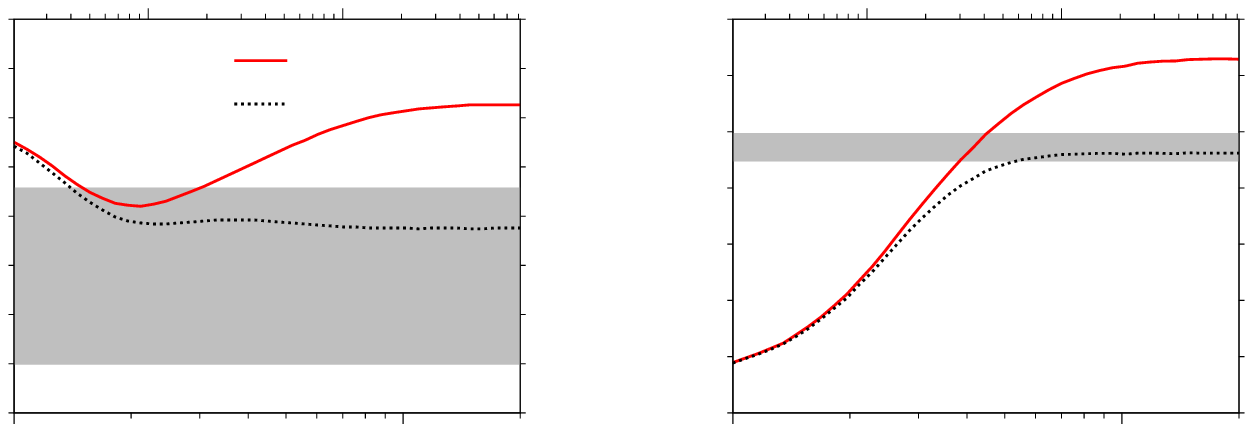}}%
    \gplfronttext
  \end{picture}%
\endgroup

%% file: Trh_vs_Yp_D_hadronic.tex
\begingroup
  \makeatletter
  \providecommand\color[2][]{%
    \GenericError{(gnuplot) \space\space\space\@spaces}{%
      Package color not loaded in conjunction with
      terminal option `colourtext'%
    }{See the gnuplot documentation for explanation.%
    }{Either use 'blacktext' in gnuplot or load the package
      color.sty in LaTeX.}%
    \renewcommand\color[2][]{}%
  }%
  \providecommand\includegraphics[2][]{%
    \GenericError{(gnuplot) \space\space\space\@spaces}{%
      Package graphicx or graphics not loaded%
    }{See the gnuplot documentation for explanation.%
    }{The gnuplot epslatex terminal needs graphicx.sty or graphics.sty.}%
    \renewcommand\includegraphics[2][]{}%
  }%
  \providecommand\rotatebox[2]{#2}%
  \@ifundefined{ifGPcolor}{%
    \newif\ifGPcolor
    \GPcolortrue
  }{}%
  \@ifundefined{ifGPblacktext}{%
    \newif\ifGPblacktext
    \GPblacktexttrue
  }{}%
  \let\gplgaddtomacro\g@addto@macro
  \gdef\gplbacktext{}%
  \gdef\gplfronttext{}%
  \makeatother
  \ifGPblacktext
    \def\colorrgb#1{}%
    \def\colorgray#1{}%
  \else
    \ifGPcolor
      \def\colorrgb#1{\color[rgb]{#1}}%
      \def\colorgray#1{\color[gray]{#1}}%
      \expandafter\def\csname LTw\endcsname{\color{white}}%
      \expandafter\def\csname LTb\endcsname{\color{black}}%
      \expandafter\def\csname LTa\endcsname{\color{black}}%
      \expandafter\def\csname LT0\endcsname{\color[rgb]{1,0,0}}%
      \expandafter\def\csname LT1\endcsname{\color[rgb]{0,1,0}}%
      \expandafter\def\csname LT2\endcsname{\color[rgb]{0,0,1}}%
      \expandafter\def\csname LT3\endcsname{\color[rgb]{1,0,1}}%
      \expandafter\def\csname LT4\endcsname{\color[rgb]{0,1,1}}%
      \expandafter\def\csname LT5\endcsname{\color[rgb]{1,1,0}}%
      \expandafter\def\csname LT6\endcsname{\color[rgb]{0,0,0}}%
      \expandafter\def\csname LT7\endcsname{\color[rgb]{1,0.3,0}}%
      \expandafter\def\csname LT8\endcsname{\color[rgb]{0.5,0.5,0.5}}%
    \else
      \def\colorrgb#1{\color{black}}%
      \def\colorgray#1{\color[gray]{#1}}%
      \expandafter\def\csname LTw\endcsname{\color{white}}%
      \expandafter\def\csname LTb\endcsname{\color{black}}%
      \expandafter\def\csname LTa\endcsname{\color{black}}%
      \expandafter\def\csname LT0\endcsname{\color{black}}%
      \expandafter\def\csname LT1\endcsname{\color{black}}%
      \expandafter\def\csname LT2\endcsname{\color{black}}%
      \expandafter\def\csname LT3\endcsname{\color{black}}%
      \expandafter\def\csname LT4\endcsname{\color{black}}%
      \expandafter\def\csname LT5\endcsname{\color{black}}%
      \expandafter\def\csname LT6\endcsname{\color{black}}%
      \expandafter\def\csname LT7\endcsname{\color{black}}%
      \expandafter\def\csname LT8\endcsname{\color{black}}%
    \fi
  \fi
    \setlength{\unitlength}{0.0500bp}%
    \ifx\gptboxheight\undefined%
      \newlength{\gptboxheight}%
      \newlength{\gptboxwidth}%
      \newsavebox{\gptboxtext}%
    \fi%
    \setlength{\fboxrule}{0.5pt}%
    \setlength{\fboxsep}{1pt}%
\begin{picture}(7200.00,5040.00)%
      \csname LTb\endcsname
      \put(3600,4760){\makebox(0,0){\strut{}}}%
    \gplgaddtomacro\gplbacktext{%
      \csname LTb\endcsname
      \put(-43,504){\makebox(0,0)[r]{\strut{}0.23}}%
      \put(-43,787){\makebox(0,0)[r]{\strut{}}}%
      \put(-43,1071){\makebox(0,0)[r]{\strut{}0.24}}%
      \put(-43,1354){\makebox(0,0)[r]{\strut{}}}%
      \put(-43,1637){\makebox(0,0)[r]{\strut{}0.25}}%
      \put(-43,1921){\makebox(0,0)[r]{\strut{}}}%
      \put(-43,2204){\makebox(0,0)[r]{\strut{}0.26}}%
      \put(-43,2488){\makebox(0,0)[r]{\strut{}}}%
      \put(-43,2771){\makebox(0,0)[r]{\strut{}0.27}}%
      \put(72,273){\makebox(0,0){\strut{}$1$}}%
      \put(2313,273){\makebox(0,0){\strut{}$10$}}%
      \put(845,2974){\makebox(0,0){\strut{}$10$}}%
      \put(1965,2974){\makebox(0,0){\strut{}$100$}}%
    }%
    \gplgaddtomacro\gplfronttext{%
      \csname LTb\endcsname
      \put(-844,1637){\rotatebox{-270}{\makebox(0,0){\strut{}\Large $Y_p$}}}%
      \put(1529,-259){\makebox(0,0){\strut{}\Large $T_{\rm RH}$ (MeV)}}%
      \put(1529,3450){\makebox(0,0){\strut{}\Large $\Gamma_\phi$\ \,(${\rm sec^{-1}}$)}}%
    }%
    \gplgaddtomacro\gplbacktext{%
      \csname LTb\endcsname
      \put(4097,504){\makebox(0,0)[r]{\strut{}1.6}}%
      \put(4097,828){\makebox(0,0)[r]{\strut{}}}%
      \put(4097,1152){\makebox(0,0)[r]{\strut{}2.0}}%
      \put(4097,1476){\makebox(0,0)[r]{\strut{}}}%
      \put(4097,1799){\makebox(0,0)[r]{\strut{}2.4}}%
      \put(4097,2123){\makebox(0,0)[r]{\strut{}}}%
      \put(4097,2447){\makebox(0,0)[r]{\strut{}2.8}}%
      \put(4097,2771){\makebox(0,0)[r]{\strut{}}}%
      \put(4212,273){\makebox(0,0){\strut{}$1$}}%
      \put(6453,273){\makebox(0,0){\strut{}$10$}}%
      \put(4985,2974){\makebox(0,0){\strut{}$10$}}%
      \put(6105,2974){\makebox(0,0){\strut{}$100$}}%
    }%
    \gplgaddtomacro\gplfronttext{%
      \csname LTb\endcsname
      \put(3497,1637){\rotatebox{-270}{\makebox(0,0){\strut{}\Large D/H $(\times 10^5)$}}}%
      \put(5669,-259){\makebox(0,0){\strut{}\Large $T_{\rm RH}$ (MeV)}}%
      \put(5669,3450){\makebox(0,0){\strut{}\Large $\Gamma_\phi$\ \,(${\rm sec^{-1}}$)}}%
      \csname LTb\endcsname
      \put(6718,1528){\makebox(0,0)[r]{\strut{}\tiny $B_h\,=\,0$\,}}%
      \csname LTb\endcsname
      \put(6718,1320){\makebox(0,0)[r]{\strut{}\tiny $m_\phi = 10$\,GeV,\,$B_h\,=\,1$\,}}%
      \csname LTb\endcsname
      \put(6718,1112){\makebox(0,0)[r]{\strut{}\tiny $m_\phi\,=\,10$\,GeV,\,$B_h\,=\,10^{-3}$\,}}%
      \csname LTb\endcsname
      \put(6718,904){\makebox(0,0)[r]{\strut{}\tiny $m_\phi\,=\,100$\,TeV,\,$B_h\,=\,1$\,}}%
      \csname LTb\endcsname
      \put(6718,696){\makebox(0,0)[r]{\strut{}\tiny $m_\phi\,=\,100$\,TeV,\,$B_h\,=\,10^{-3}$\,}}%
    }%
    \gplbacktext
    \put(0,0){\includegraphics{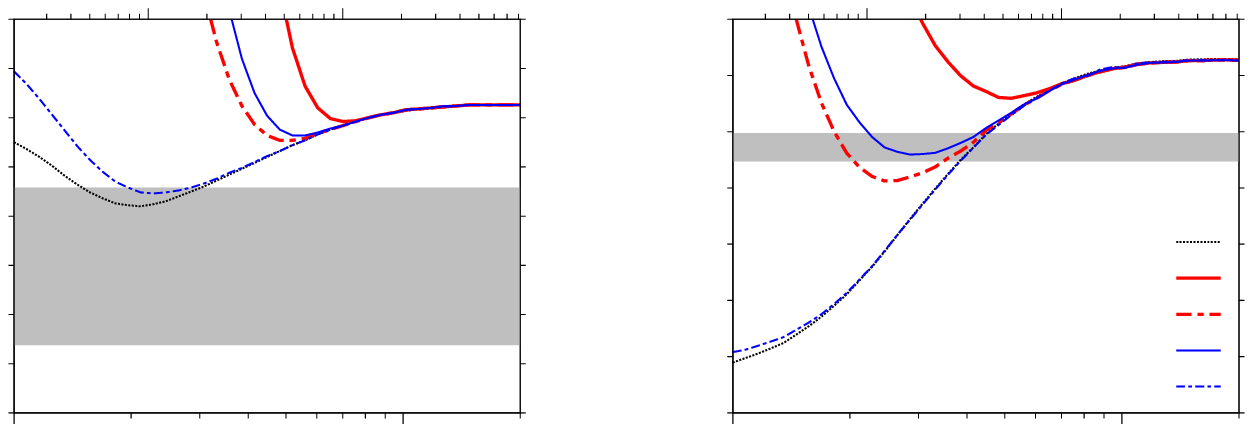}}%
    \gplfronttext
  \end{picture}%
\endgroup

%% file: result_bbncmb.tex
\section{Cosmological constraint on sterile neutrinos} \label{sec:constraint}
In this section, we summarize the cosmological constraint on sterile neutrinos,
especially focusing on the eV-scale sterile neutrinos motivated by the SBL anomaly.
Refs.~\cite{Gelmini_2004,Gelmini_2019,Gelmini_2019b} have shown that cosmological 
observations place the most stringent bound on the existence of eV-scale sterile neutrinos. 
Here we summarize the latest results of cosmological observations of light elements and 
the CMB radiation as well as ground-based neutrino experiments.

\subsection{Constraints from BBN}
Sterile neutrinos affect the synthesis of light elements as discussed 
in Sec.~I\hspace{-.1em}V. The BBN bound on sterile neutrinos 
can be obtained by requiring the agreement between theoretical predictions
and observed abundances of light elements by performing a $\chi^2$ analysis.
In this study, a $\chi^2$ function is defined as follows:
\begin{equation}
 \chi^2 \equiv \chi^2_{\rm D/H} + \chi^2_{Y_p}
 = \frac{\{{\rm (D/H)}_{\rm th} - {\rm (D/H)}_{\rm obs}\}^2}{\sigma^2_{\rm D,\,th} + \sigma^2_{\rm D,\,obs}}
 + \frac{\{Y_{p,\, {\rm th}}-Y_{p,\, {\rm obs}}\}^2}{\sigma^2_{Y_p,\, {\rm th}} + \sigma^2_{Y_p,\, {\rm obs}}} \,,
\end{equation}
where $\chi^2_{\rm D/H}$ and $\chi^2_{Y_p}$ are $\chi^2$ functions for each of D/H and $Y_p$. 
We use the suffix ``obs'' and ``th'' to denote the ``observational'' and ``theoretical'', respectively.
Also, $\sigma$ is the uncertainty of the light-element abundance. The theoretical prediction of 
light-element abundances is defined at each point on the three-dimensional grid of
($m_{\rm s}$, $\sin^2 2\theta$, $T_{\rm RH}$). The theoretical error on each grid point is
estimated by propagating the experimental errors in the nuclear reaction rates, the free neutron lifetime,
and the hadronic reaction rates in the Monte-Carlo calculation of BBN. For the thermal
cross-sections of the hadronic reactions, we use the results in Ref.~\cite{Reno_1988}
and assume a $30\%$ error in each cross-section. We define the region of $95$\% confidence level
as a parameter space satisfying
\begin{equation}
\chi^2 (m_{\rm s}, \sin^2 2\theta, T_{\rm RH}) < 5.991.
\end{equation}

\subsection{Constraints from CMB}
Thermalized sterile neutrinos contribute to $N_{\rm eff}$ and affect the 
recombination. Searching the signature in the CMB spectra, the {\it Planck} 
collaboration derived an upper limit on the effective mass of sterile neutrinos $m_{\rm s}^{\rm eff}$~\cite{Planck_2018}:
\begin{equation}
 m_{\rm s}^{\rm eff} < 0.65\ {\rm eV}. 
\end{equation}
As we assume an absence of the neutrino chemical potential, sterile
neutrinos are non-resonantly produced through its mixing with active
neutrinos by the Dodelson-Widrow mechanism. In this case, the physical 
mass $m_{\rm s}$ is related to the effective mass as $m_{\rm s}^{\mathrm{eff}}
= m_{\rm s} N_{{\rm eff},\,\nu_{\rm s}}$, where $N_{{\rm eff},\,\nu_{\rm s}}$ 
is the thermalization degree of sterile neutrinos.

Figure~8--10 summarize 
the constraints on sterile neutrinos in the parameter space of $(m_{\rm s}, \sin^2 2\theta)$. 
Figure~8 corresponds to the case assuming the standard 
big-bang model, while Figs.~9 and 10 
correspond to the low-reheating scenario assuming the 100\% radiative and hadronic 
decays of the parent particle $\phi$.~\footnote{
We refrain from plotting the result in the region with $m_{\rm s} > 10^{-1}$~keV in the
case of the standard big-bang model. This is because the peak production of sterile 
neutrinos with such a large mass occurs at or above $50$~MeV (see Eq.~\eqref{eq:tmax}) and
therefore collisions between neutrinos and muons or light mesons, which are not considered
in our computation, are non-negligible in the mass region.
} 
In each figure, we also show the current and future sensitivities of the ground-based 
experiments. The region denoted by R is already excluded from the reactor experiments 
(R)~\cite{An_2016,Declais_1994,Ashenfelter_2018}. In the future, KATRIN (KA)~\cite{Megas_2019} 
and PTOLEMY (P for 10~mg-yr and P2 for 100~g-yr exposures)~\cite{Betti_2019} 
will prove much smaller mixing angles. In these figures, we also show the 
95\% C.L. preferred regions of the sterile neutrino reported in Refs.~\cite{Dentler_2018} and \cite{Gariazzo_2017}.
Such sterile neutrinos that explain the SBL anomaly are excluded in the standard 
big-bang model from cosmological observations (Fig.~8), 
but the low reheating temperature scenario changes the picture. If all the energy 
of $\phi$ goes to radiation after the decay, such sterile neutrinos are still 
compatible with cosmological observations (Fig.~9), 
which is consistent with the previous studies with simplified 
treatments~\cite{Yaguna_2007,Gelmini_2004,Gelmini_2019,Gelmini_2019b}. 
In such a case, the production of sterile neutrinos is strongly suppressed, 
relaxing the BBN and CMB bounds. In contrast, if some part of the energy of 
$\phi$ goes to hadrons, injected hadrons induce additional interconversion 
between neutrons and protons, and the BBN bound gets more severe (Fig.~10), 
as we discussed in Sec.~\ref{sec:results_bbn}.
We note that in Fig.~10 we intentionally choose the mass and the hadronic 
branching ratio of $\phi$ to make the effect of the hadronic decay clear.
For a heavier $\phi$, $m_\phi > 10$~GeV, or a smaller hadronic branching ratio, 
$B_h < 1$, the effect of the hadronic decay should be smaller (See 
Refs.~\cite{Reno_1988,Kawasaki_2000,Hasegawa_2019a} for further discussions). 

In light of our results, the existence of the light sterile neutrino explaining
the SBL anomaly is still compatible with the observations of BBN and CMB, 
although it depends on the reheating temperature $T_{\rm RH}$, the mass of the 
parent particle $m_\phi$, and the hadronic branching ratio $B_h$ of the decay. 
In the future, such light sterile neutrinos could be detected by direct detection
experiments of the cosmic neutrino background such as the PTOLEMY project~\cite{Betti_2019}.
Also, energy spectra of active neutrinos are sensitive to the reheating temperature
as well as the mass and the mixing angle of sterile neutrinos. A direct
detection of the cosmic neutrino background should bring us reliable information
on the values of these parameters, associated with the theory beyond the standard
model of particle physics.

\newpage
 
\begin{figure}[!t]
\begin{center}
\resizebox{15cm}{!}{\input{allowed_sterile_std}}
\end{center}
 \caption{
 Constraints on sterile neutrinos in the parameter space of $(m_{\rm s}, \sin^22\theta$) 
 assuming the standard big-bang model, for the case of $\nu_e$--$\nu_{\rm s}$ mixing. 
 The 95\% C.L. bound based on the BBN calculation is shown by the blue-hatched region 
 while that on the CMB by the red region. We fix the reheating temperature $T_{\rm RH} = 5$~MeV 
 to plot the CMB bound. Other excluded regions come from Daya Bay~\cite{An_2016},
 Bugey-3~\cite{Declais_1994}, and PROSPECT~\cite{Ashenfelter_2018}~(R, gray region), the KATRIN 
 neutrino mass experiment (KA, solid-black line)~\cite{Megas_2019}. The narrow vertically-long 
 region colored in red corresponds to the best-fit region in the $\nu_e$ disappearance only 
 (left, Ref.~\cite{Dentler_2018}) and the global (right, Ref.~\cite{Gariazzo_2017}) data analysis. 
 The sensitivities with the future cosmic neutrino background experiment~\cite{Ashenfelter_2018} 
 are also shown (P and P2, green region).
} \label{fig:constraint_std_cosmo}
\end{figure}
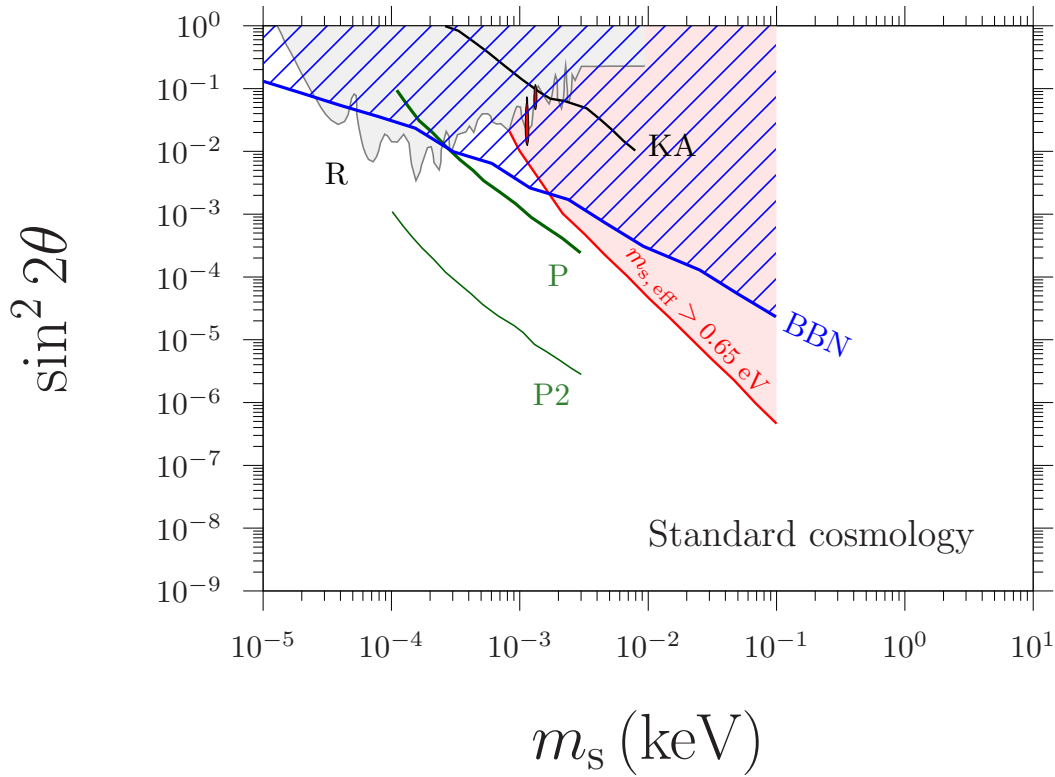

\begin{figure}[!t]
\begin{center}
\resizebox{15cm}{!}{\input{allowed_sterile_radiative}}
\end{center}
 \caption{
 The same as Fig.~8 but for the low reheating temperature case 
 assuming 100\% radiative decay of the parent particle $\phi$.
 } \label{fig:constraint_radiative}
\end{figure}
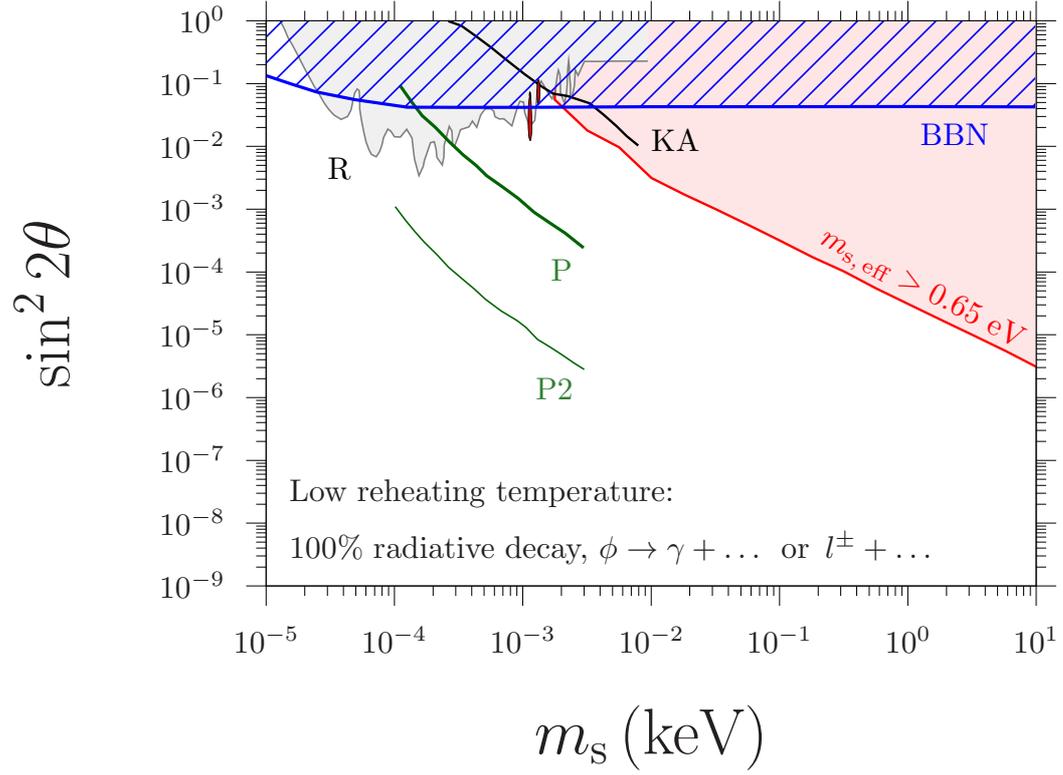
 
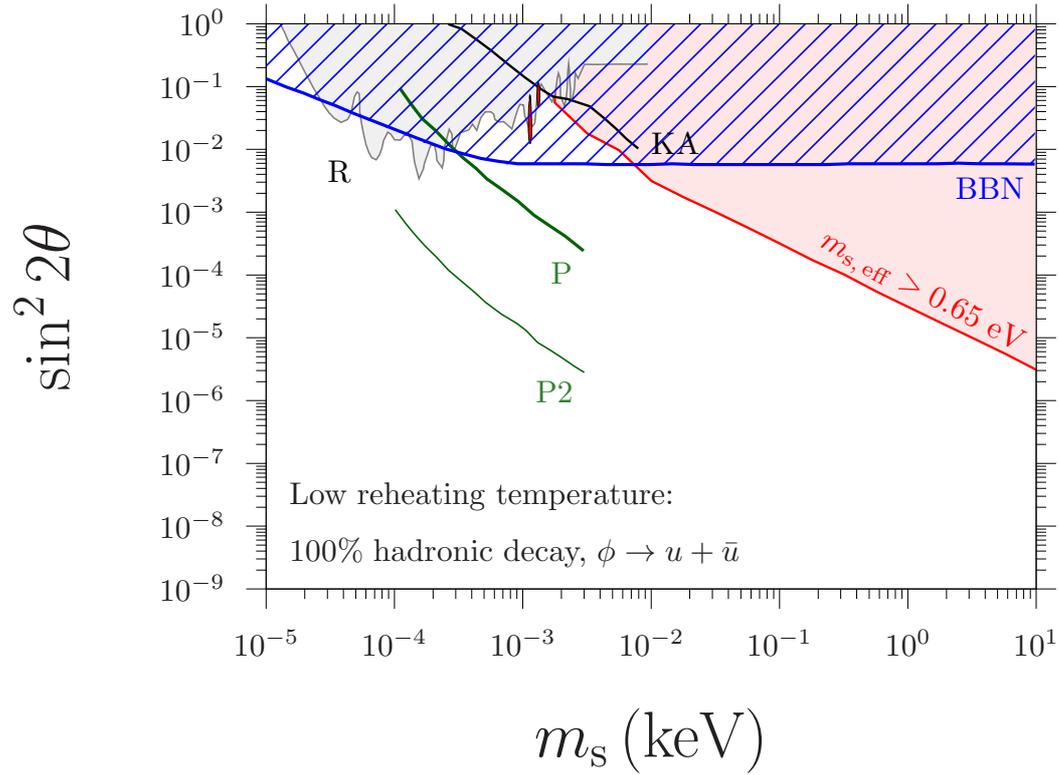
\begin{figure}[!t]
\begin{center}
\resizebox{15cm}{!}{\input{allowed_sterile_10GeV_1}}
\end{center}
 \caption{
 The same as Fig.~9 but for the case corresponding to the 100\% 
 hadronic decay of the parent particle $\phi$ with $m_\phi=10$~GeV. Compared to Fig.~9, 
 the BBN bound is more severe and the preferred regions in Refs.~\cite{Dentler_2018} and \cite{Gariazzo_2015}
 are excluded.
 } \label{fig:constraint_hadronic}
\end{figure}

%% file: allowed_sterile_std.tex
\begingroup
  \makeatletter
  \providecommand\color[2][]{%
    \GenericError{(gnuplot) \space\space\space\@spaces}{%
      Package color not loaded in conjunction with
      terminal option `colourtext'%
    }{See the gnuplot documentation for explanation.%
    }{Either use 'blacktext' in gnuplot or load the package
      color.sty in LaTeX.}%
    \renewcommand\color[2][]{}%
  }%
  \providecommand\includegraphics[2][]{%
    \GenericError{(gnuplot) \space\space\space\@spaces}{%
      Package graphicx or graphics not loaded%
    }{See the gnuplot documentation for explanation.%
    }{The gnuplot epslatex terminal needs graphicx.sty or graphics.sty.}%
    \renewcommand\includegraphics[2][]{}%
  }%
  \providecommand\rotatebox[2]{#2}%
  \@ifundefined{ifGPcolor}{%
    \newif\ifGPcolor
    \GPcolortrue
  }{}%
  \@ifundefined{ifGPblacktext}{%
    \newif\ifGPblacktext
    \GPblacktexttrue
  }{}%
  \let\gplgaddtomacro\g@addto@macro
  \gdef\gplbacktext{}%
  \gdef\gplfronttext{}%
  \makeatother
  \ifGPblacktext
    \def\colorrgb#1{}%
    \def\colorgray#1{}%
  \else
    \ifGPcolor
      \def\colorrgb#1{\color[rgb]{#1}}%
      \def\colorgray#1{\color[gray]{#1}}%
      \expandafter\def\csname LTw\endcsname{\color{white}}%
      \expandafter\def\csname LTb\endcsname{\color{black}}%
      \expandafter\def\csname LTa\endcsname{\color{black}}%
      \expandafter\def\csname LT0\endcsname{\color[rgb]{1,0,0}}%
      \expandafter\def\csname LT1\endcsname{\color[rgb]{0,1,0}}%
      \expandafter\def\csname LT2\endcsname{\color[rgb]{0,0,1}}%
      \expandafter\def\csname LT3\endcsname{\color[rgb]{1,0,1}}%
      \expandafter\def\csname LT4\endcsname{\color[rgb]{0,1,1}}%
      \expandafter\def\csname LT5\endcsname{\color[rgb]{1,1,0}}%
      \expandafter\def\csname LT6\endcsname{\color[rgb]{0,0,0}}%
      \expandafter\def\csname LT7\endcsname{\color[rgb]{1,0.3,0}}%
      \expandafter\def\csname LT8\endcsname{\color[rgb]{0.5,0.5,0.5}}%
    \else
      \def\colorrgb#1{\color{black}}%
      \def\colorgray#1{\color[gray]{#1}}%
      \expandafter\def\csname LTw\endcsname{\color{white}}%
      \expandafter\def\csname LTb\endcsname{\color{black}}%
      \expandafter\def\csname LTa\endcsname{\color{black}}%
      \expandafter\def\csname LT0\endcsname{\color{black}}%
      \expandafter\def\csname LT1\endcsname{\color{black}}%
      \expandafter\def\csname LT2\endcsname{\color{black}}%
      \expandafter\def\csname LT3\endcsname{\color{black}}%
      \expandafter\def\csname LT4\endcsname{\color{black}}%
      \expandafter\def\csname LT5\endcsname{\color{black}}%
      \expandafter\def\csname LT6\endcsname{\color{black}}%
      \expandafter\def\csname LT7\endcsname{\color{black}}%
      \expandafter\def\csname LT8\endcsname{\color{black}}%
    \fi
  \fi
    \setlength{\unitlength}{0.0500bp}%
    \ifx\gptboxheight\undefined%
      \newlength{\gptboxheight}%
      \newlength{\gptboxwidth}%
      \newsavebox{\gptboxtext}%
    \fi%
    \setlength{\fboxrule}{0.5pt}%
    \setlength{\fboxsep}{1pt}%
\begin{picture}(7200.00,5040.00)%
    \gplgaddtomacro\gplbacktext{%
      \csname LTb\endcsname
      \put(1342,1162){\makebox(0,0)[r]{\strut{}$10^{-9}$}}%
      \put(1342,1562){\makebox(0,0)[r]{\strut{}$10^{-8}$}}%
      \put(1342,1961){\makebox(0,0)[r]{\strut{}$10^{-7}$}}%
      \put(1342,2361){\makebox(0,0)[r]{\strut{}$10^{-6}$}}%
      \put(1342,2761){\makebox(0,0)[r]{\strut{}$10^{-5}$}}%
      \put(1342,3160){\makebox(0,0)[r]{\strut{}$10^{-4}$}}%
      \put(1342,3560){\makebox(0,0)[r]{\strut{}$10^{-3}$}}%
      \put(1342,3960){\makebox(0,0)[r]{\strut{}$10^{-2}$}}%
      \put(1342,4359){\makebox(0,0)[r]{\strut{}$10^{-1}$}}%
      \put(1342,4759){\makebox(0,0)[r]{\strut{}$10^{0}$}}%
      \put(1603,812){\makebox(0,0){\strut{}$10^{-5}$}}%
      \put(2413,812){\makebox(0,0){\strut{}$10^{-4}$}}%
      \put(3223,812){\makebox(0,0){\strut{}$10^{-3}$}}%
      \put(4034,812){\makebox(0,0){\strut{}$10^{-2}$}}%
      \put(4844,812){\makebox(0,0){\strut{}$10^{-1}$}}%
      \put(5654,812){\makebox(0,0){\strut{}$10^{0}$}}%
      \put(6464,812){\makebox(0,0){\strut{}$10^{1}$}}%
      \put(1992,3824){\makebox(0,0)[l]{\strut{}\textcolor{black}{R}}}%
      \put(3402,3176){\makebox(0,0)[l]{\strut{}\textcolor[rgb]{0.2,0.5,0.2}{P}}}%
      \put(3304,2421){\makebox(0,0)[l]{\strut{}\textcolor[rgb]{0.2,0.5,0.2}{P2}}}%
    }%
    \gplgaddtomacro\gplfronttext{%
      \csname LTb\endcsname
      \put(240,2960){\rotatebox{-270}{\makebox(0,0){\strut{}\huge $\sin^2 2\theta$}}}%
      \put(4033,196){\makebox(0,0){\strut{}\huge $m_{\rm s}$\,(keV)}}%
      \csname LTb\endcsname
      \put(4908,2889){\rotatebox{-25}{\makebox(0,0)[l]{\strut{}\textcolor{blue}{BBN}}}}%
      \put(4034,4004){\makebox(0,0)[l]{\strut{}\textcolor{black}{KA}}}%
      \put(3936,3356){\rotatebox{-47}{\makebox(0,0)[l]{\strut{}\small \textcolor{red}{\scriptsize $m_{\rm s,\,eff} > 0.65~{\rm eV}$}}}}%
      \put(4034,1522){\makebox(0,0)[l]{\strut{}\large Standard cosmology}}%
    }%
    \gplbacktext
    \put(0,0){\includegraphics{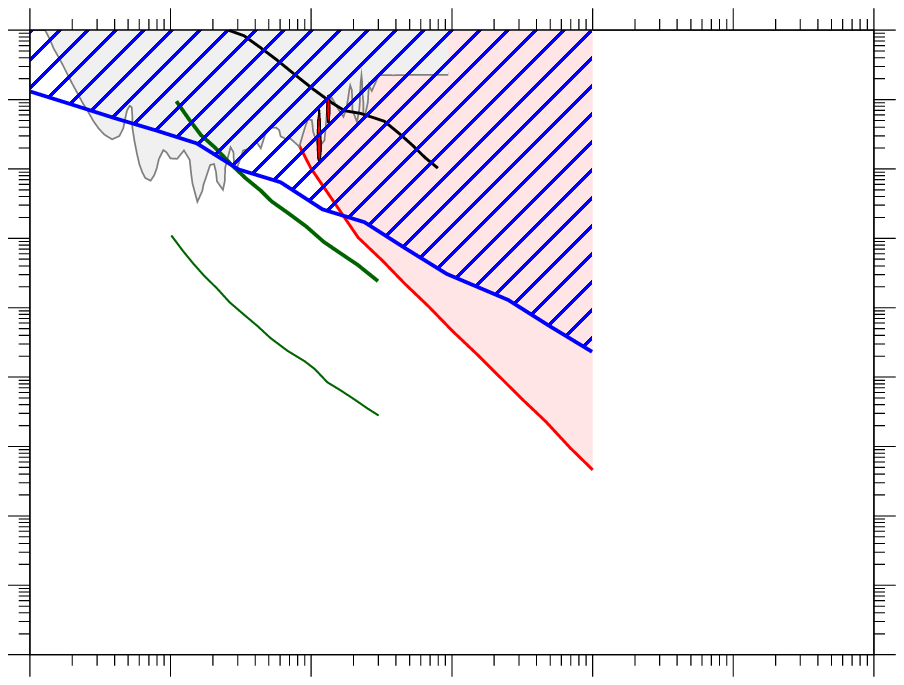}}%
    \gplfronttext
  \end{picture}%
\endgroup

%% file: allowed_sterile_radiative.tex
\begingroup
  \makeatletter
  \providecommand\color[2][]{%
    \GenericError{(gnuplot) \space\space\space\@spaces}{%
      Package color not loaded in conjunction with
      terminal option `colourtext'%
    }{See the gnuplot documentation for explanation.%
    }{Either use 'blacktext' in gnuplot or load the package
      color.sty in LaTeX.}%
    \renewcommand\color[2][]{}%
  }%
  \providecommand\includegraphics[2][]{%
    \GenericError{(gnuplot) \space\space\space\@spaces}{%
      Package graphicx or graphics not loaded%
    }{See the gnuplot documentation for explanation.%
    }{The gnuplot epslatex terminal needs graphicx.sty or graphics.sty.}%
    \renewcommand\includegraphics[2][]{}%
  }%
  \providecommand\rotatebox[2]{#2}%
  \@ifundefined{ifGPcolor}{%
    \newif\ifGPcolor
    \GPcolortrue
  }{}%
  \@ifundefined{ifGPblacktext}{%
    \newif\ifGPblacktext
    \GPblacktexttrue
  }{}%
  \let\gplgaddtomacro\g@addto@macro
  \gdef\gplbacktext{}%
  \gdef\gplfronttext{}%
  \makeatother
  \ifGPblacktext
    \def\colorrgb#1{}%
    \def\colorgray#1{}%
  \else
    \ifGPcolor
      \def\colorrgb#1{\color[rgb]{#1}}%
      \def\colorgray#1{\color[gray]{#1}}%
      \expandafter\def\csname LTw\endcsname{\color{white}}%
      \expandafter\def\csname LTb\endcsname{\color{black}}%
      \expandafter\def\csname LTa\endcsname{\color{black}}%
      \expandafter\def\csname LT0\endcsname{\color[rgb]{1,0,0}}%
      \expandafter\def\csname LT1\endcsname{\color[rgb]{0,1,0}}%
      \expandafter\def\csname LT2\endcsname{\color[rgb]{0,0,1}}%
      \expandafter\def\csname LT3\endcsname{\color[rgb]{1,0,1}}%
      \expandafter\def\csname LT4\endcsname{\color[rgb]{0,1,1}}%
      \expandafter\def\csname LT5\endcsname{\color[rgb]{1,1,0}}%
      \expandafter\def\csname LT6\endcsname{\color[rgb]{0,0,0}}%
      \expandafter\def\csname LT7\endcsname{\color[rgb]{1,0.3,0}}%
      \expandafter\def\csname LT8\endcsname{\color[rgb]{0.5,0.5,0.5}}%
    \else
      \def\colorrgb#1{\color{black}}%
      \def\colorgray#1{\color[gray]{#1}}%
      \expandafter\def\csname LTw\endcsname{\color{white}}%
      \expandafter\def\csname LTb\endcsname{\color{black}}%
      \expandafter\def\csname LTa\endcsname{\color{black}}%
      \expandafter\def\csname LT0\endcsname{\color{black}}%
      \expandafter\def\csname LT1\endcsname{\color{black}}%
      \expandafter\def\csname LT2\endcsname{\color{black}}%
      \expandafter\def\csname LT3\endcsname{\color{black}}%
      \expandafter\def\csname LT4\endcsname{\color{black}}%
      \expandafter\def\csname LT5\endcsname{\color{black}}%
      \expandafter\def\csname LT6\endcsname{\color{black}}%
      \expandafter\def\csname LT7\endcsname{\color{black}}%
      \expandafter\def\csname LT8\endcsname{\color{black}}%
    \fi
  \fi
    \setlength{\unitlength}{0.0500bp}%
    \ifx\gptboxheight\undefined%
      \newlength{\gptboxheight}%
      \newlength{\gptboxwidth}%
      \newsavebox{\gptboxtext}%
    \fi%
    \setlength{\fboxrule}{0.5pt}%
    \setlength{\fboxsep}{1pt}%
\begin{picture}(7200.00,5040.00)%
    \gplgaddtomacro\gplbacktext{%
      \csname LTb\endcsname
      \put(1342,1162){\makebox(0,0)[r]{\strut{}$10^{-9}$}}%
      \put(1342,1562){\makebox(0,0)[r]{\strut{}$10^{-8}$}}%
      \put(1342,1961){\makebox(0,0)[r]{\strut{}$10^{-7}$}}%
      \put(1342,2361){\makebox(0,0)[r]{\strut{}$10^{-6}$}}%
      \put(1342,2761){\makebox(0,0)[r]{\strut{}$10^{-5}$}}%
      \put(1342,3160){\makebox(0,0)[r]{\strut{}$10^{-4}$}}%
      \put(1342,3560){\makebox(0,0)[r]{\strut{}$10^{-3}$}}%
      \put(1342,3960){\makebox(0,0)[r]{\strut{}$10^{-2}$}}%
      \put(1342,4359){\makebox(0,0)[r]{\strut{}$10^{-1}$}}%
      \put(1342,4759){\makebox(0,0)[r]{\strut{}$10^{0}$}}%
      \put(1603,812){\makebox(0,0){\strut{}$10^{-5}$}}%
      \put(2413,812){\makebox(0,0){\strut{}$10^{-4}$}}%
      \put(3223,812){\makebox(0,0){\strut{}$10^{-3}$}}%
      \put(4034,812){\makebox(0,0){\strut{}$10^{-2}$}}%
      \put(4844,812){\makebox(0,0){\strut{}$10^{-1}$}}%
      \put(5654,812){\makebox(0,0){\strut{}$10^{0}$}}%
      \put(6464,812){\makebox(0,0){\strut{}$10^{1}$}}%
      \put(1992,3824){\makebox(0,0)[l]{\strut{}\textcolor{black}{R}}}%
      \put(3402,3176){\makebox(0,0)[l]{\strut{}\textcolor[rgb]{0.2,0.5,0.2}{P}}}%
      \put(3304,2421){\makebox(0,0)[l]{\strut{}\textcolor[rgb]{0.2,0.5,0.2}{P2}}}%
    }%
    \gplgaddtomacro\gplfronttext{%
      \csname LTb\endcsname
      \put(240,2960){\rotatebox{-270}{\makebox(0,0){\strut{}\huge $\sin^2 2\theta$}}}%
      \put(4033,196){\makebox(0,0){\strut{}\huge $m_{\rm s}$\,(keV)}}%
      \csname LTb\endcsname
      \put(5735,4040){\makebox(0,0)[l]{\strut{}\textcolor{blue}{BBN}}}%
      \put(4034,4004){\makebox(0,0)[l]{\strut{}\textcolor{black}{KA}}}%
      \put(5127,3410){\rotatebox{-28}{\makebox(0,0)[l]{\strut{}\small \textcolor{red}{$m_{\rm s,\,eff} > 0.65~{\rm eV}$}}}}%
      \put(1749,1773){\makebox(0,0)[l]{\strut{}Low reheating temperature:}}%
      \put(1749,1414){\makebox(0,0)[l]{\strut{}100\% radiative decay, $\phi \rightarrow \gamma + \dots\ \,{\rm or}\ \,l^\pm + \dots$}}%
    }%
    \gplbacktext
    \put(0,0){\includegraphics{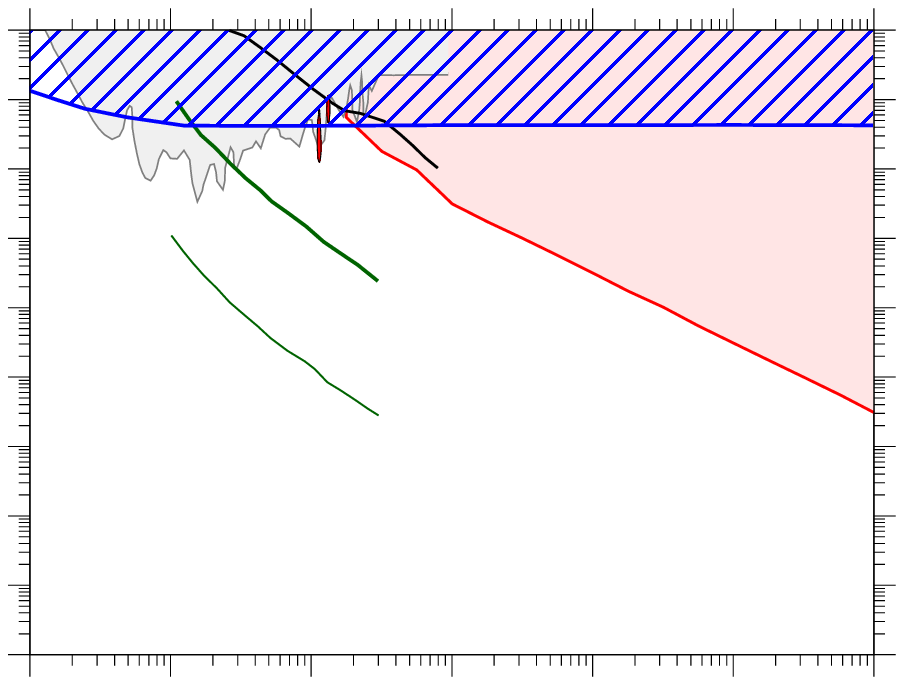}}%
    \gplfronttext
  \end{picture}%
\endgroup

%% file: allowed_sterile_10GeV_1.tex
\begingroup
  \makeatletter
  \providecommand\color[2][]{%
    \GenericError{(gnuplot) \space\space\space\@spaces}{%
      Package color not loaded in conjunction with
      terminal option `colourtext'%
    }{See the gnuplot documentation for explanation.%
    }{Either use 'blacktext' in gnuplot or load the package
      color.sty in LaTeX.}%
    \renewcommand\color[2][]{}%
  }%
  \providecommand\includegraphics[2][]{%
    \GenericError{(gnuplot) \space\space\space\@spaces}{%
      Package graphicx or graphics not loaded%
    }{See the gnuplot documentation for explanation.%
    }{The gnuplot epslatex terminal needs graphicx.sty or graphics.sty.}%
    \renewcommand\includegraphics[2][]{}%
  }%
  \providecommand\rotatebox[2]{#2}%
  \@ifundefined{ifGPcolor}{%
    \newif\ifGPcolor
    \GPcolortrue
  }{}%
  \@ifundefined{ifGPblacktext}{%
    \newif\ifGPblacktext
    \GPblacktexttrue
  }{}%
  \let\gplgaddtomacro\g@addto@macro
  \gdef\gplbacktext{}%
  \gdef\gplfronttext{}%
  \makeatother
  \ifGPblacktext
    \def\colorrgb#1{}%
    \def\colorgray#1{}%
  \else
    \ifGPcolor
      \def\colorrgb#1{\color[rgb]{#1}}%
      \def\colorgray#1{\color[gray]{#1}}%
      \expandafter\def\csname LTw\endcsname{\color{white}}%
      \expandafter\def\csname LTb\endcsname{\color{black}}%
      \expandafter\def\csname LTa\endcsname{\color{black}}%
      \expandafter\def\csname LT0\endcsname{\color[rgb]{1,0,0}}%
      \expandafter\def\csname LT1\endcsname{\color[rgb]{0,1,0}}%
      \expandafter\def\csname LT2\endcsname{\color[rgb]{0,0,1}}%
      \expandafter\def\csname LT3\endcsname{\color[rgb]{1,0,1}}%
      \expandafter\def\csname LT4\endcsname{\color[rgb]{0,1,1}}%
      \expandafter\def\csname LT5\endcsname{\color[rgb]{1,1,0}}%
      \expandafter\def\csname LT6\endcsname{\color[rgb]{0,0,0}}%
      \expandafter\def\csname LT7\endcsname{\color[rgb]{1,0.3,0}}%
      \expandafter\def\csname LT8\endcsname{\color[rgb]{0.5,0.5,0.5}}%
    \else
      \def\colorrgb#1{\color{black}}%
      \def\colorgray#1{\color[gray]{#1}}%
      \expandafter\def\csname LTw\endcsname{\color{white}}%
      \expandafter\def\csname LTb\endcsname{\color{black}}%
      \expandafter\def\csname LTa\endcsname{\color{black}}%
      \expandafter\def\csname LT0\endcsname{\color{black}}%
      \expandafter\def\csname LT1\endcsname{\color{black}}%
      \expandafter\def\csname LT2\endcsname{\color{black}}%
      \expandafter\def\csname LT3\endcsname{\color{black}}%
      \expandafter\def\csname LT4\endcsname{\color{black}}%
      \expandafter\def\csname LT5\endcsname{\color{black}}%
      \expandafter\def\csname LT6\endcsname{\color{black}}%
      \expandafter\def\csname LT7\endcsname{\color{black}}%
      \expandafter\def\csname LT8\endcsname{\color{black}}%
    \fi
  \fi
    \setlength{\unitlength}{0.0500bp}%
    \ifx\gptboxheight\undefined%
      \newlength{\gptboxheight}%
      \newlength{\gptboxwidth}%
      \newsavebox{\gptboxtext}%
    \fi%
    \setlength{\fboxrule}{0.5pt}%
    \setlength{\fboxsep}{1pt}%
\begin{picture}(7200.00,5040.00)%
    \gplgaddtomacro\gplbacktext{%
      \csname LTb\endcsname
      \put(1342,1162){\makebox(0,0)[r]{\strut{}$10^{-9}$}}%
      \put(1342,1562){\makebox(0,0)[r]{\strut{}$10^{-8}$}}%
      \put(1342,1961){\makebox(0,0)[r]{\strut{}$10^{-7}$}}%
      \put(1342,2361){\makebox(0,0)[r]{\strut{}$10^{-6}$}}%
      \put(1342,2761){\makebox(0,0)[r]{\strut{}$10^{-5}$}}%
      \put(1342,3160){\makebox(0,0)[r]{\strut{}$10^{-4}$}}%
      \put(1342,3560){\makebox(0,0)[r]{\strut{}$10^{-3}$}}%
      \put(1342,3960){\makebox(0,0)[r]{\strut{}$10^{-2}$}}%
      \put(1342,4359){\makebox(0,0)[r]{\strut{}$10^{-1}$}}%
      \put(1342,4759){\makebox(0,0)[r]{\strut{}$10^{0}$}}%
      \put(1603,812){\makebox(0,0){\strut{}$10^{-5}$}}%
      \put(2413,812){\makebox(0,0){\strut{}$10^{-4}$}}%
      \put(3223,812){\makebox(0,0){\strut{}$10^{-3}$}}%
      \put(4034,812){\makebox(0,0){\strut{}$10^{-2}$}}%
      \put(4844,812){\makebox(0,0){\strut{}$10^{-1}$}}%
      \put(5654,812){\makebox(0,0){\strut{}$10^{0}$}}%
      \put(6464,812){\makebox(0,0){\strut{}$10^{1}$}}%
      \put(1992,3824){\makebox(0,0)[l]{\strut{}\textcolor{black}{R}}}%
      \put(3402,3176){\makebox(0,0)[l]{\strut{}\textcolor[rgb]{0.2,0.5,0.2}{P}}}%
      \put(3304,2421){\makebox(0,0)[l]{\strut{}\textcolor[rgb]{0.2,0.5,0.2}{P2}}}%
    }%
    \gplgaddtomacro\gplfronttext{%
      \csname LTb\endcsname
      \put(240,2960){\rotatebox{-270}{\makebox(0,0){\strut{}\huge $\sin^2 2\theta$}}}%
      \put(4033,196){\makebox(0,0){\strut{}\huge $m_{\rm s}$\,(keV)}}%
      \csname LTb\endcsname
      \put(5954,3716){\makebox(0,0)[l]{\strut{}\textcolor{blue}{BBN}}}%
      \put(4034,4004){\makebox(0,0)[l]{\strut{}\textcolor{black}{KA}}}%
      \put(5127,3410){\rotatebox{-28}{\makebox(0,0)[l]{\strut{}\small \textcolor{red}{$m_{\rm s,\,eff} > 0.65~{\rm eV}$}}}}%
      \put(1749,1773){\makebox(0,0)[l]{\strut{}Low reheating temperature:}}%
      \put(1749,1414){\makebox(0,0)[l]{\strut{}100\% hadronic decay, $\phi \rightarrow u + \bar{u}$}}%
    }%
    \gplbacktext
    \put(0,0){\includegraphics{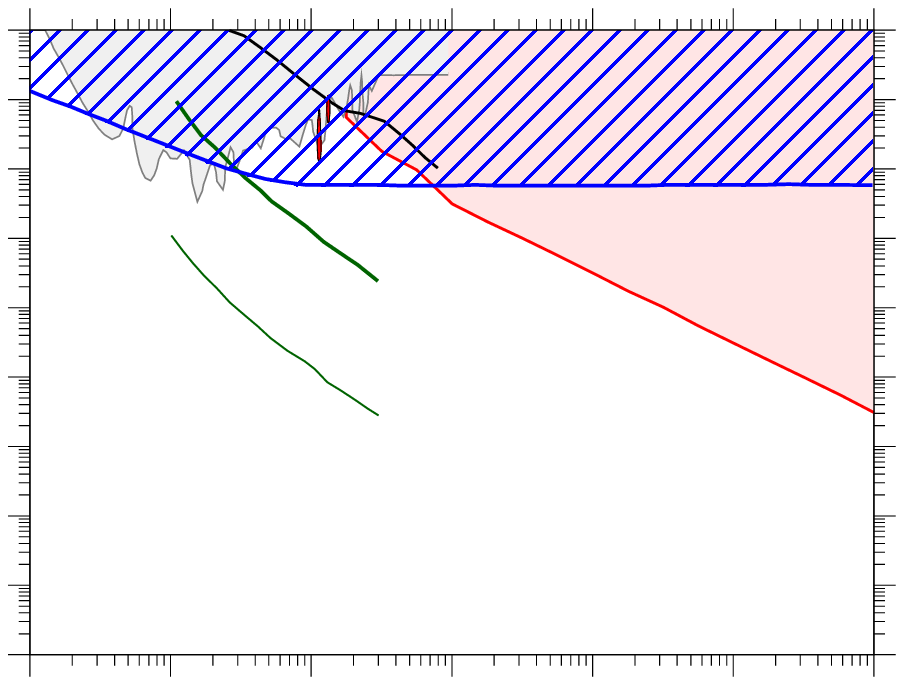}}%
    \gplfronttext
  \end{picture}%
\endgroup

%% file: conclusion.tex
\section{Conclusions} \label{sec:conclusion}
 
We have investigated the cosmological production of light sterile neutrinos with 
masses and mixings consistent with those needed to explain the anomaly in 
short-baseline neutrino experiments, assuming a low reheating temperature of the 
Universe $T_{\rm RH} \sim {\cal O}(1)$ MeV. Considering the sterile neutrino 
production through the combination of scatterings and non-resonant oscillations, 
we have numerically solved its evolution and found that the existence of such 
sterile neutrinos becomes consistent with Big Bang nucleosynthesis if the parent 
particle responsible for reheating decays exclusively into electromagnetically 
interacting radiation. In contrast, if the parent particle mainly decays into 
hadrons, the BBN bound gets tighter and the preferred regions for explaining the 
anomaly are excluded for a wide range of the mass and the hadronic branching 
ratio of the parent particle.

%% file: acknowledgement.tex
\begin{acknowledgments}
  Numerical computations were carried out on PC clusters at Center for
  Computational Astrophysics, National Astronomical Observatory of
  Japan (NAOJ) and Computing Research Center, High Energy Accelerator
  Research Organization (KEK). The work of NH has been partially
  supported by JSPS KAKENHI Grants No.~JP19K23446.  The work of KK has
  been supported by JSPS KAKENHI Grants No.~JP17H01131, MEXT
  Grant-in-Aid for Scientific Research on Innovative Areas JP15H05889,
  JP18H04594, JP19H05114, and JP20H04750. The work of RSLH is
  supported by Danmarks Frie Forskningsfond (Project No. 8049-00038B).
\end{acknowledgments}

%% file: LowReheat_sterile.bbl
\providecommand{\href}[2]{#2}\begingroup\raggedright\begin{thebibliography}{10}

\bibitem{Athanassopoulos_1997}
{\scshape LSND} collaboration, \emph{{Evidence for nu(mu) $\rightarrow$ nu(e)
  neutrino oscillations from LSND}},
  \href{https://doi.org/10.1103/PhysRevLett.81.1774}{\emph{Phys. Rev. Lett.}
  {\bfseries 81} (1998) 1774}
  [\href{https://arxiv.org/abs/nucl-ex/9709006}{{\ttfamily nucl-ex/9709006}}].

\bibitem{AguilarArevalo_2010}
{\scshape MiniBooNE} collaboration, \emph{{Event Excess in the MiniBooNE Search
  for $\bar \nu_\mu \rightarrow \bar \nu_e$ Oscillations}},
  \href{https://doi.org/10.1103/PhysRevLett.105.181801}{\emph{Phys. Rev. Lett.}
  {\bfseries 105} (2010) 181801}
  [\href{https://arxiv.org/abs/1007.1150}{{\ttfamily 1007.1150}}].

\bibitem{MiniBooNE_2018}
{\scshape MiniBooNE} collaboration, \emph{{Observation of a Significant Excess
  of Electron-Like Events in the MiniBooNE Short-Baseline Neutrino
  Experiment}},  \href{https://arxiv.org/abs/1805.12028}{{\ttfamily
  1805.12028}}.

\bibitem{DayaBay_2016}
{\scshape Daya Bay} collaboration, \emph{{Measurement of the Reactor
  Antineutrino Flux and Spectrum at Daya Bay}},
  \href{https://doi.org/10.1103/PhysRevLett.116.061801,
  10.1103/PhysRevLett.118.099902}{\emph{Phys. Rev. Lett.} {\bfseries 116}
  (2016) 061801} [\href{https://arxiv.org/abs/1508.04233}{{\ttfamily
  1508.04233}}].

\bibitem{DoubleChooz_2011}
{\scshape Double Chooz} collaboration, \emph{{Indication of Reactor
  $\bar{\nu}_e$ Disappearance in the Double Chooz Experiment}},
  \href{https://doi.org/10.1103/PhysRevLett.108.131801}{\emph{Phys. Rev. Lett.}
  {\bfseries 108} (2012) 131801}
  [\href{https://arxiv.org/abs/1112.6353}{{\ttfamily 1112.6353}}].

\bibitem{SAGE_1996}
D.~Abdurashitov et~al., \emph{{The Russian-American gallium experiment (SAGE)
  Cr neutrino source measurement}},
  \href{https://doi.org/10.1103/PhysRevLett.77.4708}{\emph{Phys. Rev. Lett.}
  {\bfseries 77} (1996) 4708}.

\bibitem{SAGE_1998}
{\scshape SAGE} collaboration, \emph{{Measurement of the response of the
  Russian-American gallium experiment to neutrinos from a Cr-51 source}},
  \href{https://doi.org/10.1103/PhysRevC.59.2246}{\emph{Phys. Rev.} {\bfseries
  C59} (1999) 2246} [\href{https://arxiv.org/abs/hep-ph/9803418}{{\ttfamily
  hep-ph/9803418}}].

\bibitem{SAGE_2005}
J.~Abdurashitov et~al., \emph{{Measurement of the response of a Ga solar
  neutrino experiment to neutrinos from an Ar-37 source}},
  \href{https://doi.org/10.1103/PhysRevC.73.045805}{\emph{Phys. Rev.}
  {\bfseries C73} (2006) 045805}
  [\href{https://arxiv.org/abs/nucl-ex/0512041}{{\ttfamily nucl-ex/0512041}}].

\bibitem{Gallex_1995}
{\scshape GALLEX} collaboration, \emph{{First results from the Cr-51 neutrino
  source experiment with the GALLEX detector}},
  \href{https://doi.org/10.1016/0370-2693(94)01586-2}{\emph{Phys. Lett.}
  {\bfseries B342} (1995) 440}.

\bibitem{Gallex_1997}
{\scshape GALLEX} collaboration, \emph{{Final results of the Cr-51 neutrino
  source experiments in GALLEX}},
  \href{https://doi.org/10.1016/S0370-2693(97)01562-1}{\emph{Phys. Lett.}
  {\bfseries B420} (1998) 114}.

\bibitem{Gallex_2010}
F.~Kaether, W.~Hampel, G.~Heusser, J.~Kiko and T.~Kirsten, \emph{{Reanalysis of
  the GALLEX solar neutrino flux and source experiments}},
  \href{https://doi.org/10.1016/j.physletb.2010.01.030}{\emph{Phys. Lett.}
  {\bfseries B685} (2010) 47}
  [\href{https://arxiv.org/abs/1001.2731}{{\ttfamily 1001.2731}}].

\bibitem{MINOS_2017}
{\scshape MINOS+} collaboration, \emph{{Search for sterile neutrinos in MINOS
  and MINOS+ using a two-detector fit}},
  \href{https://doi.org/10.1103/PhysRevLett.122.091803}{\emph{Phys. Rev. Lett.}
  {\bfseries 122} (2019) 091803}
  [\href{https://arxiv.org/abs/1710.06488}{{\ttfamily 1710.06488}}].

\bibitem{NOvA_2017}
{\scshape NOvA} collaboration, \emph{{Search for active-sterile neutrino mixing
  using neutral-current interactions in NOvA}},
  \href{https://doi.org/10.1103/PhysRevD.96.072006}{\emph{Phys. Rev.}
  {\bfseries D96} (2017) 072006}
  [\href{https://arxiv.org/abs/1706.04592}{{\ttfamily 1706.04592}}].

\bibitem{IceCube_2016}
{\scshape IceCube} collaboration, \emph{{Searches for Sterile Neutrinos with
  the IceCube Detector}},
  \href{https://doi.org/10.1103/PhysRevLett.117.071801}{\emph{Phys. Rev. Lett.}
  {\bfseries 117} (2016) 071801}
  [\href{https://arxiv.org/abs/1605.01990}{{\ttfamily 1605.01990}}].

\bibitem{SBN_2019}
P.~Machado, O.~Palamara and D.~Schmitz, \emph{{The Short-Baseline Neutrino
  Program at Fermilab}}, {\emph{Ann. Rev. Nucl. Part. Sci.} {\bfseries 69}
  (2019) } [\href{https://arxiv.org/abs/1903.04608}{{\ttfamily 1903.04608}}].

\bibitem{JSNS2_2017}
J.~Park, \emph{{Searching for a Sterile Neutrino at J-PARC MLF: $JSNS^2$
  experiment}}, \href{https://doi.org/10.22323/1.314.0128}{\emph{PoS}
  {\bfseries EPS-HEP2017} (2017) 128}.

\bibitem{Hamann_2011}
J.~Hamann, S.~Hannestad, G.~Raffelt and Y.~Wong, \emph{{Sterile neutrinos with
  eV masses in cosmology: How disfavoured exactly?}},
  \href{https://doi.org/10.1088/1475-7516/2011/09/034}{\emph{JCAP} {\bfseries
  1109} (2011) 034} [\href{https://arxiv.org/abs/1108.4136}{{\ttfamily
  1108.4136}}].

\bibitem{Hannestad_2012}
S.~Hannestad, I.~Tamborra and T.~Tram, \emph{{Thermalisation of light sterile
  neutrinos in the early universe}},
  \href{https://doi.org/10.1088/1475-7516/2012/07/025}{\emph{JCAP} {\bfseries
  1207} (2012) 025} [\href{https://arxiv.org/abs/1204.5861}{{\ttfamily
  1204.5861}}].

\bibitem{Gariazzo_2019}
S.~Gariazzo, P.~F. de~Salas and S.~Pastor, \emph{{Thermalisation of sterile
  neutrinos in the early Universe in the 3+1 scheme with full mixing matrix}},
  \href{https://doi.org/10.1088/1475-7516/2019/07/014}{\emph{JCAP} {\bfseries
  1907} (2019) 014} [\href{https://arxiv.org/abs/1905.11290}{{\ttfamily
  1905.11290}}].

\bibitem{Hagstotz_2020}
S.~Hagstotz, P.~F. de~Salas, S.~Gariazzo, M.~Gerbino, M.~Lattanzi, S.~Vagnozzi
  et~al., \emph{{Bounds on light sterile neutrino mass and mixing from
  cosmology and laboratory searches}},
  \href{https://arxiv.org/abs/2003.02289}{{\ttfamily 2003.02289}}.

\bibitem{Abazajian_2004}
K.~Abazajian, N.~Bell, G.~Fuller and Y.~Wong, \emph{{Cosmological lepton
  asymmetry, primordial nucleosynthesis, and sterile neutrinos}},
  \href{https://doi.org/10.1103/PhysRevD.72.063004}{\emph{Phys. Rev.}
  {\bfseries D72} (2005) 063004}
  [\href{https://arxiv.org/abs/astro-ph/0410175}{{\ttfamily
  astro-ph/0410175}}].

\bibitem{Mirizzi_2012}
A.~Mirizzi, N.~Saviano, G.~Miele and P.~Serpico, \emph{{Light sterile neutrino
  production in the early universe with dynamical neutrino asymmetries}},
  \href{https://doi.org/10.1103/PhysRevD.86.053009}{\emph{Phys. Rev.}
  {\bfseries D86} (2012) 053009}
  [\href{https://arxiv.org/abs/1206.1046}{{\ttfamily 1206.1046}}].

\bibitem{Saviano_2013}
N.~Saviano, A.~Mirizzi, O.~Pisanti, P.~Serpico, G.~Mangano and G.~Miele,
  \emph{{Multi-momentum and multi-flavour active-sterile neutrino oscillations
  in the early universe: role of neutrino asymmetries and effects on
  nucleosynthesis}},
  \href{https://doi.org/10.1103/PhysRevD.87.073006}{\emph{Phys. Rev.}
  {\bfseries D87} (2013) 073006}
  [\href{https://arxiv.org/abs/1302.1200}{{\ttfamily 1302.1200}}].

\bibitem{Hannestad_2013_2}
S.~Hannestad, R.~Hansen and T.~Tram, \emph{{How Self-Interactions can Reconcile
  Sterile Neutrinos with Cosmology}},
  \href{https://doi.org/10.1103/PhysRevLett.112.031802}{\emph{Phys. Rev. Lett.}
  {\bfseries 112} (2014) 031802}
  [\href{https://arxiv.org/abs/1310.5926}{{\ttfamily 1310.5926}}].

\bibitem{Saviano_2014}
N.~Saviano, O.~Pisanti, G.~Mangano and A.~Mirizzi, \emph{{Unveiling secret
  interactions among sterile neutrinos with big-bang nucleosynthesis}},
  \href{https://doi.org/10.1103/PhysRevD.90.113009}{\emph{Phys. Rev.}
  {\bfseries D90} (2014) 113009}
  [\href{https://arxiv.org/abs/1409.1680}{{\ttfamily 1409.1680}}].

\bibitem{Archidiacono_2014}
M.~Archidiacono, S.~Hannestad, R.~Hansen and T.~Tram, \emph{{Cosmology with
  self-interacting sterile neutrinos and dark matter - A pseudoscalar model}},
  \href{https://doi.org/10.1103/PhysRevD.91.065021}{\emph{Phys. Rev.}
  {\bfseries D91} (2015) 065021}
  [\href{https://arxiv.org/abs/1404.5915}{{\ttfamily 1404.5915}}].

\bibitem{Archidiacono_2016}
M.~Archidiacono, S.~Gariazzo, C.~Giunti, S.~Hannestad, R.~Hansen, M.~Laveder
  et~al., \emph{{Pseudoscalar―sterile neutrino interactions: reconciling the
  cosmos with neutrino oscillations}},
  \href{https://doi.org/10.1088/1475-7516/2016/08/067}{\emph{JCAP} {\bfseries
  1608} (2016) 067} [\href{https://arxiv.org/abs/1606.07673}{{\ttfamily
  1606.07673}}].

\bibitem{Chu_2015}
X.~Chu, B.~Dasgupta and J.~Kopp, \emph{{Sterile neutrinos with secret
  interactions―lasting friendship with cosmology}},
  \href{https://doi.org/10.1088/1475-7516/2015/10/011}{\emph{JCAP} {\bfseries
  1510} (2015) 011} [\href{https://arxiv.org/abs/1505.02795}{{\ttfamily
  1505.02795}}].

\bibitem{Chu_2018}
X.~Chu, B.~Dasgupta, M.~Dentler, J.~Kopp and N.~Saviano, \emph{{Sterile
  neutrinos with secret interactions―cosmological discord?}},
  \href{https://doi.org/10.1088/1475-7516/2018/11/049}{\emph{JCAP} {\bfseries
  1811} (2018) 049} [\href{https://arxiv.org/abs/1806.10629}{{\ttfamily
  1806.10629}}].

\bibitem{Gelmini_2004}
G.~Gelmini, S.~Palomares-Ruiz and S.~Pascoli, \emph{{Low reheating temperature
  and the visible sterile neutrino}},
  \href{https://doi.org/10.1103/PhysRevLett.93.081302}{\emph{Phys. Rev. Lett.}
  {\bfseries 93} (2004) 081302}
  [\href{https://arxiv.org/abs/astro-ph/0403323}{{\ttfamily
  astro-ph/0403323}}].

\bibitem{Gelmini_2008}
G.~Gelmini, E.~Osoba, S.~Palomares-Ruiz and S.~Pascoli, \emph{{MeV sterile
  neutrinos in low reheating temperature cosmological scenarios}},
  \href{https://doi.org/10.1088/1475-7516/2008/10/029}{\emph{JCAP} {\bfseries
  0810} (2008) 029} [\href{https://arxiv.org/abs/0803.2735}{{\ttfamily
  0803.2735}}].

\bibitem{Gelmini_2019}
G.~Gelmini, P.~Lu and V.~Takhistov, \emph{{Visible Sterile Neutrinos as the
  Earliest Relic Probes of Cosmology}},
  \href{https://arxiv.org/abs/1909.04168}{{\ttfamily 1909.04168}}.

\bibitem{Gelmini_2019b}
G.~Gelmini, P.~Lu and V.~Takhistov, \emph{{Cosmological Dependence of
  Non-resonantly Produced Sterile Neutrinos}},
  \href{https://arxiv.org/abs/1909.13328}{{\ttfamily 1909.13328}}.

\bibitem{Yaguna_2007}
C.~Yaguna, \emph{{Sterile neutrino production in models with low reheating
  temperatures}},
  \href{https://doi.org/10.1088/1126-6708/2007/06/002}{\emph{JHEP} {\bfseries
  06} (2007) 002} [\href{https://arxiv.org/abs/0706.0178}{{\ttfamily
  0706.0178}}].

\bibitem{Abazajian_2017}
K.~Abazajian, \emph{{Sterile neutrinos in cosmology}},
  \href{https://doi.org/10.1016/j.physrep.2017.10.003}{\emph{Phys. Rept.}
  {\bfseries 711-712} (2017) 1}
  [\href{https://arxiv.org/abs/1705.01837}{{\ttfamily 1705.01837}}].

\bibitem{Salas_2015}
P.~de~Salas, M.~Lattanzi, G.~Mangano, G.~Miele, S.~Pastor and O.~Pisanti,
  \emph{{Bounds on very low reheating scenarios after Planck}},
  \href{https://doi.org/10.1103/PhysRevD.92.123534}{\emph{Phys. Rev.}
  {\bfseries D92} (2015) 123534}
  [\href{https://arxiv.org/abs/1511.00672}{{\ttfamily 1511.00672}}].

\bibitem{Hasegawa_2019a}
T.~Hasegawa, N.~Hiroshima, K.~Kohri, R.~Hansen, T.~Tram and S.~Hannestad,
  \emph{{MeV-scale reheating temperature and thermalization of oscillating
  neutrinos by radiative and hadronic decays of massive particles}},
  \href{https://doi.org/10.1088/1475-7516/2019/12/012}{\emph{JCAP} {\bfseries
  12} (2019) 012} [\href{https://arxiv.org/abs/1908.10189}{{\ttfamily
  1908.10189}}].

\bibitem{Bell_1999}
N.~Bell, R.~Volkas and Y.~Wong, \emph{{Relic neutrino asymmetry evolution from
  first principles}},
  \href{https://doi.org/10.1103/PhysRevD.59.113001}{\emph{Phys. Rev.}
  {\bfseries D59} (1999) 113001}
  [\href{https://arxiv.org/abs/hep-ph/9809363}{{\ttfamily hep-ph/9809363}}].

\bibitem{Venumadhav_2016}
T.~Venumadhav, F.~Cyr-Racine, K.~Abazajian and C.~Hirata, \emph{{Sterile
  neutrino dark matter: Weak interactions in the strong coupling epoch}},
  \href{https://doi.org/10.1103/PhysRevD.94.043515}{\emph{Phys. Rev.}
  {\bfseries D94} (2016) 043515}
  [\href{https://arxiv.org/abs/1507.06655}{{\ttfamily 1507.06655}}].

\bibitem{Dodelson_1993}
S.~Dodelson and L.~Widrow, \emph{{Sterile-neutrinos as dark matter}},
  \href{https://doi.org/10.1103/PhysRevLett.72.17}{\emph{Phys. Rev. Lett.}
  {\bfseries 72} (1994) 17}
  [\href{https://arxiv.org/abs/hep-ph/9303287}{{\ttfamily hep-ph/9303287}}].

\bibitem{PDG2018}
{\scshape Particle Data Group} collaboration, \emph{{Review of Particle
  Physics}}, \href{https://doi.org/10.1103/PhysRevD.98.030001}{\emph{Phys.
  Rev.} {\bfseries D98} (2018) 030001}.

\bibitem{Akhmedov_2017}
E.~Akhmedov, \emph{{Do non-relativistic neutrinos oscillate?}},
  \href{https://doi.org/10.1007/JHEP07(2017)070}{\emph{JHEP} {\bfseries 07}
  (2017) 070} [\href{https://arxiv.org/abs/1703.08169}{{\ttfamily
  1703.08169}}].

\bibitem{McKellar_1994}
B.~McKellar and M.~Thomson, \emph{{Oscillating doublet neutrinos in the early
  universe}}, \href{https://doi.org/10.1103/PhysRevD.49.2710}{\emph{Phys. Rev.}
  {\bfseries D49} (1994) 2710}.

\bibitem{Sigl_1993}
G.~Sigl and G.~Raffelt, \emph{{General kinetic description of relativistic
  mixed neutrinos}},
  \href{https://doi.org/10.1016/0550-3213(93)90175-O}{\emph{Nucl. Phys.}
  {\bfseries B406} (1993) 423}.

\bibitem{Shi_1998}
X.~Shi and G.~Fuller, \emph{{A New dark matter candidate: Nonthermal sterile
  neutrinos}}, \href{https://doi.org/10.1103/PhysRevLett.82.2832}{\emph{Phys.
  Rev. Lett.} {\bfseries 82} (1999) 2832}
  [\href{https://arxiv.org/abs/astro-ph/9810076}{{\ttfamily
  astro-ph/9810076}}].

\bibitem{Hannestad_2013}
S.~Hannestad, R.~Hansen and T.~Tram, \emph{{Can active-sterile neutrino
  oscillations lead to chaotic behavior of the cosmological lepton
  asymmetry?}},
  \href{https://doi.org/10.1088/1475-7516/2013/04/032}{\emph{JCAP} {\bfseries
  1304} (2013) 032} [\href{https://arxiv.org/abs/1302.7279}{{\ttfamily
  1302.7279}}].

\bibitem{Planck_2018}
{\scshape Planck} collaboration, \emph{{Planck 2018 results. VI. Cosmological
  parameters}},  \href{https://arxiv.org/abs/1807.06209}{{\ttfamily
  1807.06209}}.

\bibitem{Hannestad_2015}
S.~Hannestad, R.~Hansen, T.~Tram and Y.~Wong, \emph{{Active-sterile neutrino
  oscillations in the early Universe with full collision terms}},
  \href{https://doi.org/10.1088/1475-7516/2015/08/019}{\emph{JCAP} {\bfseries
  1508} (2015) 019} [\href{https://arxiv.org/abs/1506.05266}{{\ttfamily
  1506.05266}}].

\bibitem{Li_2005}
X.~S. Li, \emph{An overview of superlu: Algorithms, implementation, and user
  interface}, \href{https://doi.org/10.1145/1089014.1089017}{\emph{ACM Trans.
  Math. Softw.} {\bfseries 31} (2005) 302}.

\bibitem{Demmel_1999}
J.~W. Demmel, J.~R. Gilbert and X.~S. Li, \emph{An asynchronous parallel
  supernodal algorithm for sparse gaussian elimination},
  \href{https://doi.org/10.1137/S0895479897317685}{\emph{SIAM J. Matrix Anal.
  Appl.} {\bfseries 20} (1999) 915}.

\bibitem{Dentler_2018}
M.~Dentler, A.~Hernandez-Cabezudo, J.~Kopp, P.~Machado, M.~Maltoni,
  I.~Martinez-Soler et~al., \emph{{Updated Global Analysis of Neutrino
  Oscillations in the Presence of eV-Scale Sterile Neutrinos}},
  \href{https://doi.org/10.1007/JHEP08(2018)010}{\emph{JHEP} {\bfseries 08}
  (2018) 010} [\href{https://arxiv.org/abs/1803.10661}{{\ttfamily
  1803.10661}}].

\bibitem{Kainulainen_1990}
K.~Kainulainen, \emph{{Light Singlet Neutrinos and the Primordial
  Nucleosynthesis}},
  \href{https://doi.org/10.1016/0370-2693(90)90054-A}{\emph{Phys. Lett.}
  {\bfseries B244} (1990) 191}.

\bibitem{Kawasaki_1999}
M.~Kawasaki, K.~Kohri and N.~Sugiyama, \emph{{Cosmological constraints on late
  time entropy production}},
  \href{https://doi.org/10.1103/PhysRevLett.82.4168}{\emph{Phys. Rev. Lett.}
  {\bfseries 82} (1999) 4168}
  [\href{https://arxiv.org/abs/astro-ph/9811437}{{\ttfamily
  astro-ph/9811437}}].

\bibitem{Kawasaki_2000}
M.~Kawasaki, K.~Kohri and N.~Sugiyama, \emph{{MeV scale reheating temperature
  and thermalization of neutrino background}},
  \href{https://doi.org/10.1103/PhysRevD.62.023506}{\emph{Phys. Rev.}
  {\bfseries D62} (2000) 023506}
  [\href{https://arxiv.org/abs/astro-ph/0002127}{{\ttfamily
  astro-ph/0002127}}].

\bibitem{Aver_2015}
E.~Aver, K.~Olive and E.~Skillman, \emph{{The effects of He I $\lambda$10830 on
  helium abundance determinations}},
  \href{https://doi.org/10.1088/1475-7516/2015/07/011}{\emph{JCAP} {\bfseries
  1507} (2015) 011} [\href{https://arxiv.org/abs/1503.08146}{{\ttfamily
  1503.08146}}].

\bibitem{Izotov:2014fga}
Y.~Izotov, T.~Thuan and N.~Guseva, \emph{{A new determination of the primordial
  He abundance using the He i $\lambda$10830 ^^c3^^85 emission line:
  cosmological implications}},
  \href{https://doi.org/10.1093/mnras/stu1771}{\emph{Mon. Not. Roy. Astron.
  Soc.} {\bfseries 445} (2014) 778}
  [\href{https://arxiv.org/abs/1408.6953}{{\ttfamily 1408.6953}}].

\bibitem{Zavarygin_2018}
E.~Zavarygin, J.~Webb, S.~Riemer-S^^c3^^b8rensen and V.~Dumont,
  \emph{{Primordial deuterium abundance at $z_{abs}$ = 2:504 towards
  Q1009+2956}}, \href{https://doi.org/10.1088/1742-6596/1038/1/012012}{\emph{J.
  Phys. Conf. Ser.} {\bfseries 1038} (2018) 012012}
  [\href{https://arxiv.org/abs/1801.04704}{{\ttfamily 1801.04704}}].

\bibitem{Cooke:2017cwo}
R.~J. Cooke, M.~Pettini and C.~C. Steidel, \emph{{One Percent Determination of
  the Primordial Deuterium Abundance}},
  \href{https://doi.org/10.3847/1538-4357/aaab53}{\emph{Astrophys. J.}
  {\bfseries 855} (2018) 102}
  [\href{https://arxiv.org/abs/1710.11129}{{\ttfamily 1710.11129}}].

\bibitem{Kawano_1992}
L.~Kawano, \emph{{Let's go: Early universe. 2. Primordial nucleosynthesis: The
  Computer way}}, .

\bibitem{Kernan:1994je}
P.~J. Kernan and L.~M. Krauss, \emph{{Refined big bang nucleosynthesis
  constraints on Omega (baryon) and N (neutrino)}},
  \href{https://doi.org/10.1103/PhysRevLett.72.3309}{\emph{Phys. Rev. Lett.}
  {\bfseries 72} (1994) 3309}
  [\href{https://arxiv.org/abs/astro-ph/9402010}{{\ttfamily
  astro-ph/9402010}}].

\bibitem{Angulo:1999zz}
C.~Angulo et~al., \emph{{A compilation of charged-particle induced
  thermonuclear reaction rates}},
  \href{https://doi.org/10.1016/S0375-9474(99)00030-5}{\emph{Nucl. Phys. A}
  {\bfseries 656} (1999) 3}.

\bibitem{Serpico:2004gx}
P.~D. Serpico, S.~Esposito, F.~Iocco, G.~Mangano, G.~Miele and O.~Pisanti,
  \emph{{Nuclear reaction network for primordial nucleosynthesis: A Detailed
  analysis of rates, uncertainties and light nuclei yields}},
  \href{https://doi.org/10.1088/1475-7516/2004/12/010}{\emph{JCAP} {\bfseries
  12} (2004) 010} [\href{https://arxiv.org/abs/astro-ph/0408076}{{\ttfamily
  astro-ph/0408076}}].

\bibitem{Pisanti:2007hk}
O.~Pisanti, A.~Cirillo, S.~Esposito, F.~Iocco, G.~Mangano, G.~Miele et~al.,
  \emph{{PArthENoPE: Public Algorithm Evaluating the Nucleosynthesis of
  Primordial Elements}},
  \href{https://doi.org/10.1016/j.cpc.2008.02.015}{\emph{Comput. Phys. Commun.}
  {\bfseries 178} (2008) 956}
  [\href{https://arxiv.org/abs/0705.0290}{{\ttfamily 0705.0290}}].

\bibitem{Cyburt:2008up}
R.~H. Cyburt and B.~Davids, \emph{{Evaluation of Modern He-3(alpha,gamma)Be-7
  Data}}, \href{https://doi.org/10.1103/PhysRevC.78.064614}{\emph{Phys. Rev. C}
  {\bfseries 78} (2008) 064614}
  [\href{https://arxiv.org/abs/0809.3240}{{\ttfamily 0809.3240}}].

\bibitem{Kawasaki_2018}
M.~Kawasaki, K.~Kohri, T.~Moroi and Y.~Takaesu, \emph{{Revisiting Big-Bang
  Nucleosynthesis Constraints on Long-Lived Decaying Particles}},
  \href{https://doi.org/10.1103/PhysRevD.97.023502}{\emph{Phys. Rev.}
  {\bfseries D97} (2018) 023502}
  [\href{https://arxiv.org/abs/1709.01211}{{\ttfamily 1709.01211}}].

\bibitem{Hannestad_2004}
S.~Hannestad, \emph{{What is the lowest possible reheating temperature?}},
  \href{https://doi.org/10.1103/PhysRevD.70.043506}{\emph{Phys. Rev.}
  {\bfseries D70} (2004) 043506}
  [\href{https://arxiv.org/abs/astro-ph/0403291}{{\ttfamily
  astro-ph/0403291}}].

\bibitem{Reno_1988}
M.~Reno and D.~Seckel, \emph{{Primordial Nucleosynthesis: The Effects of
  Injecting Hadrons}},
  \href{https://doi.org/10.1103/PhysRevD.37.3441}{\emph{Phys. Rev.} {\bfseries
  D37} (1988) 3441}.

\bibitem{Kohri:2001jx}
K.~Kohri, \emph{{Primordial nucleosynthesis and hadronic decay of a massive
  particle with a relatively short lifetime}},
  \href{https://doi.org/10.1103/PhysRevD.64.043515}{\emph{Phys. Rev.}
  {\bfseries D64} (2001) 043515}
  [\href{https://arxiv.org/abs/astro-ph/0103411}{{\ttfamily
  astro-ph/0103411}}].

\bibitem{Kawasaki:2004qu}
M.~Kawasaki, K.~Kohri and T.~Moroi, \emph{{Big-Bang nucleosynthesis and
  hadronic decay of long-lived massive particles}},
  \href{https://doi.org/10.1103/PhysRevD.71.083502}{\emph{Phys. Rev.}
  {\bfseries D71} (2005) 083502}
  [\href{https://arxiv.org/abs/astro-ph/0408426}{{\ttfamily
  astro-ph/0408426}}].

\bibitem{Sjostrand_2006}
T.~Sjostrand, S.~Mrenna and P.~Z. Skands, \emph{{PYTHIA 6.4 Physics and
  Manual}}, \href{https://doi.org/10.1088/1126-6708/2006/05/026}{\emph{JHEP}
  {\bfseries 05} (2006) 026}
  [\href{https://arxiv.org/abs/hep-ph/0603175}{{\ttfamily hep-ph/0603175}}].

\bibitem{Sjostrand_2007}
T.~Sjostrand, S.~Mrenna and P.~Z. Skands, \emph{{A Brief Introduction to PYTHIA
  8.1}}, \href{https://doi.org/10.1016/j.cpc.2008.01.036}{\emph{Comput. Phys.
  Commun.} {\bfseries 178} (2008) 852}
  [\href{https://arxiv.org/abs/0710.3820}{{\ttfamily 0710.3820}}].

\bibitem{An_2016}
{\scshape Daya Bay} collaboration, \emph{{Improved Search for a Light Sterile
  Neutrino with the Full Configuration of the Daya Bay Experiment}},
  \href{https://doi.org/10.1103/PhysRevLett.117.151802}{\emph{Phys. Rev. Lett.}
  {\bfseries 117} (2016) 151802}
  [\href{https://arxiv.org/abs/1607.01174}{{\ttfamily 1607.01174}}].

\bibitem{Declais_1994}
Y.~Declais et~al., \emph{{Search for neutrino oscillations at 15-meters,
  40-meters, and 95-meters from a nuclear power reactor at Bugey}},
  \href{https://doi.org/10.1016/0550-3213(94)00513-E}{\emph{Nucl. Phys.}
  {\bfseries B434} (1995) 503}.

\bibitem{Ashenfelter_2018}
{\scshape PROSPECT} collaboration, \emph{{First search for short-baseline
  neutrino oscillations at HFIR with PROSPECT}},
  \href{https://doi.org/10.1103/PhysRevLett.121.251802}{\emph{Phys. Rev. Lett.}
  {\bfseries 121} (2018) 251802}
  [\href{https://arxiv.org/abs/1806.02784}{{\ttfamily 1806.02784}}].

\bibitem{Megas_2019}
F.~Megas, \emph{eV-scale Sterile Neutrino Investigation with the First Tritium
  KATRIN Data}, Ph.D. thesis, MSc Thesis, Munich University, 2019.

\bibitem{Betti_2019}
{\scshape PTOLEMY} collaboration, \emph{{Neutrino physics with the PTOLEMY
  project: active neutrino properties and the light sterile case}},
  \href{https://doi.org/10.1088/1475-7516/2019/07/047}{\emph{JCAP} {\bfseries
  1907} (2019) 047} [\href{https://arxiv.org/abs/1902.05508}{{\ttfamily
  1902.05508}}].

\bibitem{Gariazzo_2017}
S.~Gariazzo, C.~Giunti, M.~Laveder and Y.~F. Li, \emph{{Updated Global 3+1
  Analysis of Short-BaseLine Neutrino Oscillations}},
  \href{https://doi.org/10.1007/JHEP06(2017)135}{\emph{JHEP} {\bfseries 06}
  (2017) 135} [\href{https://arxiv.org/abs/1703.00860}{{\ttfamily
  1703.00860}}].

\bibitem{Gariazzo_2015}
S.~Gariazzo, C.~Giunti, M.~Laveder, Y.~F. Li and E.~M. Zavanin, \emph{{Light
  sterile neutrinos}},
  \href{https://doi.org/10.1088/0954-3899/43/3/033001}{\emph{J. Phys.}
  {\bfseries G43} (2016) 033001}
  [\href{https://arxiv.org/abs/1507.08204}{{\ttfamily 1507.08204}}].

\end{thebibliography}\endgroup
